\documentclass[traditabstract]{aa}
\pdfoutput=1
\usepackage{graphicx}
\usepackage{amsmath,amsfonts,amssymb,tabu}
\usepackage{txfonts}
\usepackage[breaklinks,colorlinks,citecolor=blue,pdfa=true]{hyperref}
\usepackage{color}
\usepackage{fixltx2e}
\usepackage{natbib,twoopt}
\usepackage{url}
\usepackage{multirow}
\usepackage{epsf}
\usepackage{epsfig}
\usepackage{longtable}
\usepackage{float}
\usepackage{subfig}
\usepackage{caption}
\usepackage{ifthen}
\usepackage[T1]{fontenc}
\usepackage{lmodern}
\usepackage{ifxetex,ifluatex}
\usepackage{latexsym}
\usepackage{pdfpages}
\usepackage{dblfloatfix}
\usepackage{morefloats}
\usepackage{caption}
\bibpunct{(}{)}{;}{a}{}{,} 
\makeatletter
\newcommandtwoopt{\citeads}[3][][]{\href{http://adsabs.harvard.edu/abs/#3}%
{\def\hyper@linkstart##1##2{}%
\let\hyper@linkend\@empty\citealp[#1][#2]{#3}}}
\newcommandtwoopt{\citepads}[3][][]{\href{http://adsabs.harvard.edu/abs/#3}%
{\def\hyper@linkstart##1##2{}%
\let\hyper@linkend\@empty\citep[#1][#2]{#3}}}
\newcommandtwoopt{\citetads}[3][][]{\href{http://adsabs.harvard.edu/abs/#3}%
{\def\hyper@linkstart##1##2{}%
\let\hyper@linkend\@empty\citet[#1][#2]{#3}}}
\newcommandtwoopt{\citeyearads}[3][][]%
{\href{http://adsabs.harvard.edu/abs/#3}
{\def\hyper@linkstart##1##2{}%
\let\hyper@linkend\@empty\citeyear[#1][#2]{#3}}}
\makeatother

\usepackage{graphicx}
\usepackage{txfonts}

\begin{document}

   \title{Giant radio galaxies in the LOFAR Two-metre Sky Survey-I}
%
   \titlerunning{GRGs in LoTSS DR1}
%
%

\author{P. Dabhade,\inst{1,2}\thanks{E-mail: pratik@strw.leidenuniv.nl} 
\and H. J. A. R\"{o}ttgering\inst{1}
\and J . Bagchi\inst{2}
\and T. W. Shimwell\inst{1,3}
\and M. J. Hardcastle\inst{4} 
\and S. Sankhyayan\inst{5}
\and R. Morganti \inst{3,6}
\and M. Jamrozy\inst{7}
\and A. Shulevski\inst{8}
\and K. J. Duncan\inst{1}
}
  
\authorrunning{Dabhade et al}
\institute{$^{1}$Leiden Observatory, Leiden University, P.O. Box 9513, NL-2300 RA, Leiden, The Netherlands \\
$^{2}$Inter University Centre for Astronomy and Astrophysics (IUCAA), Pune 411007, India.\\ 
$^{3}$ASTRON, The Netherlands Institute for Radio Astronomy, Postbus 2, 7990 AA, Dwingeloo, The Netherlands \\
$^{4}$Centre for Astrophysics Research, School of Physics, Astronomy and Mathematics, University of Hertfordshire, College Lane, Hatfield AL10 9AB, UK \\
$^{5}$Indian Institute of Science Education and Research (IISER), Dr. Homi Bhabha Road, Pashan, Pune 411008, India \\
$^{6}$Kapteyn Astronomical Institute, University of Groningen, PO Box 800, 9700 AV, Groningen, The Netherlands \\
$^{7}$Astronomical Observatory, Jagiellonian University, ul. Orla 171, 30–244 Kraków, Poland \\
$^{8}$Anton Pannekoek Institute for Astronomy, University of Amsterdam, Postbus 94249, 1090 GE Amsterdam, The Netherlands}

 \date{\today} 
 
  \abstract
   {Giant radio galaxies (GRGs) are a subclass of radio galaxies which have grown to megaparsec scales. GRGs are much rarer than normal sized radio galaxies (< 0.7 Mpc) and the reason for their 
gigantic sizes is still debated. Here, we report the biggest sample of GRGs identified to date. These objects were found in the LOFAR Two-metre Sky Survey (LoTSS) first data release images, which 
cover a 424 deg$^{2}$ region. Of the 239 GRGs found, 225 are new discoveries. The GRGs in our sample have sizes ranging from 0.7 to 3.5 Mpc and have redshifts ($z$) between 0.1 and 2.3. 
Seven GRGs have sizes above 2 Mpc 
and  one has a size of $\sim$ 3.5 Mpc. The sample contains 40 GRGs hosted by spectroscopically confirmed quasars. 
Here, we present the search techniques employed and the resulting catalogue 
of the newly discovered large sample of GRGs along with their radio properties. We, here also show for the first 
time that the spectral index of GRGs is similar to that of normal sized radio galaxies, indicating that most of the GRG population is not dead or is not like remnant type radio galaxy. We find 
20/239 GRGs in our sample are located at the centres of clusters and present our analysis on their cluster environment and radio morphology.}

   \keywords{galaxies: jets -- galaxies: active -- radio continuum: galaxies  -- quasars: general}

   \maketitle
%

\section{Introduction}
\label{sec:intro}
A radio galaxy normally contains a radio core, jets and lobes powered by an active galactic nucleus (AGN). A radio galaxy that has grown to Mpc scales is traditionally defined as a giant radio galaxy 
(GRG) 
\citep{willis74, ishwar2000}. Here, the total size is defined as the largest angular separation between the end of the two radio lobes. This subclass of radio galaxies is among the largest single structures 
known in the universe along with the cluster radio relics \citep{huub-a3667,bagchisci,vanweeren11}. 
Born in the active nucleus of a galaxy or a quasar, radio galaxies/quasars eject collimated and bipolar relativistic jets \citep{1969Natur.223..690L,1984RvMP...56..255B}. The driving engine for these 
jets is  an accreting  super massive black hole (SMBH) with a typical mass of $10^{8} - 10^{10}$~$M_{\odot}$. SMBHs that drive powerful jets reside in elliptical galaxies and only a handful are 
found in spiral galaxies \citep{2011MNRAS.417L..36H,2014ApJ...788..174B}. 

Morphologically, radio galaxies have been historically divided into two classes, Fanaroff-Riley type I (FR-I) and Fanaroff-Riley type II (FR-II). The lower radio luminosity FR-I sources have their 
brightest regions  closer to the nucleus and their jets fade with distance from core.
For the more powerful Fanaroff-Riley type II (FR-II) radio galaxies \citep{FR74}, the jet remains relativistic all the way from the central AGN to the hotspots in lobes. 

For all radio galaxies the ejection of a collimated jet is dependent on the availability of fuel.
Assuming GRGs grow to an enormous size due to a prolonged period activity, this would require either an unusually large reservoir of fuel or a very efficient jet formation mechanism.

Radio galaxies were first discovered about six decades ago \citep{1953Natur.172..996J} and since then hundreds of thousands of radio galaxies have been found. In contrast to the large number of radio 
galaxies, only  $\sim$ 350 (\citet{2017MNRAS.469.2886D,grscat} $\&$ references therein) or so GRGs have been found of which only a small fraction  has	 been studied in detail. These 
giants, when associated with quasars as their AGN, are called giant radio quasars (GRQs) and only around 70 GRQs are known so far \citep{grscat}.The term `GRQ' is used 
here to emphasize the fact that the GRG has quasar as the powering AGN.

The hypotheses proposed (which are not mutually exclusive) to explain the enormous sizes of GRGs include:
\begin{enumerate}
\item GRGs possess exceptionally powerful radio jets when compared to normal radio galaxies and these provide the necessary thrust to reach Mpc scales \citep{wita-grg-agn}.
\item GRGs are very old radio galaxies  and have had sufficient time to expand over large distances \citep{ravilongages}.
\item GRGs grow in low density environments, \citep{mack-ages,Malarecki-grg-env,Saripalli-env1} enabling them to grow comparatively fast.
\end{enumerate}

None of the above hypotheses have been tested using large uniform samples of GRGs. In smaller samples, contradictory results have 
been found. For example, 
\citet{mack-ages} found evidence that the ages of GRGs in their sample are similar to that of normal sized radio galaxies which is contradictory to the above mentioned second point. Also, point 
three has been contradicted by findings of  \citealt{komberg09,2017MNRAS.469.2886D}, where they have reported a number of GRGs to be located in cluster environments.

The exceptionally large lobes of  GRGs makes them excellent laboratories for studying the evolution of the particle and magnetic field energy density, acceleration of high energy 
cosmic rays and can also be used to probe large scale environments \citep{kronberg04,safouris09,Malarecki-grg-env,isobe15}. 

Until the mid 1990s, GRGs were mostly discovered serendipitously. Only after the advent of deep and large sky radio surveys like  Faint Images of the Radio Sky at Twenty-cm (FIRST) \citep{1995ApJ...450..559B},  NRAO VLA Sky Survey (NVSS) \citep{1998AJ....115.1693C},  Westerbork Northern Sky Survey (WENSS) \citep{wenss97} and Sydney University Molonglo Sky Survey (SUMSS) \citep{sumss99}  were systematic searches for GRGs carried out. 
\citet{lara01grgs}, \citet{Machalski01grgs} and \citet{2017MNRAS.469.2886D} searched the higher frequency survey such as  NVSS (1400 MHz) for GRGs, and \citet{Saripalli-sumssgrgs} used SUMSS, whereas  
\citet{SchoenmakersGRGs01} used lower frequency surveys such as WENSS (327 MHz). 

The lobes of the GRGs have steep spectral indices and hence are bright at low radio frequencies. 
\citet{cotter96grg} made a sample of GRGs using the 151 MHz   7C survey \citep{7csurvey}. The 7C survey has a low resolution of 70 $\times$ 70 cosec ($\delta$) arcsec$^{2}$ 
and a noise level  $\sim$ 15 mJy beam$^{-1}$ (1 $\sigma$) and as a consequence contained only a few GRGs. The WENSS has better resolution 
(54 $\times$ 54 cosec ($\delta$) arcsec$^{2}$) and better sensitivity (RMS noise (1 $\sigma$) $\sim$ 3 mJy beam$^{-1}$) and also covers a somewhat larger area $\sim$ 8100 deg$^{2}$  ( 7C survey $\sim$ 5580 deg$^{2}$). This enabled \citet{SchoenmakersGRGs01} to compile a large sample of 47 GRGs from the WENSS.

In recent years, four large low frequency surveys have been carried out, namely :
\begin{itemize}
\item 119-158 MHz Multifrequency Snapshot Sky Survey (MSSS) \citep{msss}.
\item 150 MHz TIFR GMRT SKY SURVEY- Alternative data release-1 (TGSS-ADR1) \citep{intema-tgss-17}.
\item 72-231 MHz GaLactic and Extragalactic All-sky Murchison Widefield Array (GLEAM) survey \citep{gleam-walker17}.
\item 120 - 168 MHz LOFAR Two-metre Sky Survey (LoTSS) \citep{lotss-pdr,shimwell-lotss}.
\end{itemize}

In the past 20 years, large surveys have also been carried out at optical wavelengths. These surveys include the Sloan Digital Sky Survey (SDSS) \citep{sdssyork}, the 2 degree 
Field Galaxy Redshift Survey (2dFGRS) \citep{2dfcolles}, the 2MASS Redshift Survey (2MRS) \citep{huchra2mass}, the 6 degree Field Galaxy Survey (6dFGS) \citep{6dfjones} and most recently the deep photometric survey called Panoramic Survey Telescope and 
Rapid Response System (Pan-STARRS; \citealt{kaiser02pan,kaiser10pan,panstars-chambers}). The data from these surveys has allowed the  identification of many new GRGs as shown in \citet{2017MNRAS.469.2886D}.

High sensitivity to low surface brightness features and high spatial resolution to decipher the morphologies are key requirements in identifying GRGs. LoTSS provides combination of both these properties for the first time and hence, combining it with the SDSS/Pan-STARRS  optical surveys, we use it to search for new GRGs in order to form a statistically significant sample.
Our study of GRGs will be presented in 2 papers:

\begin{enumerate}
\item Paper I (this paper) reports the methodology used for the systematic search scheme implemented for the discovery of new GRGs/GRQs from the LoTSS and presents the sample's radio properties.

\item Paper II will focus on studying the host AGN and galaxy properties of the GRGs/GRQs sample and comparing them with another sample (also from LoTSS) of normal sized radio galaxies (NRGs) matched in redshift and optical/radio luminosity to the GRG sample.

\end{enumerate} 

Throughout the paper, we adopt the flat $\Lambda$CDM cosmological model based on the latest Planck results ($H_o$ = 67.8 km $s^{-1}$ $Mpc^{-1}$, $\Omega_m$ =0.308) \citep{2016A&A...594A..13P}, which 
gives a scale of 4.6 kpc/\arcsec  for the redshift of 0.3. In this paper, radio galaxies with projected linear sizes $\ge$ 0.7 Mpc, computed using the above mentioned cosmological 
parameters are called GRGs.
All images are in the J2000 coordinate system. We use the convention $S_{\nu}\propto \nu^{-\alpha}$, where $S_{\nu}$ is flux at frequency $\nu$ and $\alpha$ is the spectral index.

\begin{figure*}
\centering
\includegraphics[width=0.75\textwidth]{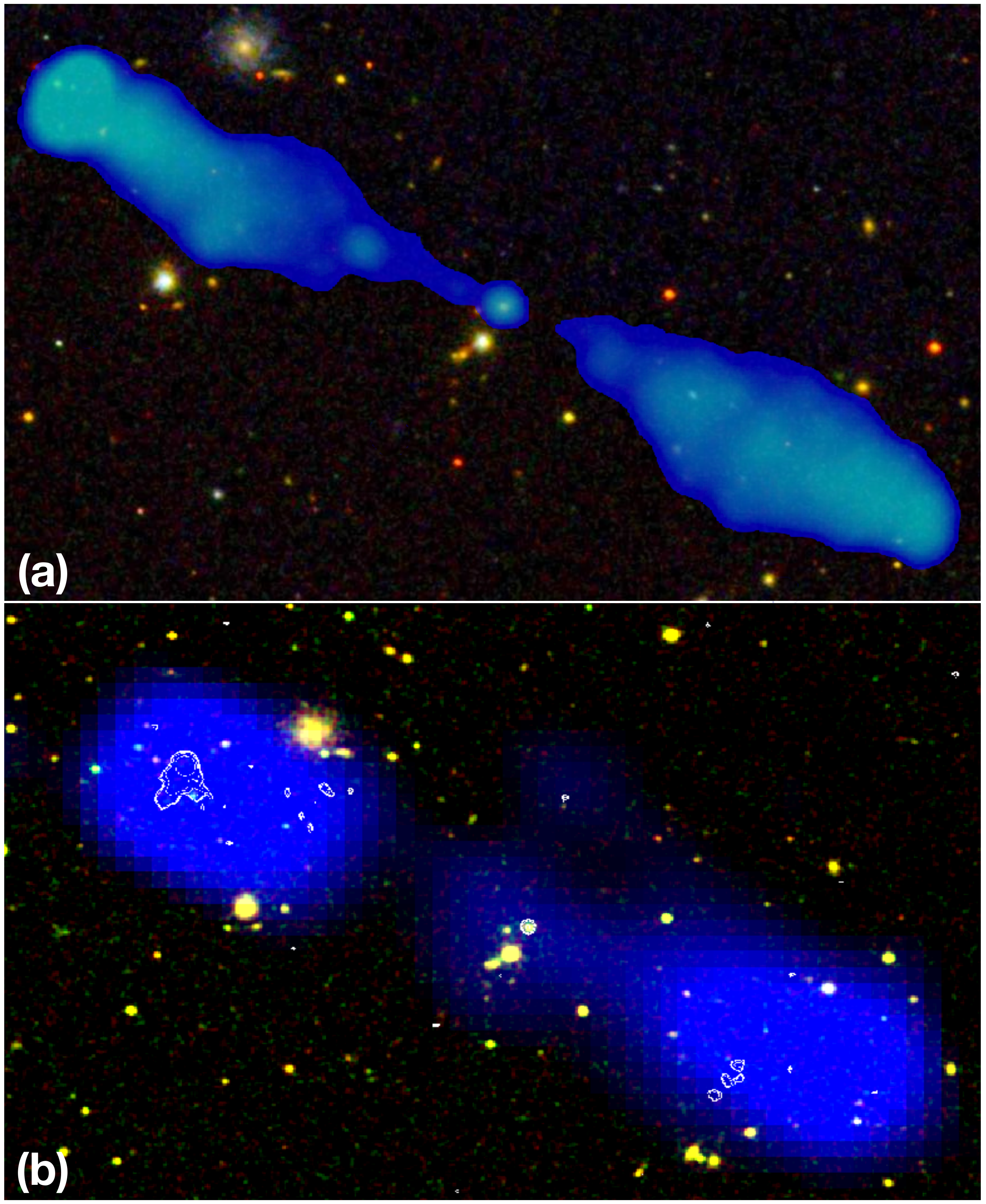} 
\caption{(a)A colour composite image of 1.86 Mpc long  GRG J105817.90+514017.70 made using LoTSS-DR1 144 MHz radio and optical SDSS image. (b) A colour composite image of the same 
object with optical and radio overlay, where blue represents 1400 MHz NVSS image with 45\arcsec resolution and white contours from 1400 MHz FIRST survey having 5\arcsec resolution
superimposed on optical SDSS image. The LoTSS 144 MHz 6$\arcsec$ resolution  image shown in top
panel 
clearly resolves the core and jet and also highlights the diffuse parts of the lobes, which is missed by the FIRST and unresolved in NVSS.}
\label{fig:color-composite}
\end{figure*}
 
\section{Identifying new GRGs in LoTSS} \label{sec:identity}
\subsection{The LoTSS first data release}
LoTSS is a 120-168 MHz survey that is being conducted with the high-band antennas  (HBA) of LOFAR and will eventually cover the whole northern sky. 
\citet{hardcastle-hatlas} have already demonstrated the potential of LoTSS deep observations for discovering GRGs and found seven in the \textit{Herschel} ATLAS North Galactic Pole survey area 
(142 deg$^{2}$).
Here, we focus on the LoTSS first data release (LoTSS DR1; \citealt{shimwell-lotss}).
The LoTSS DR1 spans (J2000.0 epoch) right ascension 10h45m to 15h30m and declination $45^\circ 00\arcmin$ to $57^\circ 00\arcmin$ (HETDEX:\textit{Hobby-Eberly} Telescope Dark Energy Experiment Spring 
field region) covering an area of 424 deg$^{2}$ with a median noise level across the mosaic of 71 $\mu$Jy beam$^{-1}$ and $\sim$6\arcsec  resolution. In Fig.\ \ref{fig:color-composite}, we 
see a comparison between the LoTSS and other radio surveys like NVSS and FIRST for a GRG from our sample. The top image (a) is an optical-radio overlay with blue colour indicating  
LoTSS low frequency 6\arcsec resolution map on the optical SDSS tri-colour image. In the bottom image (b), the same source is shown as it is observed in NVSS and FIRST. NVSS, though highly sensitive 
to large 
scale diffuse emission, fails to reveal the finer details across the source and cannot properly resolve the core due to it coarser resolution of 45\arcsec. FIRST survey on the other hand has 
high resolution, which manages to resolve the core and hence help in identifying the host AGN/galaxy but misses out on almost all the diffuse emission of the lobes and hence it alone cannot be 
used to identify RGs. The LoTSS data provides with both high resolution as well as high sensitivity and does not resolve out structures revealing finer details of the emission of large scale jets. 
Both images (Fig.\ \ref{fig:color-composite} and source 7 of Fig.\ \ref{fig:lotsshighres1}) show clearly the radio core and jets feeding the giant radio lobes and the hotspots. This clearly 
illustrates the excellence of LoTSS and its great potential in unveiling interesting sources.

Using the LoTSS DR1 radio data and optical-infrared data, a Value Added Catalogue\footnote{\url{https://www.lofar-surveys.org/}} (VAC) of 318520 radio sources has been created  
(\citet{williamslotss}; DR1-II). The host galaxies/quasars were 
identified using the Pan-STARRS and Wide-field Infrared Survey Explorer (WISE; \citep{wright10-wise}). The Pan-STARRS-AllWISE catalogue was cross matched with LoTSS survey using a likelihood 
ratio method. Furthermore, human visual inspection was used for the final classification of complex radio sources using the LOFAR Galaxy Zoo (LGZ), the details of which are given in 
\citet{williamslotss} (DR1-II). The  photometric redshift and rest-frame colour estimates for all hosts (galaxies/quasars) of the matched radio sources are presented in \citet{duncanlotss} (DR1-III). The VAC lists the radio properties, identification methods and optical properties where available.

\subsection{Semi-automated search for GRGs}
\begin{figure*}
\centering
\includegraphics[scale=0.3,width=\textwidth]{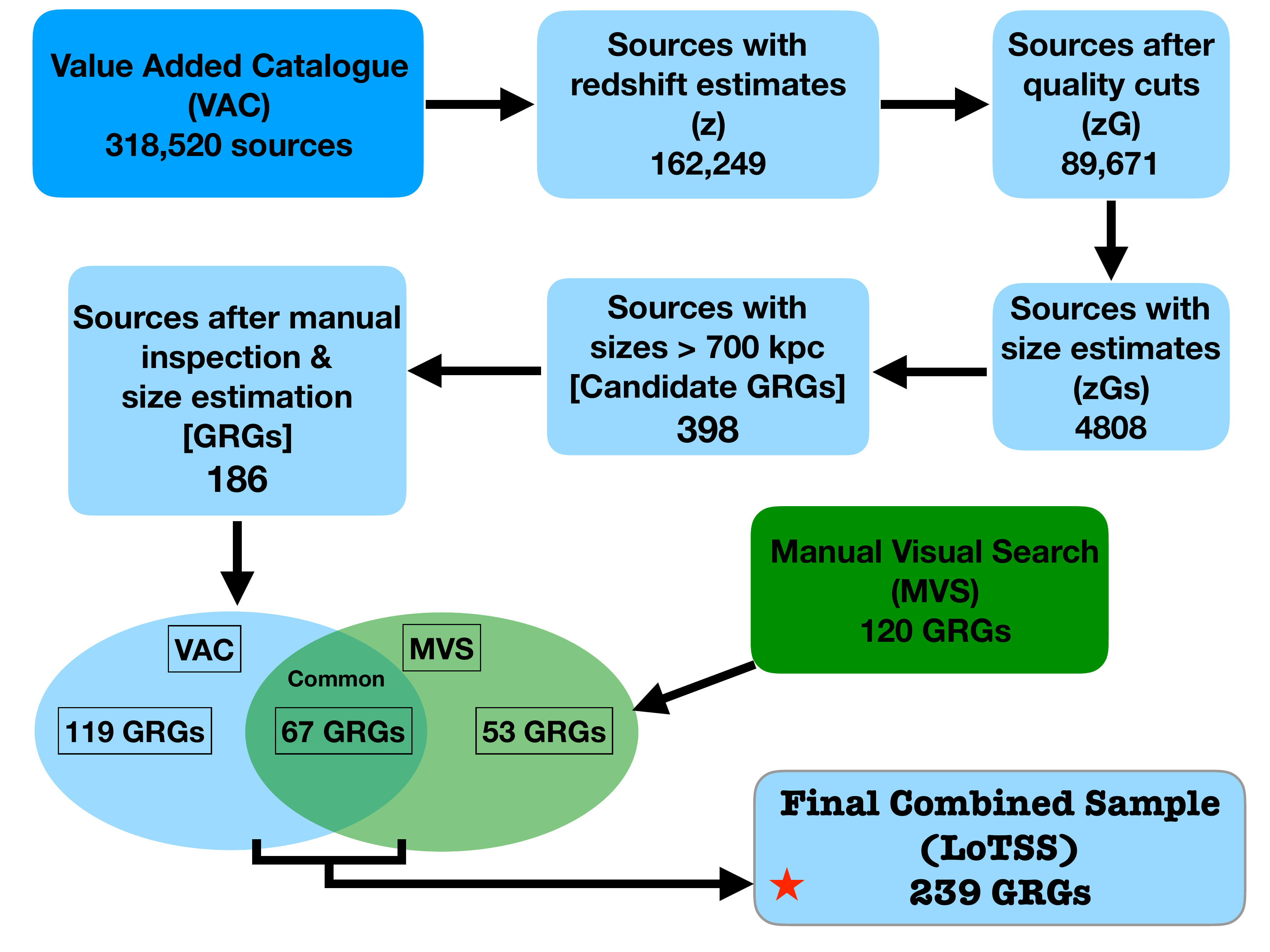} 
\caption{The above figure shows schematics for finding GRGs from LoTSS DR1 using VAC and MVS in steps. More details are presented in Sect.\ \ref{sec:identity}.}
\label{fig:flowchart}
\end{figure*}

The methodology of identifying GRGs and forming the final catalogue is presented in flow chart as seen in Fig.\ \ref{fig:flowchart} and is described below:
\begin{enumerate}
\item The VAC of 318520 radio sources was at first refined based on the point source completeness, which is 90\% at an integrated  flux density of 0.5 mJy for LoTSS DR1 \citep{shimwell-lotss}. We 
apply a flux density cut at this level. This results in the total number of sources reducing to 239845.

\item Secondly, only objects with optical identification and redshift estimates were selected resulting in 162249 sources.

\item  The redshift information in the VAC is compiled mainly using the SDSS spectroscopic data. For the sources which do not have spectroscopic redshift information from the SDSS or any 
other spectroscopic survey, \citet{duncanlotss} have estimated photometric redshifts using multi-band photometry. For our work, we have imposed a further photometric quality cut on the estimated 
photometric redshifts. Only sources satisfying the condition $\Delta z/(1+z) < 0.1$ were selected, where $\Delta z$ is the half width of the 80\% credible interval. This results in reduction of 
the sample from 162249 sources to 
89671 sources.

\item \citet{williamslotss} inspected all the extended complex radio sources visually via the LGZ program and estimated the angular size for them. A 	semi-automated way was adopted for 
this work by \citet{williamslotss}. This resulted in angular size estimation of total 13222 sources in the VAC.

We find that, of the 89671 sources (which have reliable redshifts), only 4808 sources have angular size estimates by the above mentioned method (LGZ).  Or, in other words, 4808 sources have angular 
size estimates as well 
have passed the photometric quality cut (point 3).

\item For all the 4808 sources, we computed the projected linear size (kpc)  and only those that had an extension above 700 kpc were considered for further analysis. This resulted in a sample of 398 candidate GRGs.

\item  Further, the candidate GRGs were visually inspected by us to identify and remove those with uncertainty in the host, large asymmetry or  a high degree of bending or narrow angle tailed 
morphology.  
The angular size measurement was refined by taking the distance between the farthest points of the 3$\sigma$ contours of the source and the projected linear size was recomputed. We have 
measured the sizes of the sources using the LoTSS DR1- low resolution maps (20\arcsec) as it shows the entire structure of the GRGs and also does not miss any diffuse emission.
.
The final sample size of GRGs from the VAC is 186 GRGs.

\end{enumerate}

\subsection{Manual Visual Search from LoTSS DR1}
An independent manual visual search (MVS) was also carried out to search for  GRGs in LoTSS DR1 to look for additional GRGs that were missed in the semi-automated method employed on the VAC. In this 
approach, all mosaics from LoTSS DR1 were scanned for extended double lobed structures (candidate GRGs). The radio core of the candidate GRGs were searched in the optical band (SDSS and 
PAN-STARRS)  and IR band (WISE) for counterparts (host galaxies). We also used other radio surveys like FIRST, TGSS and WENSS for additional information and consistency checks. This method (MVS) 
yielded 53 additional GRGs. 
These 53 GRGs were missed  by the selection criteria because of the photometric redshift cuts we imposed or because the complex nature of the source structures lead them to being not fully 
characterised in the VAC.

\subsection{Final catalogue: VAC+MVS}
We combined both GRG samples (VAC and MVS) to form the final GRG catalogue of 239 GRGs (Table.~\ref{tab:maingrg}).
The final catalogue of 239 GRGs was cross matched with the GRG catalog of \citet{grscat}, which is a complete compendium of GRGs published till 2018 along with other literature search, and we find 13 
of our 239 GRGs to be already known (listed in $13^{th}$ column of 
Table~\ref{tab:maingrg}). The high-resolution 6\arcsec ~ LoTSS images at 144 MHz  of GRGs can be found in the appendix~\ref{sec:appendix} from Fig.\ \ref{fig:lotsshighres1} to Fig.\ 
\ref{fig:lotsshighres8}, where the white contours represent emission seen in the LoTSS low resolution 20$\arcsec$  map at 144 MHz. The position of host galaxy is marked with a white cross 
`+'. 
 
 
\section{Results : The LoTSS catalogue of GRGs}
\begin{figure}
\includegraphics[width=0.48\textwidth]{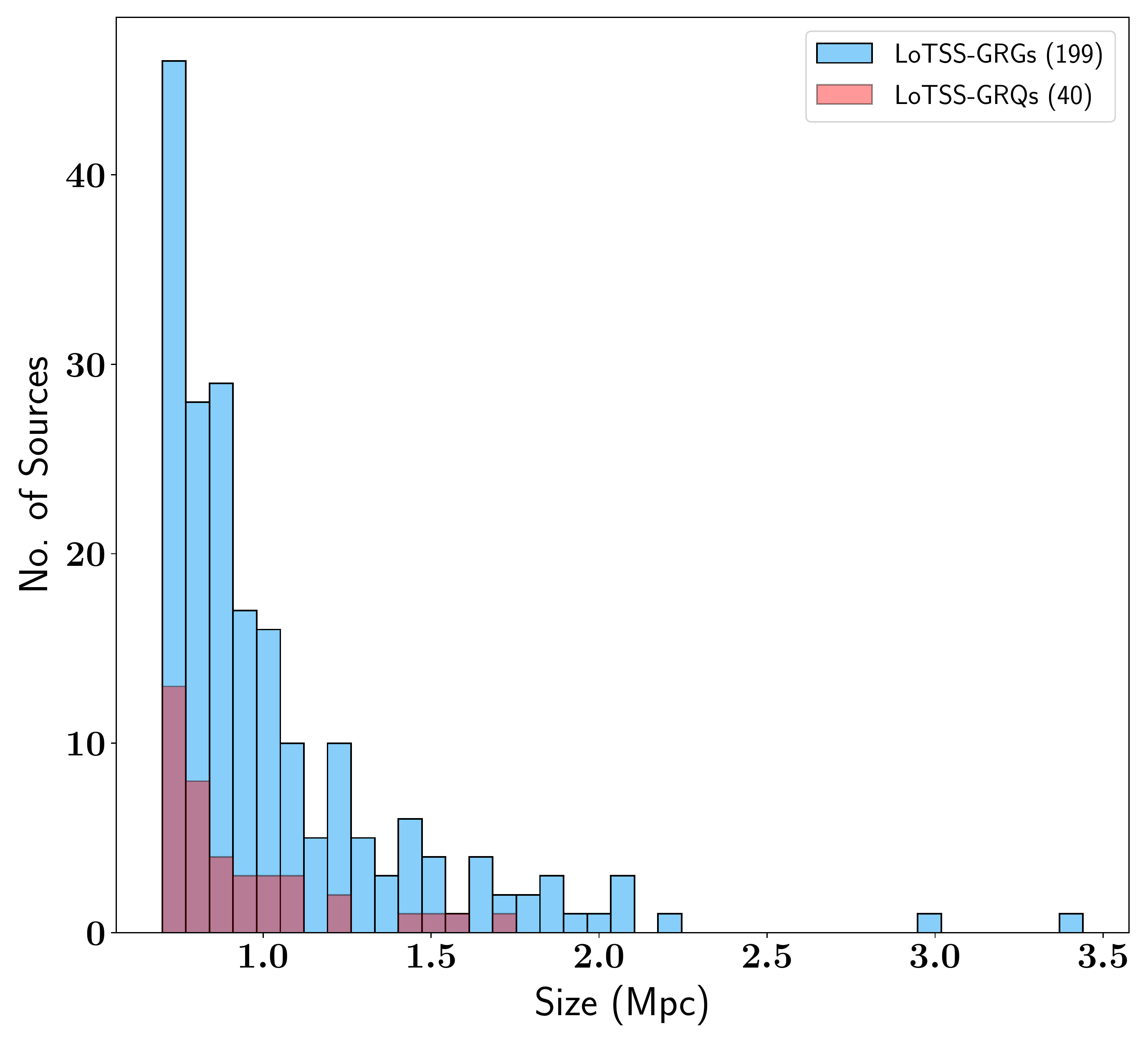} 
\caption{Histogram for sizes of GRGs and GRQs from the LoTSS sample.}
\label{fig:size-hist}
\end{figure}

Our search of LoTSS DR1 has enabled us to construct a catalog\footnote{Before the precise measurement of $H_o$, originally GRGs were defined as RGs with projected linear sizes $\ge$ 1 Mpc 
using $H_o$ = 50 km $s^{-1}$ $Mpc^{-1}$. If we convert the original definition with the latest precise measurement of $H_o$ = 67.8 km $s^{-1}$ $Mpc^{-1}$ with $\Omega_m$ = 0.308, then the lower limit 
of projected linear size of GRGs is $\sim$ 0.74 Mpc. In our sample, 28 GRGs have projected linear sizes between 0.7 Mpc to 0.74 Mpc.} of 239 GRGs. With the high sensitivity of LoTSS, we are able to 
detect GRGs as faint as $\sim$ 2.5 mJy in total flux at 144 MHz.
Using the available optical data and radio data, we have computed the radio powers, spectral indices and classified the morphological types for the sample of 239 GRGs (Table.~\ref{tab:maingrg}). The 
GRGs from our sample have sizes in the range of 0.7 Mpc to $\sim$ 3.5 Mpc (Fig.\ \ref{fig:size-hist}) with median size of 0.89 Mpc and mean size of 1.02 Mpc. GRGs with sizes greater than 2 Mpc are 
very rare within the GRG population and in our sample we find 7/239 GRGs having sizes $\geq$ 2 Mpc.

\subsection{Optical host properties}
\begin{figure}
\includegraphics[width=0.48\textwidth]{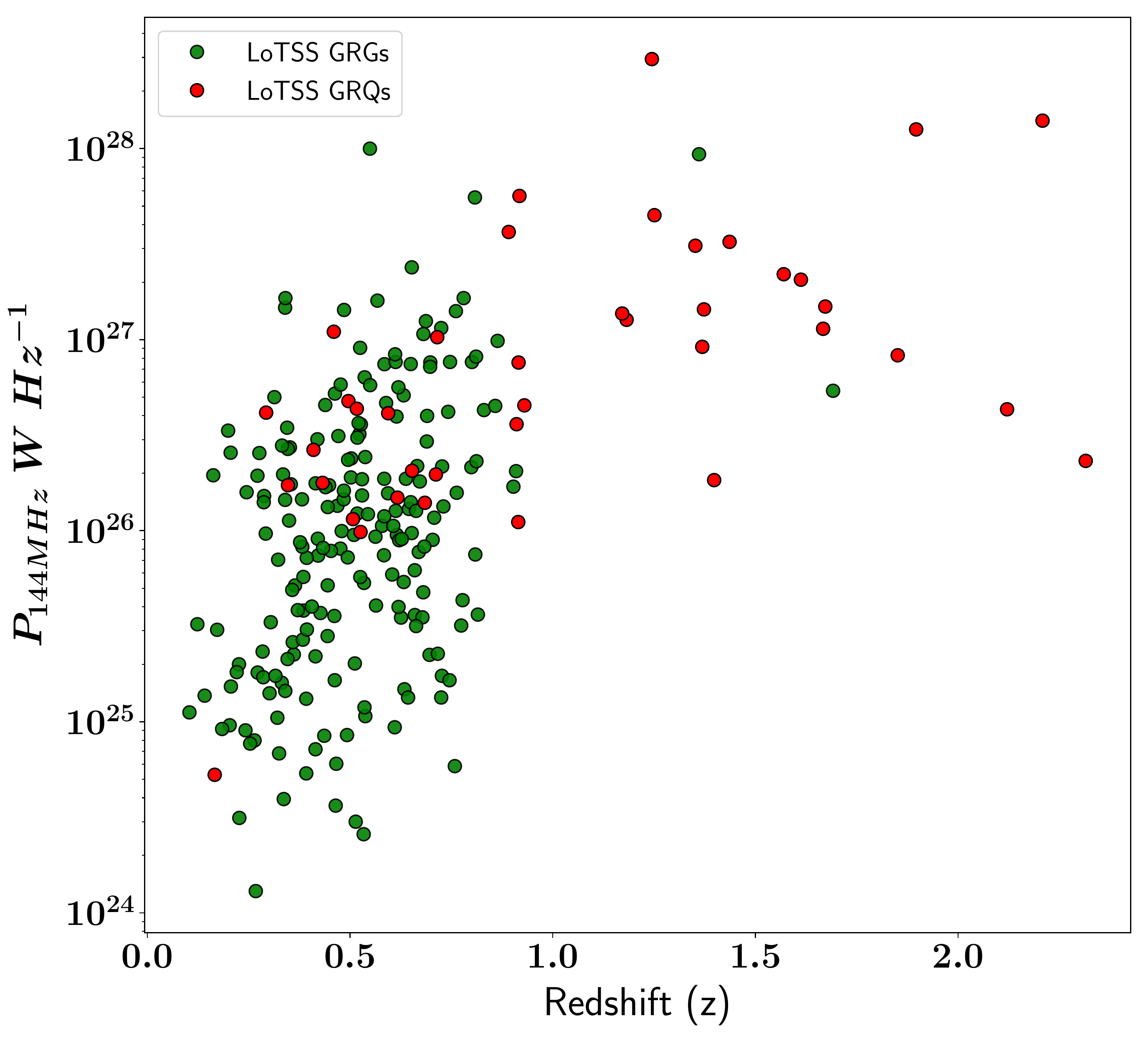} 
\caption{Radio power and redshift distribution for GRGs and GRQs from LoTSS. The two galaxies marked in green colour with $z$ $\ge$1 are quasar candidates as explaind in Sec.\ \ref{sec:grq}.}
\label{fig:pz-diag}
\end{figure}

The GRGs in our sample span a wide redshift  range from 0.1 to 2.3 (Fig.\ \ref{fig:pz-diag}). GRGs that are hosted by galaxies are not selected beyond $z$ $\sim$ 1 due to sensitivity limits of the 
optical surveys. We use the SDSS DR14 Quasar catalog \citep{parisqso} and SDSS to identify the GRGs hosted by quasars in our sample. 151/239 GRGs 
have optical spectroscopic redshifts from SDSS. Of the 151 GRGs with optical spectroscopic redshifts 40 are hosted by quasars (GRQs).
Interestingly, based on the available optical data (SDSS and Pan-STARRS) and the Galaxy Zoo catalog of spiral galaxies \citep{spiralgzoo}, we find that none of the GRGs is hosted by a spiral galaxy. 

Most of the redshifts of the hosts of GRGs were obtained from the VAC, which has made use of the spectroscopic data from the SDSS and has estimated photometric redshifts for the sources which do  
not have spectroscopic data in SDSS or in other literature. The uncertainties in the photometric redshifts are explained in section 3 (especially section 3.5) of \citealt{duncanlotss}.

\begin{figure}
\includegraphics[width=0.48\textwidth]{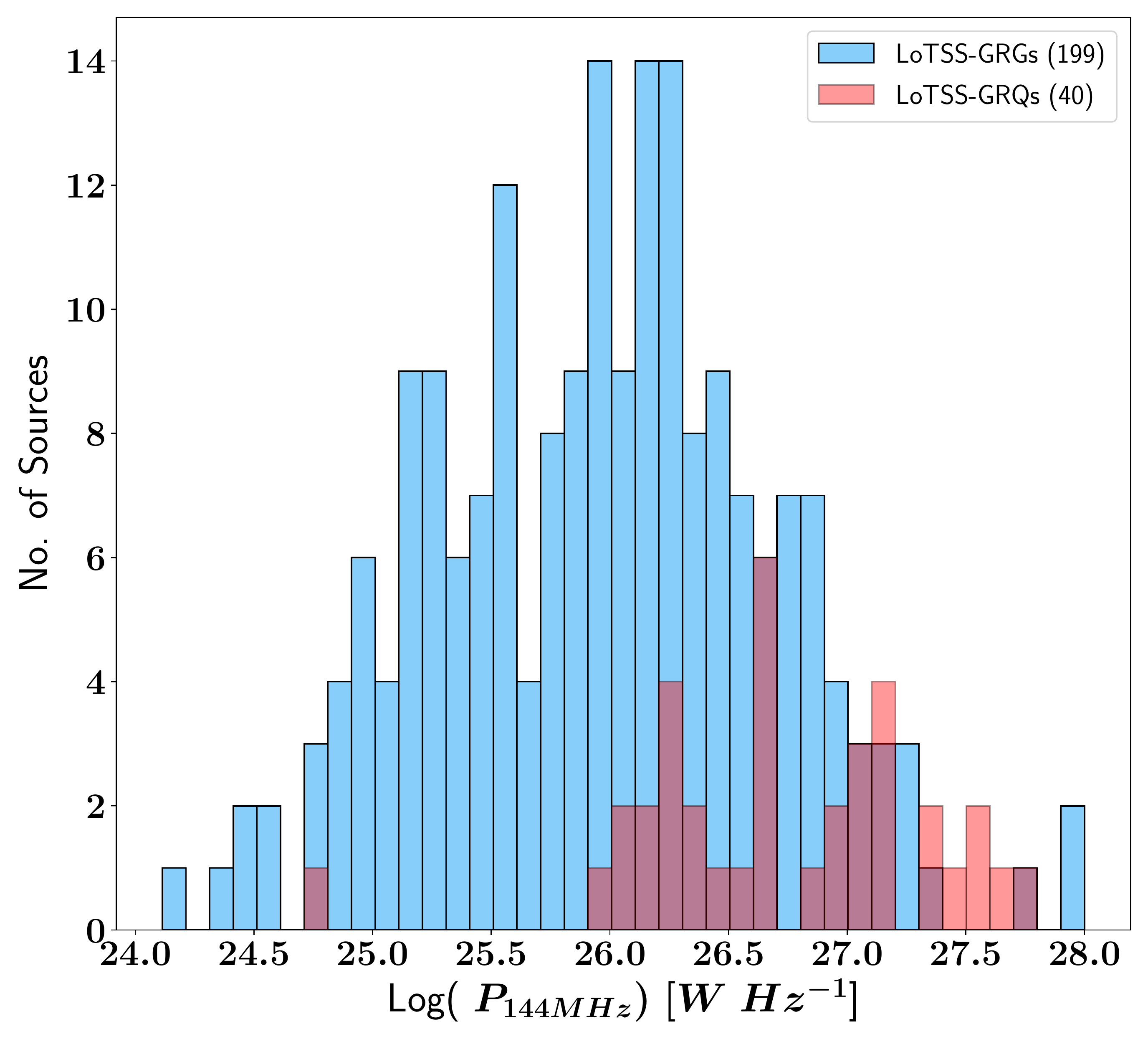}
\caption{Histogram of radio power at 144 MHz of GRGs and GRQs.}
\label{fig:histpower}
\end{figure}

\subsection{Spectral Index ($\alpha^{1400}_{144}$)}\label{sec:si}
The LoTSS DR1 provides maps of the radio sky (centered at 144 MHz) at two resolutions, high (6\arcsec) and low (20\arcsec) which enables us to identify compact cores, jets and hotspots as well as see the extent of diffuse emission. To compute the flux at 144 MHz, we use the 20\arcsec ~ low resolution maps of LoTSS DR 1 (column 8 of Table~\ref{tab:maingrg}).
Measurement in the flux errors is done taking a 20\% calibration error for LoTSS DR1 \citep{shimwell-lotss}.
By combining high frequency (1400 MHz) NVSS and low frequency (144 MHz) LoTSS, we have computed the integrated spectral index ($\alpha^{1400}_{144}$) for GRGs in our sample (column 10 of 
Table~\ref{tab:maingrg}). The NVSS radio map cutouts for GRGs were obtained from its server\footnote{\url{https://www.cv.nrao.edu/nvss/postage.shtml}}, from which we have measured the flux 
of the source. 

Following steps were adopted to obtain  spectral index:
\begin{itemize}
\item Convolve the LoTSS low resolution images (cutouts of GRGs from main mosaics) to the same resolution of NVSS (45\arcsec) and regrid the LoTSS images to match with NVSS.
\item Make automated masks (regions to extract flux ) using the package PyBDSF\footnote{\url{http://www.astron.nl/citt/pybdsf/}} (Python Blob Detection and Source Finder ) of \citet{pybdsf}.
\item The 45\arcsec ~convolved LoTSS maps were manually inspected for possible contamination (using the high resolution LoTSS maps (6$\arcsec$) and FIRST's 5$\arcsec$ maps) from other sources in the field and manual masks were made. The flux was obtained by considering only the region in manual masks from the automated masks.

\item Finally, using the fluxes obtained from NVSS and LoTSS,  integrated spectral index was computed for the whole source.
\end{itemize}

A total of 37/239 sources were contaminated by other nearby radio sources in the low resolution convolved LoTSS maps and NVSS maps, for these sources (marked with `-' in column 10 of 
Table~\ref{tab:maingrg}) a spectral index was not computed. For sources with no detection in NVSS, upper limits on the flux were computed and a spectral index limit was obtained (indicated with $<$ 
sign in column 10 of Table~\ref{tab:maingrg}).

\begin{figure}
\includegraphics[width=0.48\textwidth]{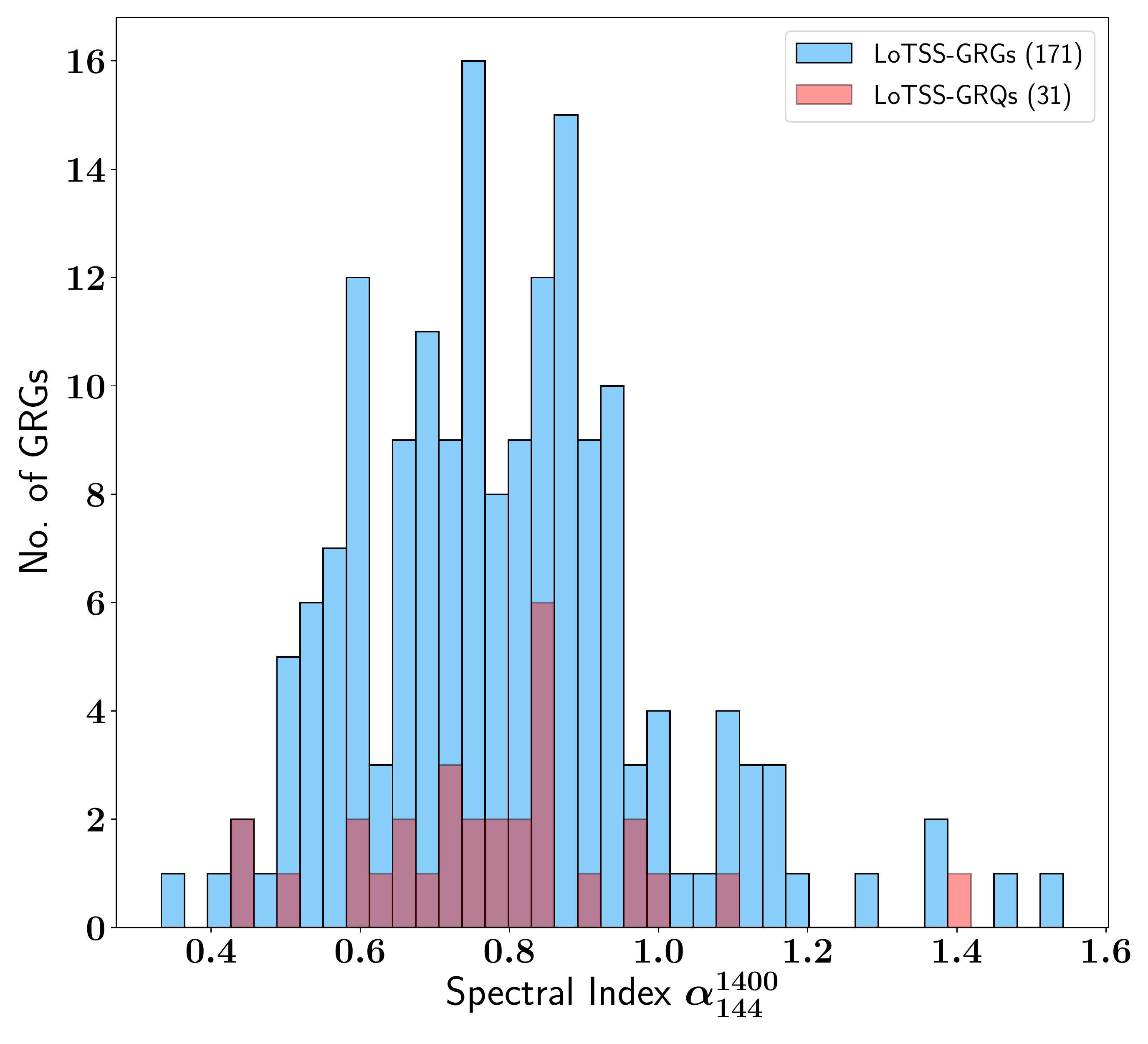}
\caption{Histogram of Spectral index ($\alpha^{1400}_{144}$) of GRGs and GRQs in LoTSS DR1 sample made using LoTSS and NVSS.}
\label{fig:histSI}
\end{figure}

Fig.\ \ref{fig:histSI} shows the spectral index ($\alpha^{1400}_{144}$) distribution of 171 GRGs  and 31 GRQs. The median and mean values for spectral index of  GRGs are 0.77 and 0.79, respectively. 
Similarly for GRQs, the median and mean values are 0.78 and 0.76, respectively. 
These mean spectral index values of GRGs and GRQs are similar to those of normal sized radio galaxies \citep{Oort88,Gruppioni97,Kapahi98,ishwar2010,Mahony16}. In the work of \citet{Oort88}, 
they have surveyed the Lynx field with the Westerbork Synthesis Radio Telescope (WSRT) at 325 MHz and 1400 MHz. \citet{Gruppioni97} surveyed the Marano field using the Australia Telescope Compact 
Array (ATCA) at 1.4 GHz and 2.4 GHz. \citet{Kapahi98} surveyed the Molonglo Radio Catalogue sources with the VLA at L and S bands.
\citet{ishwar2010} surveyed the LBDS-Lynx field (LBDS: Leiden-Berkeley  Deep  Survey) using the 150-MHz band of Giant Metrewave Radio Telescope and other archival data from 
GMRT at 610 MHz and 325 MHz along with data from other surveys like WENSS, FIRST and NVSS.   \citet{Mahony16} surveyed the Lockman Hole field using LOFAR 150 MHz and WSRT 1.4 GHz. All the above work 
obtained spectral index values in the range of $\sim$ 0.7 to 0.8 and therefore 0.75 value is often assumed as the average spectral index value for radio galaxies in absence of any multi-frequency 
observations.

The above mentioned result implies that RGs and GRGs do not differ in terms of their spectral index properties and most of the GRG population are active radio galaxies and not dead or 
remnant radio galaxies. The values for spectral index of GRGs given by us are integrated values for the whole source and not just one component of GRGs (like core or lobes). It is found that for some 
GRGs, the lobes come out to be more steeper than usual- which is obtained by multi-frequency (observations at 3 to 4 radio bands) observations of individual sources. There are examples (source number 
16,17 and 193) in our sample where the sources are core dominated (maximum flux observed in radio core) and it hence influences sources to have flatter (less than 0.5) overall spectral index.

\subsection{Notes on individual objects}
Below, we present brief notes on a selection of some of the most interesting objects in our sample.

\begin{figure}
\includegraphics[width=0.49\textwidth]{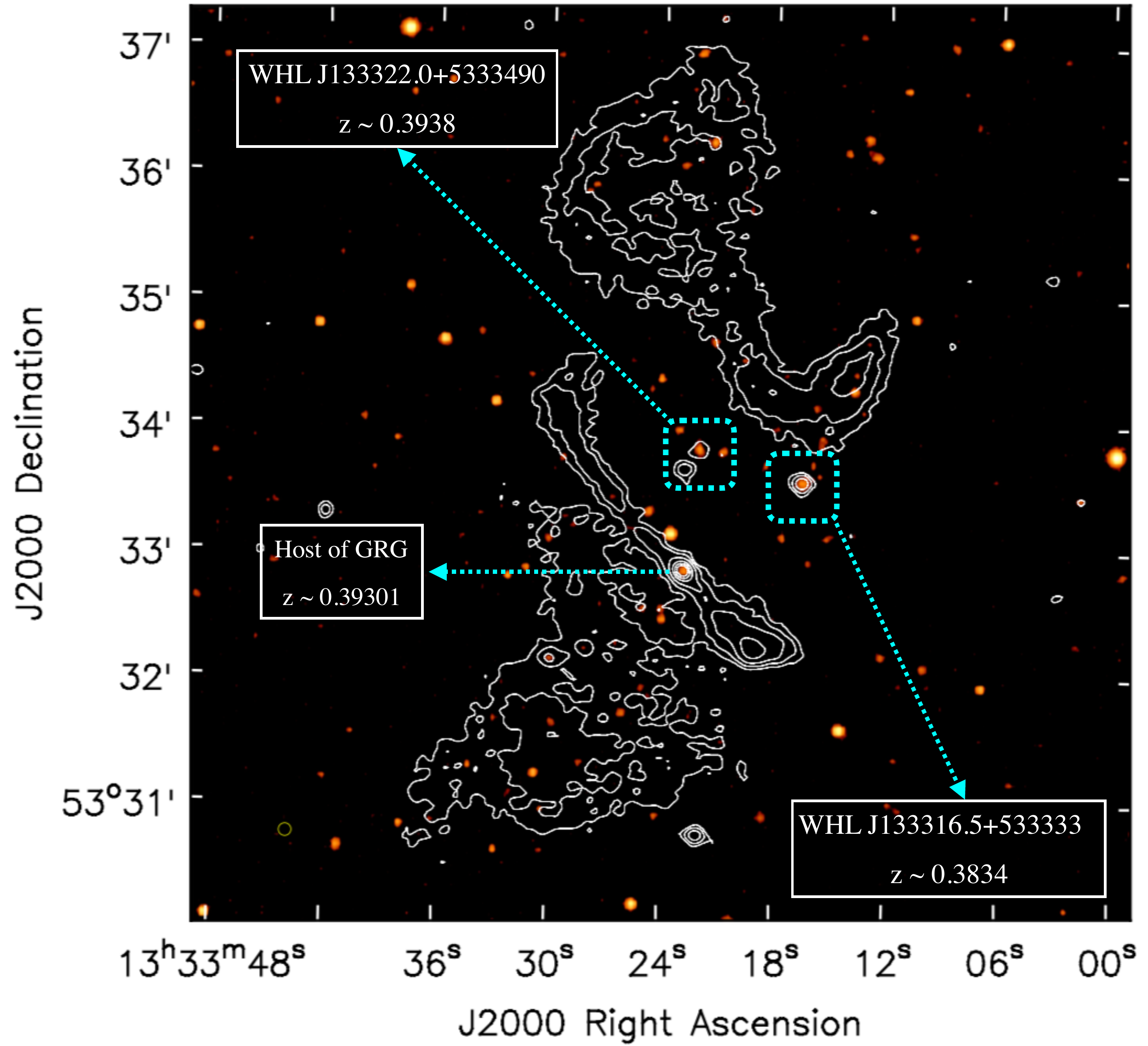}
\caption{The figure shows GRG J133322.79+533250.94 amidst possibly galaxy cluster relics. The background image in orange colour is SDSS I band optical image which is superimposed with LoTSS DR1 high 
resolution (6\arcsec) contours.}
\label{fig:moustache}
\end{figure}

\begin{itemize}

\item Source 5-  GRG J105725.96+492900.31: The host galaxy does not perfectly coincide with the bight core like feature which are separated by $\sim$5\arcsec distance from one another. It is 
most likely a knot in the jet, which is located right next to a faint radio core. There is no other galaxy present in the immediate vicinity and no other alternate compact radio emission along with 
axis of the source, which could possibly be an alternate host of this source.

\item Source 10- GRG J110433.11+464225.76: This unusual source has a well detected radio core but does not exhibit any jets or well formed lobes. It has diffuse emission on either side of the 
radio core and has been reported by \citet{thwala19} as a relic radio galaxy with a size of 0.86 Mpc, thereby making it a GRG. Towards north western side of the 
source, there exists a independent FR-II type radio galaxy. Using the higher 
resolution maps the independent source's flux was measured and then subtracted from the LoTSS's low resolution maps to avoid contamination of source fluxes.

\item Source 44- GRG J113931.77+472124.3: As seen in Fig.\ \ref{fig:lotsshighres2}, there is compact radio source north of the `+' marker (host galaxy-radio core), which is an 
independent source. Also, the bright compact source seen towards south eastern side is independent of the GRG J113931.77+472124.3.

\item Source 61- GRG J121555.53+512416.41 and Source 162- GRG J135628.50+524219.23:  These sources are core dominated objects at both low and high frequencies. The lobes of both sources 
are not detected in NVSS. These are 
possibly candidates of revived/rejuvenated GRGs as we only observe the diffuse lobes at low frequencies (LoTSS), which could be from the previous epoch. The presence of the bright core may indicate 
restarting activity.

\item Source 136- GRG J133322.79+533250.94: This source displays a peculiar morphology and is possibly residing in an unique environment. The VAC estimates its photometric 
redshift (z) to be 
0.3539$\pm$ 0.0344. \citet{lopes07} estimates its photometric redshift based on SDSS data to be 0.39301 $\pm$ 0.02578. 
GRG J133322.79+533250.94 is close to two galaxy clusters, namely WHL J133322.0+5333490 at redshift of 0.3938 and WHL J133316.5+533333 at a redshift of 0.3834. Based on \citet{lopes07} photometric 
redshift,  GRG J133322.79+533250.94 is plausibly to be associated with the galaxy cluster WHL J133322.0+5333490 as they are at similar redshifts and is separated by $\sim$ 1\arcmin. This GRG appears 
to be residing in an environment of a possible merger of two galaxy clusters and the associated radio relic can be seen towards north of the GRG (Fig ~\ref{fig:moustache}). 

\item Source 161- GRG J135414.72+491315.2: The radio core coinciding with a faint galaxy is detected only in the FIRST survey.

\item Source 221- GRG J145002.36+540528.27: The radio core is well detected in the FIRST survey, which coincides with a galaxy. The source is sufficiently resolved only in the LoTSS high 
resolution map as it is located right next to a bright source. The spectral index was computed by estimating an upper limit for the flux from NVSS.

\item Source 222- GRG J145057.28+530007.76: The radio core is only resolved in the FIRST survey. It is not symmetrically placed between the two lobes and is closer to the eastern lobe.

\item Source 237- GRG J151835.37+510410.70: The radio core is not symmetrically placed between the two lobes and is closer to the northern lobe. It is also well detected in the FIRST survey. 
The southern lobe shows a prominent hotspot.

\end{itemize}

\begin{figure}
\centering
\includegraphics[width=0.45\textwidth]{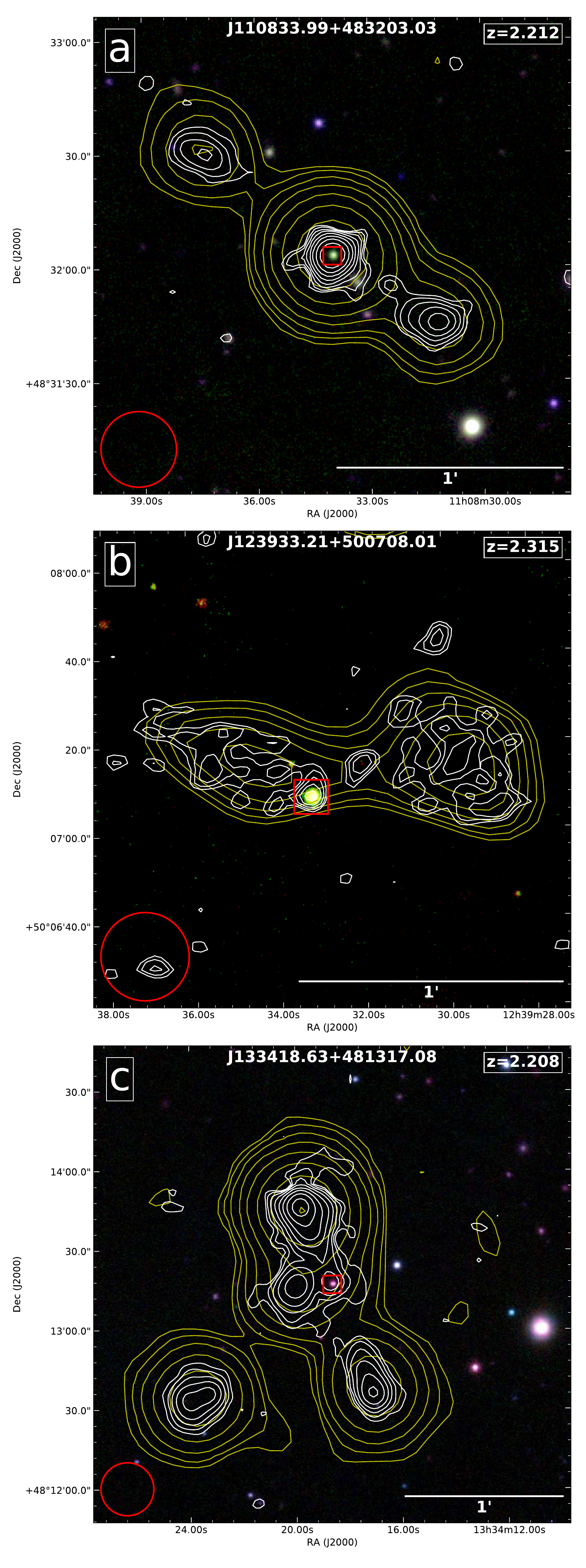}
\caption{\scriptsize{The above three images are optical-radio overlay of high redshift ($z$ > 2) GRGs, where the optical images colour composite of three bands of PanSTARRS and radio is from the 
LoTSS DR-1 at 
two resolutions. The red square indicates the host of the three GRQs. The white 
and yellow colour contours represents LoTSS high (6\arcsec$\times$6\arcsec) and low (20\arcsec$\times$20\arcsec; red circles at bottom left) resolution maps respectively, which have eight equally 
spaced relative contour levels starting 
from three times $\sigma$, where $\sigma$ denotes the local RMS noise.}}
\label{fig:highzgrg}
\end{figure}

\subsection{Giant radio quasars}\label{sec:grq}
GRGs with quasars as their AGN (GRQs) are even more rarer than the GRGs with non quasar AGN. Based on the catalogue of \cite{grscat} and candidates presented by \citet{kk18grqs}, there are 
only about $\sim$ 70 GRQs known till date and less than $\sim$ 10 GRQs known at $z\geq$ 2.  
From our LoTSS sample of 239 GRGs, 40 are confirmed GRQs and 2 are candidates GRQs ($Q_{c}$). A total of only six GRQs of the 40 confirmed GRQs were previously reported (column 13 in 
Table~\ref{tab:maingrg}) and rest are all new findings.
Our sample of $\sim$ 40 GRQs significantly increase the sample of total GRQs known till date from 70 \citep{grscat} to more than 100. The GRQs in our sample have median and mean radio powers of  $6.2 
\times 10^{26}$ W Hz$^{-1}$ and $2.4 \times 10^{27}$ W Hz$^{-1}$ at 144 MHz respectively as seen in the histogram distribution of radio powers (Fig.\ \ref{fig:histpower}).

Below we present some brief notes on individual objects-
\begin{itemize}
 \item GRQs J110833.99+483203.03, J123933.21+500708.01 and J133418.63+481317.08 are the three GRQs which we have found at very high redshift of $z\geq$2 and are shown in  Fig.\ 
\ref{fig:highzgrg}.
\item GRG J110833.99+483203.03 (16 in Table~\ref{tab:maingrg}) is a highly core dominated object at both high as well as low frequencies and hence exhibits a very flat spectral index of $\sim$ 0.2.
 \item GRGs J132554.31+551936.23 and J141408.45+484156.11 (126 and 178 in Table~\ref{tab:maingrg}) which have high redshifts and are the two quasar candidates ($Q_{c}$) as indicated 
by \citealt{Richards2009,Richards2015}. These objects have all the properties of a radio loud quasars but still are candidates because of absence of an optical spectra. Therefore they are not 
labeled as quasars (not marked in red colour) in any of the figures.	
 
\end{itemize}

\subsection{Environment analysis of GRGs}\label{sec:env}
 
We use two optically selected galaxy cluster catalogs to identify host of GRGs with Brightest Cluster Galaxies (BCGs), which are found to be at the centers of galaxy clusters. We chose the 
following two catalogues as they are made using SDSS and have an overlap with HETDEX region which is common for LoTSS DR1.

Firstly, we used the galaxy cluster catalog of \citet{whl} (here after WHL cluster catalog), which consists of 132,684 clusters. They have used  photometric data from SDSS-III to find the clusters. 
This is the biggest galaxy cluster catalog made using SDSS in the redshift range of 0.05 < $z$ < 0.8 and is $\sim~95\%$ complete for clusters with a mass of $M_{200} >  10^{14} 
M_{\odot}$ in 
the redshift range of $0.05 < z < 0.42$. A total of 71 GRGs from our sample fall in this redshift range.

We find 17 GRGs to be BCGs based on the WHL cluster catalog. 
We also used the  Gaussian Mixture Brightest Cluster Galaxy (GMBCG) catalogue \citep{gmbcg} consisting of 55,880 galaxy clusters which is more than 90$\%$ complete within the redshift range 
of 0.1 < $z$ < 0.55. We find 3 extra GRGs to be BCGs from the GMBCG catalogue. Therefore, we find a total of 20 GRGs 
(Table~\ref{tab:clusters}) from our sample of 239 to be BCGs residing in dense cluster environments and hosted by BCGs of the clusters. The mass and radius (obtained from \citealt{whl}) of 
the 17 clusters are listed in Table~\ref{tab:clusters}.


\section{Discussion}
\subsection{GRGs with sizes > 2 Mpc}
GRGs with projected linear sizes > 2 Mpc are quite rare and as observed from Fig.\ \ref{fig:size-hist}, there are only 7 GRGs with sizes greater than 2 Mpc in our sample of 239. Based on 
sample compilation (incomplete) of all the known GRGs till date of $\sim$400 sources, only $\sim$62 GRGs ($\sim$15$\%$) have sizes greater than 2 Mpc 
\citep{willis74,3crr1983,lacy93,ishwar2000,SchoenmakersGRGs01,Machalski01grgs,lara01grgs,Saripalli-sumssgrgs,Machalski07,koziel11,Solovyov14,amir15,amirgrgs,2017MNRAS.469.2886D,clarkeGRG,prescot18,
binny18}. 

If we take a complete radio sample like 3CRR \citep{3crr1983}, then we have only 12 GRGs from the total sample of 173, which is only $\sim$7$\%$ of the total population. Considering this sample of 12 
GRGs from 3CRR\footnote{\url{http://www.jb.man.ac.uk/atlas/}}, there are only 2 sources with sizes greater than 2 Mpc, 
which constitutes $\sim17\%$ of the total GRG population in 3CRR complete sample.
The conditions under which a radio galaxies grows to become are a GRG is not very well understood and studies so far based on the observations and theory seems to suggest that the gigantic size could 
be due to an interplay between the jet power and the environment.

\subsection{Morphology of GRGs}
Most known GRGs exhibit FR-II type of morphology and we very rarely see radio galaxies with Mpc scale size with FR I type of  morphology. Based on their morphologies we classify 18 GRGs as FR I type 
and 215 GRGs as FR II type. GRG J121900.76+505254.41 is the only FR-I GRG residing at cluster centre and hosted by a BCG (Table~\ref{tab:clusters}). 

\begin{figure} 
\includegraphics[width=0.48\textwidth]{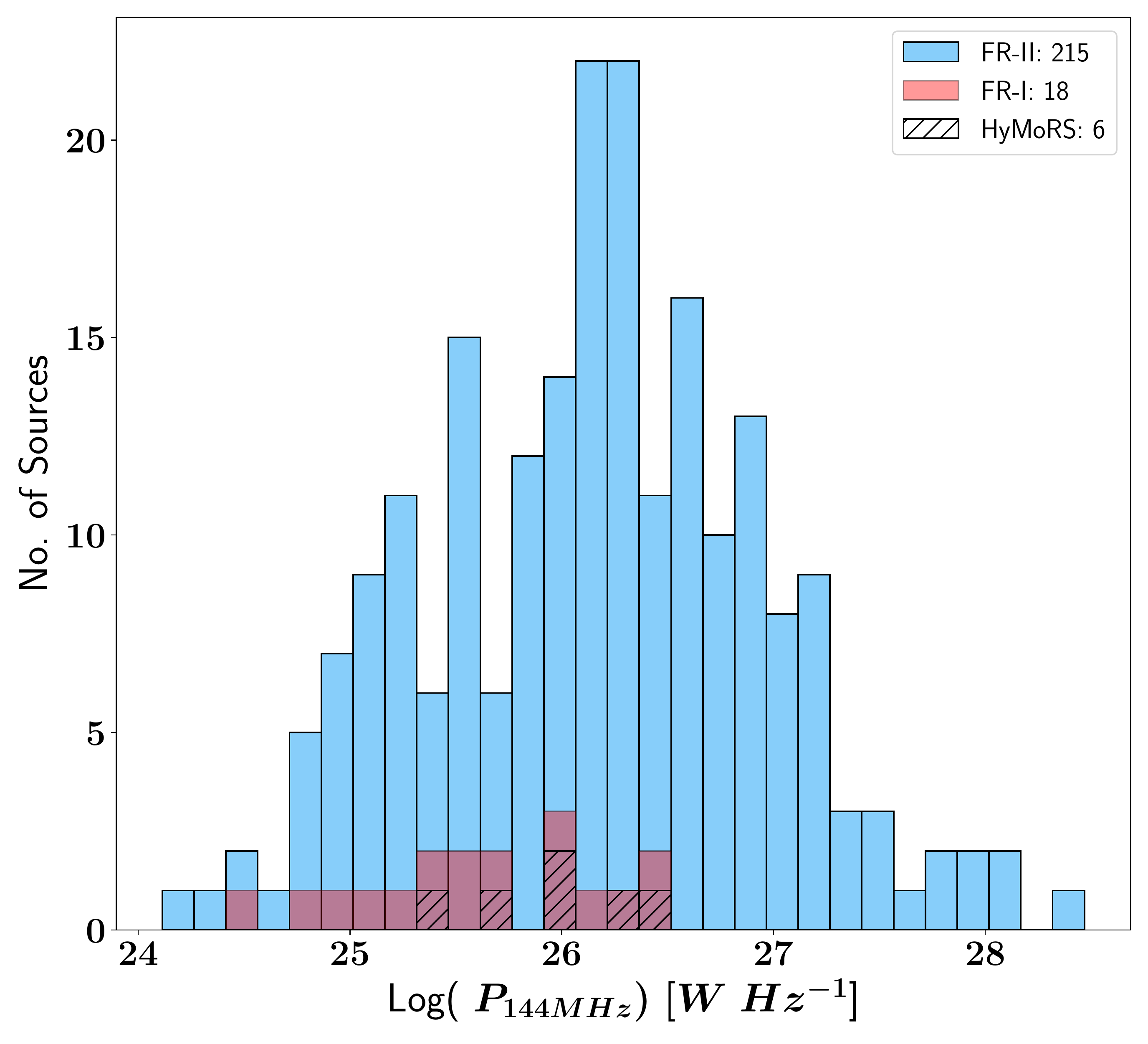}
\caption{Radio power (144 MHz) distribution of of FR-I, FR-II and HyMoRS souces in LoTSS DR1 sample of giants.}
\label{fig:FR}
\end{figure}

The total radio power of FR-I sources on average is less than that of the FRII sources \citep{ledlow-owen96,lara04}.
As seen in Fig.\ \ref{fig:FR}, the FR-I type of GRGs have a limited range in radio power $10^{24}$ $\sim$ $10^{26}$ W Hz$^{-1}$, whereas the FR-II exhibits a wide range of radio powers from $10^{24}$ 
$\sim$ $10^{28}$ W Hz$^{-1}$ at 144 MHz.

\subsubsection{GRGs with Hybrid morphology}\label{sec:hymorph}
Radio galaxies which show  FR-I morphology on one side and FR-II morphology on other side are referred as HyMoRS \citep{GK02HYMP}. The earliest example of such morphology was presented and studied by \citet{saikia-hymors}, for the radio galaxy 4C +63.07. They attribute this to an intrinsic asymmetry in either the collimation of its jets or the supply of fuel from the central black hole to opposite sides.

\citet{hymors-rare} have estimated the occurrence of HyMoRS to be as low as $\leq$ 1\% amongst the radio galaxy population. Recently, \citet{kapinska-hymors} presented 25 new candidate HyMoRS, of 
which 5 are GRGs with one being at the centre of a galaxy cluster. Therefore, one of the possible scenarios for such hybrid morphology could be attributed to different environments on each 
side of the host galaxy (radio core). 

\begin{figure}
\centering
\includegraphics[scale=0.22]{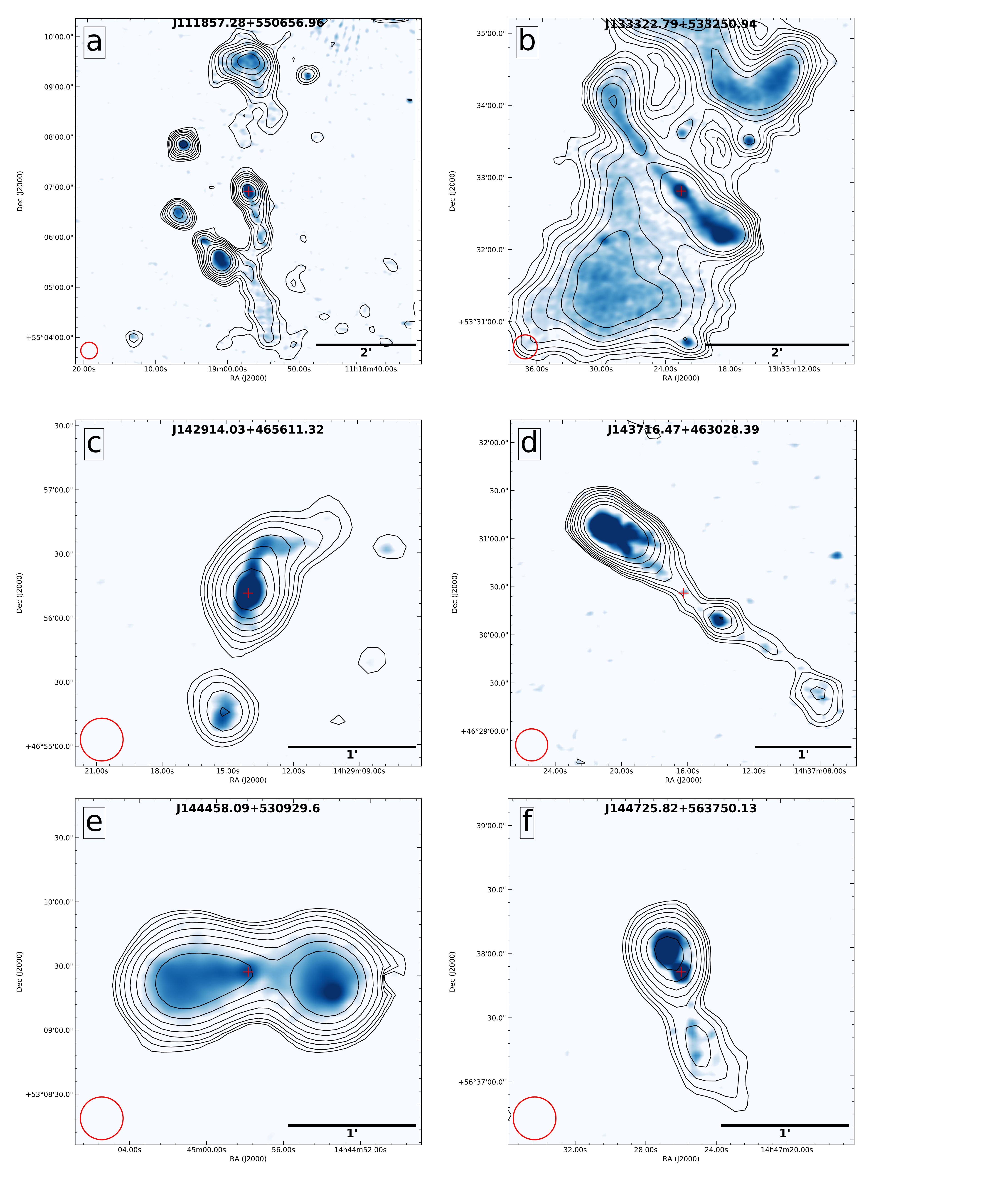}
\caption{LoTSS DR1 (6\arcsec $\times$ 6\arcsec) maps of 6 HyMoRS in inverted blue colour scale with (20\arcsec $\times$ 20\arcsec) black contours (eight equally spaced relative contour levels 
starting from three times $\sigma$, where $\sigma$ denotes the local RMS noise). The circles at bottom left corner represents the beam sizes of low resolution radio maps of LoTSS DR1. The red colour 
marker '+' indicates the location of the host galaxy.}
\label{fig:hybrid_lotss}
\end{figure}

In our sample of 239 GRGs, we find 6 examples of HyMoRS (identified via visual inspection) which are listed in column 11 (FR type) of Table ~\ref{tab:maingrg}, indicated with roman numeral 
III, 
images of these sources are presented in Fig.\ \ref{fig:hybrid_lotss}. This is by far the largest HyMoRS GRG sample reported ever. 
Environment as well as host AGN studies are needed to understand this class of radio galaxies, which can also grow to megaparsec scales in size.

\subsubsection{DDRGs: Double double radio galaxies}\label{sec:ddrgs}
DDRG are FR-II type objects with two pairs of lobes, which is indicative of their restarted nature \citep{schoenmakersDDRGs,saikiaddrgs}. The newly created jets in such 
sources travel outwards through the cocoon formed by the earlier cycle/episode of activity rather than the common intergalactic or intracluster medium, after moving through
the interstellar medium of the host galaxy. In general, the outer double lobes are aligned with the inner ones and can extend from few  kpc up to Mpc scales. It is likely that in DDRGs, due to 
an unknown mechanism or activity, the interruption of these bipolar relativistic jet flows has occurred leading to such morphologies. 

\begin{figure}
\centering
\includegraphics[scale=0.55]{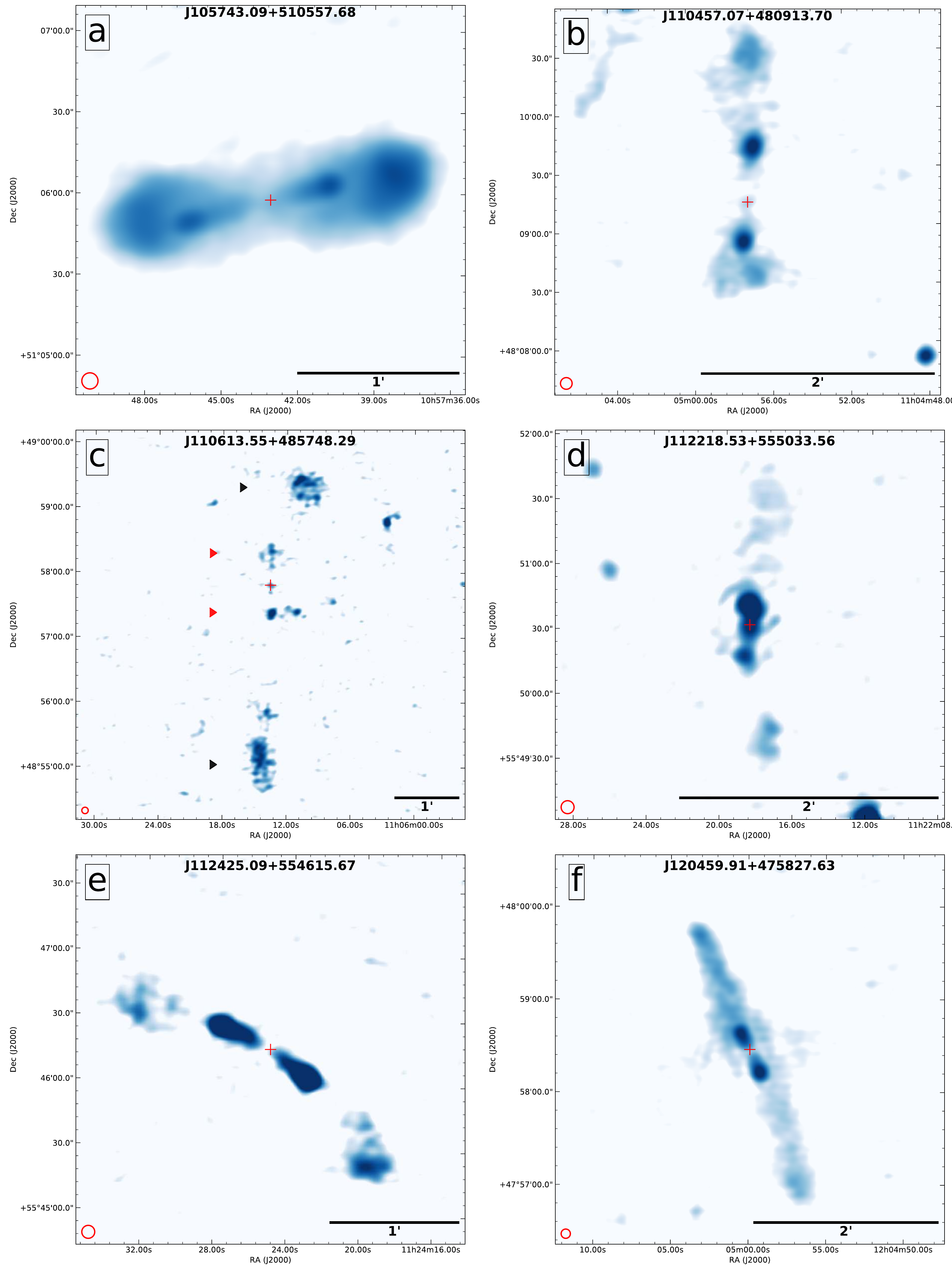}
\caption{DDRGs postage-1 (1-6 DDRGs) made using LoTSS DR1 144 MHz radio image with $6\arcsec \times 6\arcsec$ resolution in inverted blue colour scale. Refer to  Sect.\ \ref{sec:ddrgs} for more 
details. The circle at bottom left 
corner represents the beam size. The red colour marker 
`+' indicates the location of the host galaxy. In source J110613.55+485748.29 (sub-image c) the two red markers indicate the inner pair of the DDRG and the black markers indicate the out 
pair of the DDRG.}
\label{fig:ddrgs1}
\end{figure}

\begin{figure}
\centering
\includegraphics[scale=0.535]{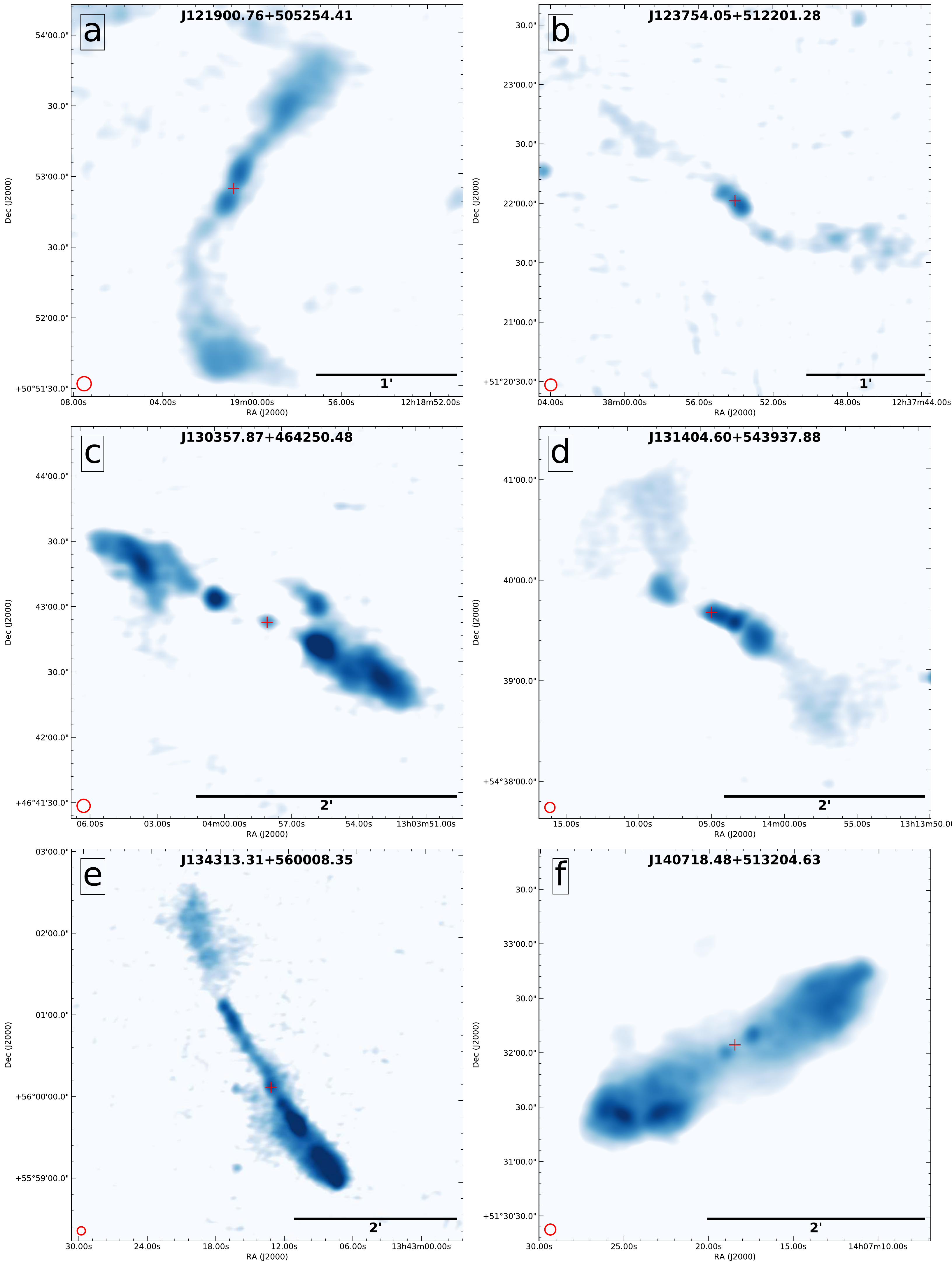}
\caption{DDRGs postage-1 (7-12 DDRGs) made using LoTSS DR1 144 MHz radio image with $6\arcsec \times 6\arcsec$ resolution in inverted blue colour scale. Refer to  Sect.\ \ref{sec:ddrgs} for more 
details.}
\label{fig:ddrgs2}
\end{figure}

In the last $\sim$20 years only about 140 DDRGs have been reported \citep{schoenmakersDDRGs,nandisaikia12,kuzmicddrgs,mahatamalotssddrgs,nandi19}.
Recently, \citealt{mahatamalotssddrgs} created a new sample of 33 DDRGs using the LoTSS DR1 and a follow up observations at higher frequency with the Jansky Very Large Array (JVLA), where 
they 
compared the optical and infrared magnitudes and colours of their host galaxies with a sample of normal radio galaxies. They find that the host galaxy properties of both DDRGs and normal radio 
galaxies are similar and suggest that the DDRG activity is a regular part of the life cycle of the radio galaxies.
We have found only 14 GRGs (largest sample reported till date) in our entire sample with DDRG type morphology as seen in Fig.\ \ref{fig:ddrgs1} and Fig.\ \ref{fig:ddrgs2}, indicating that megaparsec 
scale DDRGs are rare. Except J110457.07+480913, J110613.55+485748.29, J121900.76+505254.41, J123754.05+512201.28, J134313.31+560008.35 and J141504.70+463428.97, rest 8 
were previously classified as DDRGs in the literature (7 in \citealt{mahatamalotssddrgs} and GRGJ140718.48+513204.63 in \citealt{nandisaikia12}).

We note that sources 28 (GRG JJ112130.35+494208.14) and 221 (GRG J145002.36+540528.27) show some signs of being DDRGs but based on the current radio maps available to us they cannot be 
confirmed as DDRGs and hence are candidate DDRGs.

\begin{figure}
\centering
\includegraphics[scale=0.535]{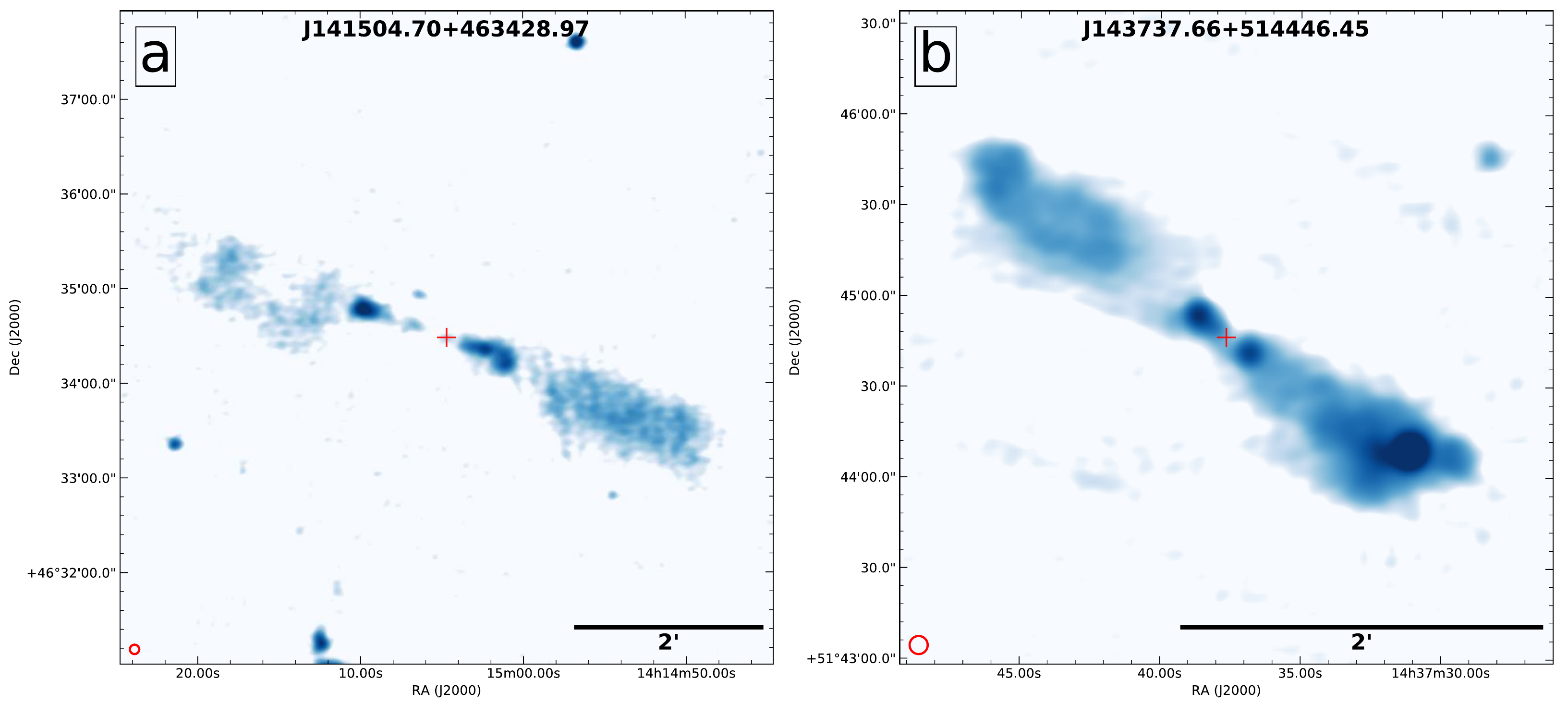}
\caption{DDRGs postage-1 (13-14 DDRGs) made using LoTSS DR1 144 MHz radio image with $6\arcsec \times 6\arcsec$ resolution in inverted blue colour scale. Refer to  Sect.\ \ref{sec:ddrgs} for more 
details.}
\label{fig:ddrgs3}
\end{figure}

These sources provide a unique opportunity to study timescales of AGN recurrent activity. As seen in 
Fig.\ \ref{fig:ddrgs1}, Fig.\ \ref{fig:ddrgs2} and Fig.\ \ref{fig:ddrgs3}, some components of the giant DDRGs are very faint and it was only possible with LoTSS's high sensitivity and resolution to 
detect and sufficiently 
resolve them. Further studies on host AGN/galaxy properties of DDRGs along with their local environments are needed to understand these peculiar sources better.

\subsection{GRGs in dense environments}
It has been hypothesized that growth of GRGs to enormous size is favoured by their location in low density environments. Using our large new sample of GRGs, it is now possible to test this hypothesis 
for objects with low redshifts ($z$ < 0.55).
We find, at least 8.4$\%$ GRGs (Table~\ref{tab:clusters}) from our sample (20/239) are located in high density environments and are hosted by BCGs. 
This low number of BCG GRGs is possibly due to absence of data for high redshift clusters in the WHL cluster catalogue and GMBCG catalogue, which are more sensitive to clusters with 
redshift less than $\sim$ 0.42 and $\sim$ 0.55, respectively  (see 
Sect. \ref{sec:env}). We have 128 GRGs with $z$ $\leq$ 0.55, therefore at least $\sim$ 16\% (20/128) GRGs are in dense cluster environments.
Based on the work of \citet{paul17}, virialized structures of mass $M_{200}$ $\geq$ 0.8 $\times$ 10$^{14}$ M$_\odot$  are classified as cluster of galaxies and non-virialized gravitionally bound 
structures consisting of few galaxies with $M_{200}$ $<$ 0.8 $\times$ 10$^{14}$ M$_\odot$ are classified as group of galaxies. Using this classification there are 14 GRGs in clusters and 3 GRGs in 
groups of galaxies (see Table ~\ref{tab:clusters}).

\citet{croston-lotss} carried out a study of environments of $\sim$ 8000 radio loud AGNs from the LoTSS, where they find that only 10\% of AGNs are associated with high density 
environments like galaxy groups/clusters and AGNs with $L_{150} > 10^{25}$ W $Hz^{-1}$ (where $L_{150}$ is the radio power at 150 MHz) are more likely to be in cluster environments. 
Similarly our 20 BCG GRGs which are also radio loud AGNs in galaxy cluster, exhibit $P_{144 MHz}$ or $L_{150} > 10^{25} W Hz^{-1}$. 

\begin{figure}
\centering
\includegraphics[width=0.50\textwidth]{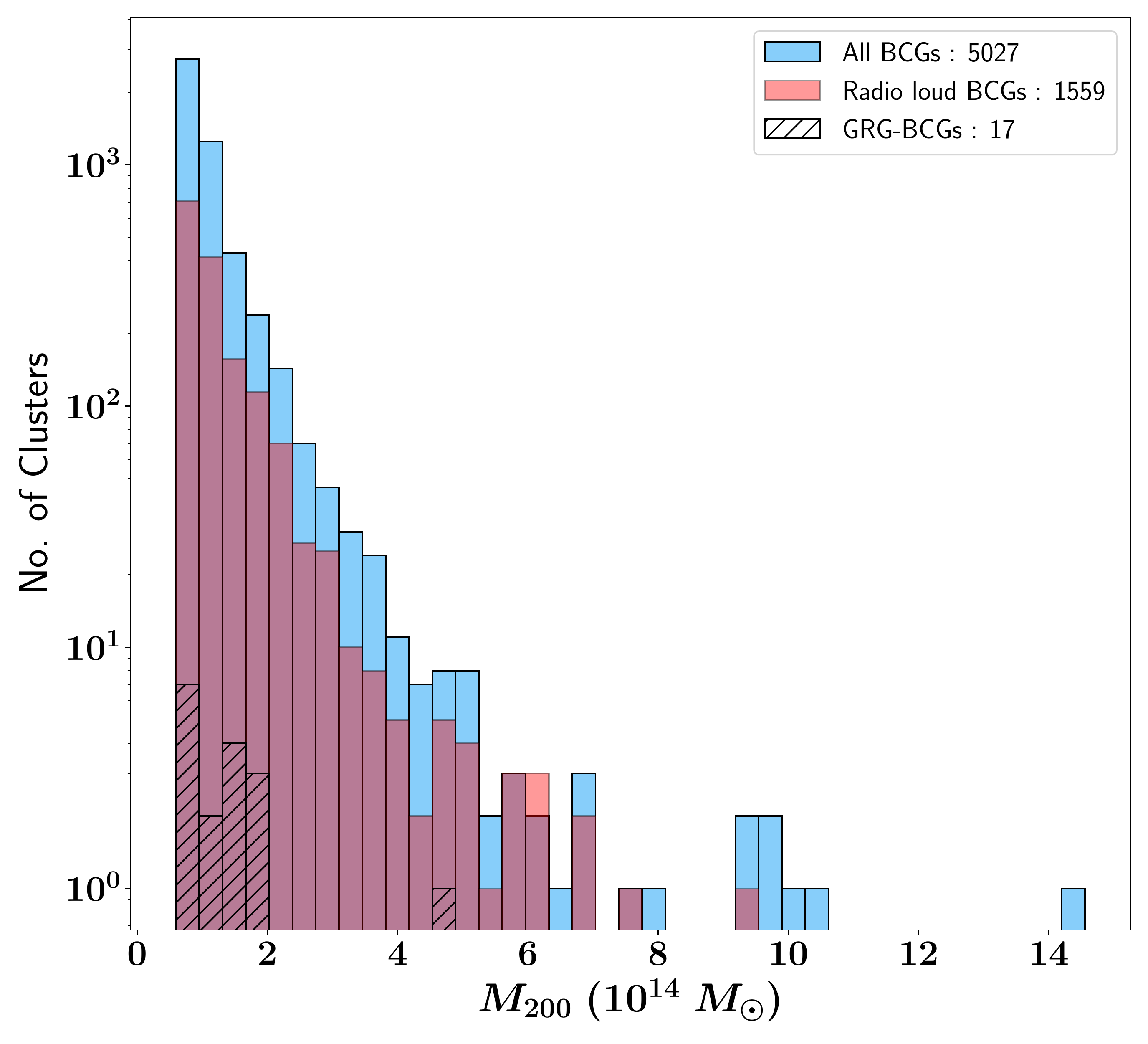} 
\caption{Distribution of $M_{200}$ (mass of the cluster within $r_{200}$) of clusters of galaxies.}
\label{fig:m200}
\end{figure}

To compare properties of galaxy clusters which have radio loud BCGs (also BCG-GRGs) and the galaxy clusters with radio quiet BCGs, we extracted all the galaxy clusters present in the LoTSS-DR1 HETDEX 
region from the WHL 
cluster catalog. This resulted in total of 5027 galaxy clusters below redshift ($z$) of 0.55 (highest redshift of BCG-GRG in our sample).
Next, we cross matched the location of BCGs (5027) from the WHL cluster with the VAC and FIRST catalog to determine all the radio loud BCGs in this region. The final number of galaxy clusters with 
radio loud BCGs is 1559 (unique sample after combining results from matches with VAC and FIRST). The 1559 radio loud BCGs also include our 17 BCG-GRGs (found in WHL; see Table~\ref{tab:clusters}).
In Fig.\ \ref{fig:m200} we have shown the distribution of $M_{200}$, which is the mass of the cluster within $r_{200}$, of clusters of galaxies, where $r_{200}$ is the radius within which the mean 
density of a cluster is 200 times of the critical density of the universe. We note that about 30$\%$ (1541) clusters have radio loud BCGs which are not giants and only $\sim$ 0.34 $\%$ of the 
clusters in WHL cluster catalogue (HETDEX region) host a GRG (or have BCG GRG).
Though the GRG sample is less here but more or less complete, it seems that GRGs which are BCGs tend to avoid very high mass 
clusters and prefer less dense clusters.

The BCG GRGs also can possibly trace the inhomogeneities in the intergalactic gas, 
which  could be one of the determining factors for the growth and evolution of these giant sources. The forward propagation of jets as well as backflow from hotspots in the lobes of GRGs are influenced by the gas they encounter.
Only 1/20 BCG GRGs in our sample have FR-I type radio morphology and rest 19 have FR-II type of radio morphology. Almost all these sources do not exhibit highly linear structures but are 
distorted or bent to certain extent, which indicates the effect of interaction between the cluster medium and the expanding radio jets of the BCG GRGs. Remarkably, despite the possible resistance 
presented by the cluster medium, these sources have managed to grow to megaparsec scales. This seems to suggest the presence of extraordinarily powerful AGN powering these sources.

\begin{table}
\caption{Short summary of classification of GRGs.}
\begin{center}
\begin{tabular}{lcll}
\hline
Classification & No. of objects \\ 
\hline
  GRQs &    40   \\
  BCGs &    20 \\
  FR-II & 215 \\
  FR-I & 18 \\
  HyMoRS & 6 \\
  DDRGs & 14 \\
\hline
\end{tabular}
\end{center}
\label{tab:sumtab}
\end{table}

\section{Summary}
A total of 239 GRGs (Table. ~\ref{tab:sumtab})  were found in $\sim$424 deg$^{2}$ area using LoTSS, which is just $\sim$2\% area of the total survey that is planned to cover the northern sky. Our 
sample of 239 GRGs represents a lower 
number estimate due to limitation of optical data, which is essential for identifying host galaxy and its corresponding redshift. Assuming the isotropy and homogeneity of the Universe, if we 
extrapolate the number of GRGs expected to be found over the final sky coverage of LoTSS  ($\sim$ 2$\pi$ steradians), then we should be able to find at least $\sim$ 12000 GRGs with LoTSS's  
sensitivity. Upcoming deep optical spectroscopic surveys like the WEAVE-LOFAR survey \citep{weave} will provide crucial data (redshifts) for identifying hundreds of more GRGs.
The summary of the paper is as follows :

\begin{enumerate}
\item Our sample of 239 GRGs is in the redshift range of 0.1 to 2.3, out of which 225 are newly found. This makes it the largest sample discovered to date.
\item About 16$\%$ (40/239) of the sample  are hosted by quasars, where three GRQs are above redshift of 2.
\item The depth and resolution of LoTSS images has enabled us to find GRGs with low powers of $\sim$ $10^{24}$ W Hz$^{-1}$ at 144 MHz.
\item We show that the spectral index of GRGs and GRQs is similar to that of their low sized counterparts (NRGs).
\item Majority (90\%) of our sample show FR-II type of morphology.
\item We have found 14 double-double GRGs, which is $\sim$ 5\% of our sample. It is the largest reporting sample of giant DDRGs till date.
\item We have found 6 GRGs with HyMoRS morphology which are very rare.
\item Based on the optical data, we find that none of the GRGs in our sample are hosted by spiral galaxies.
\item At $z$ < 0.55, at least $\sim$ 16\% of GRGs lie at the centers (BCGs) of either big galaxy groups or galaxy clusters.
\end{enumerate}


\onecolumn

\twocolumn

\clearpage
\onecolumn
\setlength{\tabcolsep}{2.35pt}
\begin{tiny}
\begin{longtable}{c c c c c c c c c c c c c}

\captionsetup{width=0.85\textwidth}
\caption{ \label{tab:maingrg} Here, we list all the 239 GRGs found from the LoTSS. The Columns (2)-RA $\&$ (3)-Dec are the right ascension $\&$ declination (J2000.0 
epoch) which indicate the center of the host galaxies of the GRG/GRQs from either 
SDSS or Pan-STARRS.   Column (4) 
`Class' represents the type of host of the GRG - G:galaxy, Q:quasar and $Q_{c}$: quasar candidate. In Column (5), $z$ (redshift) marked with \dag\ represents spectroscopic redshift from SDSS, $\star$ 
represents photometric redshift from SDSS and $\mathsection$ indicates redshifts from VAC. The one marked with superscript `a' refers to redshift obtained from \citealt{sullyGRGlotss} 
(spectroscopic) and `b' from \citealt{lopes07} (photometric), owing to higher accuracy when compared to SDSS or VAC.
Columns (6) $\&$ (7) are angular size and projected linear size of the source. Columns (8) $\&$ (9) are the integrated flux of sources at 144 MHz and its corresponding radio power. Column (10) 
$\alpha^{1400}_{144}$ is the integrated spectral index computed between 1400 MHz (NVSS) flux and 
144 MHz (LoTSS) flux. Rows marked with `-' in Column (10) are blended sources for which we do not present a spectral index. Column (11) indicates the morphological type of the GRG: I represents FR-I 
type, II represents FR-II type and III represents HyMoRS. Column (12) represents the r band magnitude of the host galaxies of the GRGs obtained from SDSS. For sources 162 and 233, in absence of reliable magnitudes from SDSS, we have given r band magnitudes from PAN-STARRS. The last column (Ref) shows references 
(see end of the table) for the GRGs already known in literature.} \\

\hline \hline
Sr.No & RA & Dec & Class & $z$ & Size & Size & $S_{144 MHz}$ & $P_{144 MHz}$ & $\alpha^{1400}_{144}$  & FR & r & Ref \\
         & (HMS) & (DMS) &  &  & (\arcsec) & (Mpc) & (mJy) & (10$^{26}$W Hz$^{-1}$)     &                                       & Type  &  Mag \\
         (1) & (2) & (3) & (4) & (5) & (6) & (7) & (8) & (9) & (10) & (11) & (12) & (13)\\
\hline
\endfirsthead
\caption{continued.}\\
\hline\hline
Sr.No & RA & Dec & Class & $z$ & Size & Size & $S_{144 MHz}$ & $P_{144 MHz}$ &  $\alpha^{1400}_{144}$  &FR & r & Ref \\
        & (HMS) & (DMS) &  &  & (\arcsec) & (Mpc) & (mJy) & (10$^{26}$W Hz$^{-1}$) &      &                                   Type &  Mag \\
         (1) & (2) & (3) & (4) & (5) & (6) & (7) & (8) & (9) & (10) & (11) & (12) & (13) \\
\hline
\endhead
\hline
\endfoot
1 & 10 57 05.43 & +53 26 24.79 & Q & 0.45992 $\pm$ 0.00003$^{\dag}$ & 153 & 0.92 & 1485 $\pm$ 297 & 11.00 $\pm$ 2.20 & 0.69 $\pm$ 0.09 & II & 19.48 $\pm$ 0.03 & - \\
2 & 10 57 09.25 & +48 40 41.03 & G & 0.27627 $\pm$ 0.00005$^{\dag}$ & 439 & 1.90 & 1062 $\pm$ 212 & 2.55 $\pm$ 0.51 & 0.79 $\pm$ 0.09 & II & 18.01 $\pm$ 0.01 & - \\
3 & 10 57 14.71 & +55 48 56.02 & Q & 1.39830 $\pm$ 0.00166$^{\dag}$ & 91 & 0.79 & 20 $\pm$ 4 & 1.84 $\pm$ 0.40 & 0.64 $\pm$ 0.11 & II & 18.86 $\pm$ 0.01 & - \\
4 & 10 57 21.22 & +53 14 22.42 & G & 0.33645 $\pm$ 0.00011$^{\dag}$ & 145 & 0.72 & 11 $\pm$ 4 & 0.04 $\pm$ 0.02 & $<$0.63 & II & 18.36 $\pm$ 0.02 & - \\
5 & 10 57 25.96 & +49 29 00.31 & G & 0.47614 $\pm$ 0.00022$^{\dag}$ & 157 & 0.96 & 94 $\pm$ 19 & 0.81 $\pm$ 0.16 & 0.86 $\pm$ 0.10 & II & 19.76 $\pm$ 0.04 & - \\
6 & 10 57 43.09 & +51 05 57.68 & G & 0.46269 $\pm$ 0.00004$^{\dag}$ & 126 & 0.76 & 657 $\pm$ 131 & 5.22 $\pm$ 1.05 & 0.85 $\pm$ 0.09 & II & 20.59 $\pm$ 0.05 & - \\
7 & 10 58 17.90 & +51 40 17.70 & G & 0.41497 $\pm$ 0.00003$^{\dag}$ & 330 & 1.86 & 298 $\pm$ 88 & 1.77 $\pm$ 0.53 & 0.74 $\pm$ 0.22 & II & 19.62 $\pm$ 0.03 & 8 \\
8 & 11 01 47.58 & +46 49 11.20 & G & 0.68096 $\pm$ 0.00017$^{\dag}$ & 150 & 1.09 & 532 $\pm$ 106 & 10.70 $\pm$ 2.15 & 0.88 $\pm$ 0.09 & II & 21.17 $\pm$ 0.12 & 3 \\
9 & 11 01 59.17 & +46 45 34.19 & G & 0.46877 $\pm$ 0.00019$^{\dag}$ & 127 & 0.77 & 168 $\pm$ 34 & 1.35 $\pm$ 0.27 & 0.80 $\pm$ 0.09 & II & 19.49 $\pm$ 0.03 & - \\
10 & 11 04 33.11 & +46 42 25.76 & G & 0.14126 $\pm$ 0.00001$^{\dag}$ & 305 & 0.78 & 253 $\pm$ 51 & 0.14 $\pm$ 0.03 & 0.68 $\pm$ 0.09 & I & 16.68 $\pm$ 0.01 & 9 \\
11 & 11 04 57.07 & +48 09 13.70 & G & 0.41471 $\pm$ 0.00011$^{\dag}$ & 137 & 0.78 & 37 $\pm$ 7 & 0.22 $\pm$ 0.04 & 0.75 $\pm$ 0.12 & II & 19.84 $\pm$ 0.03 & - \\
12 & 11 05 15.27 & +54 41 09.30 & G & 0.28425 $\pm$ 0.00010$^{\dag}$ & 169 & 0.75 & 89 $\pm$ 18 & 0.23 $\pm$ 0.05 & 0.90 $\pm$ 0.10 & II & 20.64 $\pm$ 0.05 & - \\
13 & 11 06 13.55 & +48 57 48.29 & G & 0.634 $\pm$ 0.108$^{\mathsection}$ & 292 & 2.06 & 10 $\pm$ 2 & 0.15 $\pm$ 0.07 & $<$0.63 & II & 20.25 $\pm$ 0.12 & - \\
14 & 11 08 00.77 & +51 20 28.64 & G & 0.801 $\pm$ 0.172$^{\mathsection}$ & 157 & 1.21 & 192 $\pm$ 39 & 7.64 $\pm$ 4.33 & 1.38 $\pm$ 0.13 & II & 20.47 $\pm$ 0.10 & - \\
15 & 11 08 15.18 & +50 10 03.84 & G & 0.66956 $\pm$ 0.00023$^{\dag}$ & 103 & 0.74 & 29 $\pm$ 6 & 0.77 $\pm$ 0.16 & $<$1.54 & II & 21.49 $\pm$ 0.10 & - \\
16 & 11 08 33.99 & +48 32 03.03 & Q & 2.12100 $\pm$ 0.00026$^{\dag}$ & 85 & 0.73 & 30 $\pm$ 6 & 4.32 $\pm$ 0.89 & 0.21 $\pm$ 0.09 & II & 19.64 $\pm$ 0.02 & - \\
17 & 11 09 20.21 & +48 15 00.51 & G & 0.39190 $\pm$ 0.00009$^{\dag}$ & 183 & 1.00 & 12 $\pm$ 2 & 0.05 $\pm$ 0.01 & 0.33 $\pm$ 0.12 & II & 18.31 $\pm$ 0.02 & - \\
18 & 11 09 35.40 & +51 04 02.27 & Q & 1.18230 $\pm$ 0.00049$^{\dag}$ & 106 & 0.90 & 186 $\pm$ 37 & 12.70 $\pm$ 2.58 & - & II & 19.13 $\pm$ 0.01 & 2 \\
19 & 11 09 36.42 & +53 13 48.13 & G & 0.288 $\pm$ 0.027$^{\mathsection}$ & 230 & 1.03 & 583 $\pm$ 117 & 1.52 $\pm$ 0.45 & - & II & 18.16 $\pm$ 0.01 & - \\
20 & 11 10 11.02 & +53 30 58.74 & G & 0.58483 $\pm$ 0.00018$^{\dag}$ & 142 & 0.96 & 565 $\pm$ 113 & 7.44 $\pm$ 1.50 & - & II & 20.64 $\pm$ 0.07 & - \\
21 & 11 12 59.52 & +49 42 27.11 & G & 0.502 $\pm$ 0.133$^{\mathsection}$ & 244 & 1.54 & 223 $\pm$ 45 & 1.90 $\pm$ 1.26 & 0.55 $\pm$ 0.09 & II & 21.25 $\pm$ 0.09 & - \\
22 & 11 13 31.45 & +46 22 16.04 & G & 0.589 $\pm$ 0.145$^{\mathsection}$ & 112 & 0.76 & 340 $\pm$ 68 & 4.66 $\pm$ 2.92 & 0.80 $\pm$ 0.09 & II & 21.84 $\pm$ 0.10 & - \\
23 & 11 14 00.10 & +53 22 16.57 & G & 0.727 $\pm$ 0.116$^{\mathsection}$ & 125 & 0.93 & 6 $\pm$ 1 & 0.17 $\pm$ 0.08 & $<$1.16 & II & 21.35 $\pm$ 0.15 & - \\
24 & 11 15 29.60 & +56 00 39.64 & G & 0.57880 $\pm$ 0.00007$^{\dag}$ & 107 & 0.72 & 82 $\pm$ 16 & 1.06 $\pm$ 0.21 & 0.77 $\pm$ 0.09 & II & 20.33 $\pm$ 0.06 & - \\
25 & 11 18 57.28 & +55 06 56.96 & G & 0.35159 $\pm$ 0.00007$^{\dag}$ & 347 & 1.77 & 669 $\pm$ 134 & 2.73 $\pm$ 0.55 & - & III & 19.23 $\pm$ 0.02 & - \\
26 & 11 20 12.70 & +51 32 56.34 & G & 0.72720 $\pm$ 0.00019$^{\dag}$ & 103 & 0.77 & 94 $\pm$ 19 & 2.17 $\pm$ 0.44 & 0.86 $\pm$ 0.10 & II & 21.03 $\pm$ 0.10 & - \\
27 & 11 21 26.44 & +53 44 56.71 & G & 0.10378 $\pm$ 0.00002$^{\dag}$ & 450 & 0.88 & 399 $\pm$ 80 & 0.11 $\pm$ 0.02 & 0.61 $\pm$ 0.09 & II & 15.11 $\pm$ 0.00 & - \\
28 & 11 21 30.35 & +49 42 08.14 & G & 0.494 $\pm$ 0.071$^{\mathsection}$ & 132 & 0.82 & 86 $\pm$ 17 & 0.72 $\pm$ 0.29 & 0.61 $\pm$ 0.09 & II & 20.00 $\pm$ 0.04 & - \\
29 & 11 22 18.53 & +55 50 33.56 & G & 0.90962 $\pm$ 0.00047$^{\dag}$ & 126 & 1.01 & 56 $\pm$ 11 & 2.05 $\pm$ 0.42 & - & II & 21.92 $\pm$ 0.18 & - \\
30 & 11 24 25.09 & +55 46 15.67 & G & 0.80902 $\pm$ 0.00042$^{\dag}$ & 146 & 1.13 & 26 $\pm$ 5 & 0.75 $\pm$ 0.15 & 0.77 $\pm$ 0.13 & II & 22.14 $\pm$ 0.22 & - \\
31 & 11 24 29.98 & +46 35 23.68 & Q & 0.91500 $\pm$ 0.00013$^{\dag}$ & 88 & 0.71 & 33 $\pm$ 7 & 1.11 $\pm$ 0.23 & 0.58 $\pm$ 0.10 & II & 20.12 $\pm$ 0.03 & - \\
32 & 11 24 35.86 & +49 03 25.92 & G & 0.47934 $\pm$ 0.00010$^{\dag}$ & 121 & 0.74 & 120 $\pm$ 24 & 1.00 $\pm$ 0.20 & - & II & 20.10 $\pm$ 0.04 & - \\
33 & 11 26 29.95 & +49 01 37.16 & G & 0.660 $\pm$ 0.139$^{\mathsection}$ & 111 & 0.80 & 35 $\pm$ 7 & 0.62 $\pm$ 0.34 & 0.76 $\pm$ 0.11 & II & 21.67 $\pm$ 0.12 & - \\
34 & 11 26 39.89 & +53 34 27.20 & G & 0.64523 $\pm$ 0.00024$^{\dag}$ & 125 & 0.89 & 79 $\pm$ 16 & 1.30 $\pm$ 0.26 & - & II & 20.49 $\pm$ 0.08 & - \\
35 & 11 27 13.18 & +51 13 26.35 & G & 0.36132 $\pm$ 0.00007$^{\dag}$ & 202 & 1.05 & 54 $\pm$ 11 & 0.23 $\pm$ 0.05 & 0.60 $\pm$ 0.10 & I & 18.12 $\pm$ 0.01 & - \\
36 & 11 27 42.06 & +52 19 44.68 & G & 0.72484 $\pm$ 0.00031$^{\dag}$ & 114 & 0.85 & 6 $\pm$ 1 & 0.13 $\pm$ 0.03 & $<$0.82 & II & 21.75 $\pm$ 0.12 & - \\
37 & 11 28 54.65 & +56 20 09.50 & G & 0.594 $\pm$ 0.067$^{\mathsection}$ & 188 & 1.29 & 115 $\pm$ 23 & 1.57 $\pm$ 0.53 & - & II & 21.12 $\pm$ 0.10 & - \\
38 & 11 29 43.41 & +51 24 12.11 & G & 0.79923 $\pm$ 0.00008$^{\dag}$ & 226 & 1.75 & 70 $\pm$ 14 & 2.15 $\pm$ 0.43 & 0.94 $\pm$ 0.11 & II & 21.53 $\pm$ 0.15 & - \\
39 & 11 32 02.31 & +47 28 24.14 & G & 0.26431 $\pm$ 0.00006$^{\dag}$ & 205 & 0.86 & 38 $\pm$ 8 & 0.08 $\pm$ 0.02 & 0.61 $\pm$ 0.10 & II & 17.82 $\pm$ 0.01 & - \\
40 & 11 32 50.67 & +50 57 04.68 & G & 0.35857 $\pm$ 0.00010$^{\dag}$ & 138 & 0.71 & 63 $\pm$ 13 & 0.26 $\pm$ 0.05 & 0.65 $\pm$ 0.10 & II & 18.53 $\pm$ 0.01 & - \\
41 & 11 34 21.89 & +56 34 13.28 & G & 0.626 $\pm$ 0.092$^{\mathsection}$ & 227 & 1.59 & 21 $\pm$ 4 & 0.35 $\pm$ 0.14 & 0.92 $\pm$ 0.16 & II & 20.98 $\pm$ 0.08 & - \\
42 & 11 34 35.48 & +46 08 00.45 & G & 0.704 $\pm$ 0.163$^{\mathsection}$ & 300 & 2.21 & 41 $\pm$ 8 & 0.90 $\pm$ 0.54 & 0.87 $\pm$ 0.16 & II & 21.39 $\pm$ 0.08 & - \\
43 & 11 35 03.20 & +48 26 12.12 & G & 0.22597 $\pm$ 0.00006$^{\dag}$ & 202 & 0.76 & 131 $\pm$ 26 & 0.20 $\pm$ 0.04 & 0.78 $\pm$ 0.09 & II & 17.59 $\pm$ 0.01 & - \\
44 & 11 39 31.77 & +47 21 24.3 & G & 0.51791 $\pm$ 0.00012$^{\dag}$ & 312 & 2.00 & 126 $\pm$ 25 & 1.23 $\pm$ 0.25 & 0.70 $\pm$ 0.09 & II & 20.32 $\pm$ 0.05 & - \\
45 & 11 43 05.55 & +52 27 26.89 & Q & 1.67249 $\pm$ 0.00082$^{\dag}$ & 86 & 0.75 & 100 $\pm$ 20 & 14.90 $\pm$ 3.04 & 0.72 $\pm$ 0.09 & II & 20.23 $\pm$ 0.03 & - \\
46 & 11 48 14.98 & +54 57 16.49 & G & 0.22679 $\pm$ 0.00003$^{\dag}$ & 202 & 0.76 & 22 $\pm$ 4 & 0.03 $\pm$ 0.01 & 0.51 $\pm$ 0.11 & I & 17.26 $\pm$ 0.01 & - \\
47 & 11 51 59.9 & +49 50 56.11 & Q & 0.89148 $\pm$ 0.00017$^{\dag}$ & 96 & 0.77 & 1015 $\pm$ 203 & 36.60 $\pm$ 7.37 & 0.78 $\pm$ 0.09 & II & 18.38 $\pm$ 0.01 & - \\
48 & 11 52 16.89 & +46 24 51.37 & G & 0.44454 $\pm$ 0.02004$^{\star}$ & 148 & 0.87 & 43 $\pm$ 9 & 0.28 $\pm$ 0.06 & 0.56 $\pm$ 0.10 & II & 19.68 $\pm$ 0.03 & - \\
49 & 11 58 39.12 & +46 55 45.35 & G & 0.65974 $\pm$ 0.00023$^{\dag}$ & 128 & 0.92 & 18 $\pm$ 4 & 0.36 $\pm$ 0.07 & $<$1.08 & II & 21.61 $\pm$ 0.13 & - \\
50 & 12 01 18.14 & +46 32 41.98 & G & 0.638 $\pm$ 0.088$^{\mathsection}$ & 129 & 0.91 & 111 $\pm$ 22 & 1.87 $\pm$ 0.73 & 0.83 $\pm$ 0.10 & II & 21.82 $\pm$ 0.15 & - \\
51 & 12 01 22.84 & +49 25 37.14 & G & 0.205 $\pm$ 0.045$^{\mathsection}$ & 345 & 1.20 & 1946 $\pm$ 389 & 2.56 $\pm$ 1.35 & 1.09 $\pm$ 0.09 & I & 18.14 $\pm$ 0.01 & - \\
52 & 12 02 06.86 & +51 41 34.63 & G & 0.61278 $\pm$ 0.00028$^{\dag}$ & 141 & 0.98 & 87 $\pm$ 17 & 1.27 $\pm$ 0.26 & - & II & 21.11 $\pm$ 0.10 & - \\
53 & 12 04 59.91 & +47 58 27.63 & G & 0.52963 $\pm$ 0.08487$^{\star}$ & 192 & 1.24 & 138 $\pm$ 28 & 1.53 $\pm$ 0.66 & 0.89 $\pm$ 0.09 & II & 20.96 $\pm$ 0.07 & - \\
54 & 12 05 28.50 & +56 13 42.32 & G & 0.681 $\pm$ 0.121$^{\mathsection}$ & 118 & 0.86 & 26 $\pm$ 5 & 0.48 $\pm$ 0.23 & $<$0.74 & II & 20.54 $\pm$ 0.13 & - \\
55 & 12 05 46.98 & +50 48 59.94 & G & 0.69799 $\pm$ 0.00021$^{\dag}$ & 160 & 1.18 & 352 $\pm$ 70 & 7.60 $\pm$ 1.53 & 0.90 $\pm$ 0.09 & II & 21.02 $\pm$ 0.09 & - \\
56 & 12 08 26.14 & +47 15 25.03 & G & 0.56407 $\pm$ 0.00021$^{\dag}$ & 113 & 0.75 & 34 $\pm$ 7 & 0.41 $\pm$ 0.08 & - & II & 20.76 $\pm$ 0.06 & - \\
57 & 12 10 46.06 & +53 29 22.90 & G & 0.44842 $\pm$ 0.00010$^{\dag}$ & 119 & 0.71 & 234 $\pm$ 47 & 1.73 $\pm$ 0.35 & 0.85 $\pm$ 0.09 & I & 19.87 $\pm$ 0.03 & - \\
58 & 12 10 55.67 & +49 49 39.77 & G & 0.53442 $\pm$ 0.05992$^{\star}$ & 110 & 0.72 & 49 $\pm$ 10 & 0.53 $\pm$ 0.18 & 0.79 $\pm$ 0.10 & I & 20.57 $\pm$ 0.06 & - \\
59 & 12 11 40.91 & +50 10 15.11 & G & 0.33985 $\pm$ 0.14011$^{\star}$ & 143 & 0.71 & 372 $\pm$ 74 & 1.45 $\pm$ 1.42 & 0.86 $\pm$ 0.09 & II & 21.78 $\pm$ 0.13 & - \\
60 & 12 13 04.34 & +51 18 26.38 & Q & 0.61680 $\pm$ 0.00013$^{\dag}$ & 118 & 0.82 & 102 $\pm$ 20 & 1.49 $\pm$ 0.30 & 0.72 $\pm$ 0.09 & II & 19.78 $\pm$ 0.02 & - \\
61 & 12 15 55.53 & +51 24 16.41 & G & 0.44477 $\pm$ 0.00038$^{\dag}$ & 277 & 1.63 & 74 $\pm$ 15 & 0.52 $\pm$ 0.10 & - & I & 19.17 $\pm$ 0.02 & - \\
62 & 12 17 06.80 & +51 47 34.33 & G & 0.616 $\pm$ 0.089$^{\mathsection}$ & 183 & 1.28 & 57 $\pm$ 11 & 0.95 $\pm$ 0.39 & 0.99 $\pm$ 0.12 & I & 21.61 $\pm$ 0.10 & - \\
63 & 12 17 07.40 & +48 58 22.48 & G & 0.484 $\pm$ 0.029$^{\mathsection}$ & 136 & 0.84 & 172 $\pm$ 34 & 1.46 $\pm$ 0.36 & - & I & 19.60 $\pm$ 0.03 & - \\
64 & 12 18 18.15 & +53 27 21.33 & G & 0.56754 $\pm$ 0.04447$^{\star}$ & 183 & 1.23 & 1270 $\pm$ 254 & 16.00 $\pm$ 4.41 & 0.81 $\pm$ 0.09 & II & 21.77 $\pm$ 0.17 & - \\
65 & 12 18 49.88 & +50 26 17.59 & G & 0.19920 $\pm$ 0.00002$^{\dag}$ & 210 & 0.71 & 2942 $\pm$ 588 & 3.34 $\pm$ 0.67 & 0.67 $\pm$ 0.09 & II & 17.12 $\pm$ 0.01 & 6 \\
66 & 12 19 00.76 & +50 52 54.41 & G & 0.38509 $\pm$ 0.00008$^{\dag}$ & 142 & 0.77 & 78 $\pm$ 16 & 0.38 $\pm$ 0.08 & 0.68 $\pm$ 0.10 & I & 18.35 $\pm$ 0.01 & - \\
67 & 12 19 35.92 & +46 59 29.88 & G & 0.67865 $\pm$ 0.00032$^{\dag}$ & 114 & 0.83 & 19 $\pm$ 4 & 0.35 $\pm$ 0.07 & - & II & 21.14 $\pm$ 0.09 & - \\
68 & 12 19 52.32 & +47 20 58.49 & Q & 0.65310 $\pm$ 0.00009$^{\dag}$ & 241 & 1.72 & 115 $\pm$ 23 & 2.06 $\pm$ 0.41 & 0.85 $\pm$ 0.10 & II & 19.10 $\pm$ 0.01 & - \\
69 & 12 19 52.55 & +46 54 25.74 & Q & 1.85116 $\pm$ 0.00041$^{\dag}$ & 88 & 0.76 & 60 $\pm$ 12 & 8.28 $\pm$ 1.69 & 0.43 $\pm$ 0.09 & II & 18.62 $\pm$ 0.01 & - \\
70 & 12 20 07.85 & +47 05 19.65 & G & 0.689 $\pm$ 0.159$^{\mathsection}$ & 149 & 1.09 & 109 $\pm$ 22 & 2.93 $\pm$ 1.75 & 1.37 $\pm$ 0.17 & II & 21.97 $\pm$ 0.12 & - \\
71 & 12 20 28.13 & +52 51 44.89 & G & 0.34583 $\pm$ 0.00008$^{\dag}$ & 255 & 1.29 & 58 $\pm$ 12 & 0.21 $\pm$ 0.04 & 0.55 $\pm$ 0.09 & II & 18.09 $\pm$ 0.01 & - \\
72 & 12 22 55.24 & +49 26 42.32 & G & 0.20315 $\pm$ 0.00004$^{\dag}$ & 354 & 1.22 & 84 $\pm$ 17 & 0.10 $\pm$ 0.02 & 0.48 $\pm$ 0.09 & II & 17.21 $\pm$ 0.01 & - \\
73 & 12 23 27.25 & +52 36 45.38 & G & 0.62112 $\pm$ 0.05830$^{\star}$ & 111 & 0.77 & 59 $\pm$ 12 & 0.89 $\pm$ 0.27 & 0.74 $\pm$ 0.10 & I & 21.19 $\pm$ 0.13 & - \\
74 & 12 25 03.66 & +47 23 36.40 & G & 0.774 $\pm$ 0.154$^{\mathsection}$ & 95 & 0.73 & 14 $\pm$ 3 & 0.32 $\pm$ 0.17 & 0.51 $\pm$ 0.12 & II & 21.26 $\pm$ 0.09 & - \\
75 & 12 25 31.36 & +49 46 43.95 & G & 0.32488 $\pm$ 0.00009$^{\dag}$ & 206 & 1.00 & 21 $\pm$ 4 & 0.07 $\pm$ 0.01 & 0.60 $\pm$ 0.12 & I & 18.32 $\pm$ 0.01 & - \\
76 & 12 25 44.39 & +49 49 42.93 & G & 0.63224 $\pm$ 0.08754$^{\star}$ & 164 & 1.16 & 286 $\pm$ 57 & 5.11 $\pm$ 2.00 & 1.00 $\pm$ 0.09 & II & 21.06 $\pm$ 0.06 & - \\
77 & 12 25 58.20 & +53 09 38.38 & G & 0.81100 $\pm$ 0.07937$^{\star}$ & 147 & 1.14 & 263 $\pm$ 53 & 8.15 $\pm$ 2.55 & 0.91 $\pm$ 0.09 & II & 20.25 $\pm$ 0.15 & - \\
78 & 12 28 26.35 & +52 31 01.10 & G & 0.453 $\pm$ 0.106$^{\mathsection}$ & 157 & 0.93 & 106 $\pm$ 21 & 0.78 $\pm$ 0.46 & 0.80 $\pm$ 0.09 & II & 20.65 $\pm$ 0.06 & - \\
79 & 12 29 36.25 & +50 13 04.65 & G & 0.38357 $\pm$ 0.00007$^{\dag}$ & 198 & 1.07 & 56 $\pm$ 11 & 0.27 $\pm$ 0.05 & 0.61 $\pm$ 0.10 & I & 18.53 $\pm$ 0.02 & - \\
80 & 12 29 59.59 & +53 32 47.04 & G & 0.52300 $\pm$ 0.05280$^{\star}$ & 137 & 0.88 & 296 $\pm$ 59 & 3.20 $\pm$ 1.01 & 0.90 $\pm$ 0.09 & II & 21.12 $\pm$ 0.11 & - \\
81 & 12 30 14.05 & +54 11 41.14 & G & 0.610 $\pm$ 0.087$^{\mathsection}$ & 205 & 1.42 & 7 $\pm$ 1 & 0.09 $\pm$ 0.04 & $<$0.57 & II & 20.96 $\pm$ 0.10 & - \\
82 & 12 32 04.95 & +53 06 27.31 & G & 0.20604 $\pm$ 0.00006$^{\dag}$ & 274 & 0.95 & 120 $\pm$ 24 & 0.15 $\pm$ 0.03 & 0.90 $\pm$ 0.10 & II & 17.31 $\pm$ 0.01 & - \\
83 & 12 32 50.45 & +49 06 26.14 & G & 0.69015 $\pm$ 0.00013$^{\dag}$ & 256 & 1.87 & 205 $\pm$ 41 & 3.99 $\pm$ 0.80 & 0.76 $\pm$ 0.09 & II & 20.92 $\pm$ 0.06 & - \\
84 & 12 33 05.45 & +49 02 51.93 & Q & 1.35200 $\pm$ 0.00107$^{\dag}$ & 104 & 0.89 & 332 $\pm$ 66 & 31.00 $\pm$ 6.30 & - & II & 20.55 $\pm$ 0.03 & - \\
85 & 12 35 01.52 & +53 17 55.09 & G & 0.34480 $\pm$ 0.00030$^{\dag}$ & 683 & 3.44 & 859 $\pm$ 172 & 3.46 $\pm$ 0.69 & 0.85 $\pm$ 0.09 & II & 19.24 $\pm$ 0.04 & - \\
86 & 12 35 48.77 & +50 50 36.30 & G & 0.650 $\pm$ 0.131$^{\mathsection}$ & 105 & 0.75 & 84 $\pm$ 17 & 1.41 $\pm$ 0.75 & - & II & 21.49 $\pm$ 0.10 & - \\
87 & 12 36 48.37 & +46 04 05.91 & G & 0.615 $\pm$ 0.154$^{\mathsection}$ & 125 & 0.87 & 268 $\pm$ 54 & 3.96 $\pm$ 2.53 & 0.75 $\pm$ 0.09 & II & 21.39 $\pm$ 0.10 & - \\
88 & 12 37 54.05 & +51 22 01.28 & G & 0.39184 $\pm$ 0.00018$^{\dag}$ & 164 & 0.89 & 25 $\pm$ 5 & 0.13 $\pm$ 0.03 & - & II & 19.68 $\pm$ 0.03 & - \\
89 & 12 38 07.78 & +53 25 55.91 & Q & 0.34680 $\pm$ 0.00004$^{\dag}$ & 157 & 0.80 & 455 $\pm$ 91 & 1.73 $\pm$ 0.35 & 0.61 $\pm$ 0.09 & II & 17.29 $\pm$ 0.01 & - \\
90 & 12 39 33.21 & +50 07 08.01 & Q & 2.31500 $\pm$ 0.00024$^{\dag}$ & 89 & 0.75 & 3 $\pm$ 1 & 2.32 $\pm$ 0.52 & $<$1.39 & II & 18.49 $\pm$ 0.01 & - \\
91 & 12 40 12.46 & +53 34 37.25 & Q & 0.29300 $\pm$ 0.00003$^{\dag}$ & 164 & 0.74 & 1527 $\pm$ 305 & 4.15 $\pm$ 0.83 & 0.76 $\pm$ 0.09 & II & 18.52 $\pm$ 0.01 & 1 \\
92 & 12 40 31.72 & +48 59 36.89 & G & 0.53607 $\pm$ 0.04030$^{\star}$ & 136 & 0.89 & 605 $\pm$ 121 & 6.35 $\pm$ 1.72 & 0.70 $\pm$ 0.09 & II & 21.17 $\pm$ 0.08 & - \\
93 & 12 41 42.34 & +51 35 14.32 & G & 0.52951 $\pm$ 0.00011$^{\dag}$ & 161 & 1.04 & 177 $\pm$ 36 & 1.86 $\pm$ 0.37 & 0.76 $\pm$ 0.09 & II & 19.57 $\pm$ 0.04 & - \\
94 & 12 45 25.16 & +47 07 10.11 & G & 0.51179 $\pm$ 0.00015$^{\dag}$ & 111 & 0.70 & 22 $\pm$ 5 & 0.20 $\pm$ 0.04 & 0.60 $\pm$ 0.10 & II & 19.95 $\pm$ 0.03 & - \\
95 & 12 45 31.06 & +48 50 20.16 & G & 0.25359 $\pm$ 0.00010$^{\dag}$ & 174 & 0.71 & 40 $\pm$ 8 & 0.08 $\pm$ 0.02 & 0.69 $\pm$ 0.11 & I & 17.99 $\pm$ 0.01 & - \\
96 & 12 49 03.46 & +52 40 05.29 & G & 0.663 $\pm$ 0.107$^{\mathsection}$ & 135 & 0.97 & 66 $\pm$ 13 & 1.27 $\pm$ 0.56 & 0.93 $\pm$ 0.10 & II & 21.60 $\pm$ 0.13 & - \\
97 & 12 49 13.20 & +50 00 43.68 & G & 0.34956 $\pm$ 0.00010$^{\dag}$ & 150 & 0.76 & 295 $\pm$ 59 & 1.13 $\pm$ 0.23 & 0.59 $\pm$ 0.09 & I & 18.82 $\pm$ 0.02 & - \\
98 & 12 51 42.04 & +50 34 24.65 & G & 0.54904 $\pm$ 0.00007$^{\dag}$ & 150 & 0.99 & 8355 $\pm$ 1671 & 99.90 $\pm$ 20.10 & 0.86 $\pm$ 0.09 & II & 19.35 $\pm$ 0.03 & 5 \\
99 & 12 56 24.74 & +55 28 25.99 & G & 0.808 $\pm$ 0.179$^{\mathsection}$ & 113 & 0.87 & 1765 $\pm$ 353 & 55.50 $\pm$ 32.10 & 0.94 $\pm$ 0.09 & II & 21.01 $\pm$ 0.12 & - \\
100 & 12 57 16.35 & +51 17 58.11 & Q & 0.52582 $\pm$ 0.00004$^{\dag}$ & 170 & 1.09 & 96 $\pm$ 19 & 0.98 $\pm$ 0.20 & 0.74 $\pm$ 0.10 & II & 17.69 $\pm$ 0.01 & - \\
101 & 12 58 21.37 & +54 20 29.35 & G & 0.612 $\pm$ 0.099$^{\mathsection}$ & 143 & 0.99 & 521 $\pm$ 104 & 7.64 $\pm$ 3.36 & - & II & 21.11 $\pm$ 0.09 & - \\
102 & 12 59 50.83 & +53 07 09.03 & G & 0.60663 $\pm$ 0.04364$^{\star}$ & 105 & 0.73 & 67 $\pm$ 13 & 1.06 $\pm$ 0.28 & 0.94 $\pm$ 0.11 & II & 20.93 $\pm$ 0.09 & - \\
103 & 13 01 34.99 & +54 08 09.21 & G & 0.313 $\pm$ 0.081$^{\mathsection}$ & 168 & 0.79 & 1581 $\pm$ 316 & 5.00 $\pm$ 3.15 & 0.77 $\pm$ 0.09 & II & 19.23 $\pm$ 0.02 & 1 \\
104 & 13 03 31.08 & +53 59 48.65 & G & 0.24201 $\pm$ 0.00005$^{\dag}$ & 248 & 0.97 & 50 $\pm$ 10 & 0.09 $\pm$ 0.02 & 0.84 $\pm$ 0.13 & II & 17.24 $\pm$ 0.01 & - \\
105 & 13 03 32.19 & +52 20 01.98 & G & 0.27200 $\pm$ 0.00007$^{\dag}$ & 165 & 0.71 & 83 $\pm$ 17 & 0.18 $\pm$ 0.04 & 0.56 $\pm$ 0.09 & I & 17.98 $\pm$ 0.01 & - \\
106 & 13 03 57.87 & +46 42 50.48 & G & 0.58430 $\pm$ 0.00021$^{\dag}$ & 159 & 1.08 & 137 $\pm$ 27 & 1.87 $\pm$ 0.38 & 0.83 $\pm$ 0.09 & II & 20.38 $\pm$ 0.06 & - \\
107 & 13 04 30.80 & +50 41 08.02 & Q & 0.93000 $\pm$ 0.00039$^{\dag}$ & 122 & 0.98 & 119 $\pm$ 24 & 4.53 $\pm$ 0.92 & 0.71 $\pm$ 0.09 & II & 20.33 $\pm$ 0.03 & - \\
108 & 13 05 21.32 & +49 51 42.36 & Q & 1.25098 $\pm$ 0.00026$^{\dag}$ & 88 & 0.75 & 544 $\pm$ 109 & 44.80 $\pm$ 9.07 & 0.82 $\pm$ 0.09 & II & 18.74 $\pm$ 0.01 & - \\
109 & 13 06 35.08 & +55 36 44.43 & G & 0.461 $\pm$ 0.031$^{\mathsection}$ & 136 & 0.82 & 47 $\pm$ 9 & 0.36 $\pm$ 0.09 & 0.74 $\pm$ 0.11 & II & 20.17 $\pm$ 0.03 & - \\
110 & 13 07 48.36 & +56 11 46.96 & G & 0.83043 $\pm$ 0.08339$^{\star}$ & 120 & 0.94 & 143 $\pm$ 29 & 4.28 $\pm$ 1.37 & - & II & 22.23 $\pm$ 0.29 & - \\
111 & 13 09 25.74 & +53 48 17.26 & G & 0.81187 $\pm$ 0.00032$^{\dag}$ & 118 & 0.92 & 75 $\pm$ 15 & 2.31 $\pm$ 0.47 & 0.89 $\pm$ 0.10 & II & 22.20 $\pm$ 0.26 & - \\
112 & 13 10 28.87 & +52 13 40.43 & G & 0.650 $\pm$ 0.094$^{\mathsection}$ & 197 & 1.41 & 451 $\pm$ 90 & 7.45 $\pm$ 3.01 & 0.71 $\pm$ 0.09 & II & 20.28 $\pm$ 0.06 & - \\
113 & 13 12 11.14 & +48 09 25.26 & Q & 0.71510 $\pm$ 0.00003$^{\dag}$ & 110 & 0.82 & 559 $\pm$ 112 & 10.30 $\pm$ 2.08 & 0.51 $\pm$ 0.09 & II & 17.13 $\pm$ 0.00 & - \\
114 & 13 12 16.28 & +48 47 45.40 & G & 0.36430 $\pm$ 0.00007$^{\dag}$ & 210 & 1.09 & 122 $\pm$ 24 & 0.52 $\pm$ 0.10 & 0.60 $\pm$ 0.09 & II & 18.64 $\pm$ 0.01 & - \\
115 & 13 13 55.21 & +53 04 03.22 & G & 0.632 $\pm$ 0.088$^{\mathsection}$ & 171 & 1.21 & 30 $\pm$ 6 & 0.54 $\pm$ 0.21 & $<$1.03 & II & 21.76 $\pm$ 0.12 & - \\
116 & 13 14 04.60 & +54 39 37.88 & G & 0.34645 $\pm$ 0.00008$^{\dag}$ & 210 & 1.06 & 678 $\pm$ 136 & 2.68 $\pm$ 0.54 & - & II & 17.80 $\pm$ 0.01 & - \\
117 & 13 16 34.59 & +49 32 39.67 & G & 0.563 $\pm$ 0.040$^{\mathsection}$ & 126 & 0.84 & 84 $\pm$ 17 & 0.93 $\pm$ 0.24 & 0.57 $\pm$ 0.09 & II & 21.14 $\pm$ 0.05 & - \\
118 & 13 17 22.25 & +50 45 10.03 & G & 0.758 $\pm$ 0.134$^{\mathsection}$ & 130 & 0.98 & 3 $\pm$ 1 & 0.06 $\pm$ 0.03 & $<$0.58 & II & 21.67 $\pm$ 0.10 & - \\
119 & 13 19 05.75 & +47 46 39.23 & G & 0.90300 $\pm$ 0.00010$^{\dag}$ & 262 & 2.10 & 44 $\pm$ 9 & 1.70 $\pm$ 0.34 & 0.83 $\pm$ 0.12 & II & 20.94 $\pm$ 0.04 & - \\
120 & 13 20 29.67 & +49 16 47.11 & Q & 0.68447 $\pm$ 0.00009$^{\dag}$ & 113 & 0.82 & 74 $\pm$ 15 & 1.40 $\pm$ 0.28 & - & II & 19.08 $\pm$ 0.01 & - \\
121 & 13 22 29.07 & +50 48 44.53 & G & 0.18439 $\pm$ 0.00004$^{\dag}$ & 326 & 1.04 & 94 $\pm$ 19 & 0.09 $\pm$ 0.02 & - & II & 17.23 $\pm$ 0.01 & - \\
122 & 13 23 36.36 & +47 29 49.68 & G & 0.440 $\pm$ 0.035$^{\mathsection}$ & 180 & 1.05 & 249 $\pm$ 50 & 1.69 $\pm$ 0.46 & - & II & 19.36 $\pm$ 0.03 & - \\
123 & 13 23 54.44 & +46 31 42.89 & G & 0.35737 $\pm$ 0.00022$^{\dag}$ & 136 & 0.70 & 118 $\pm$ 24 & 0.49 $\pm$ 0.10 & 0.69 $\pm$ 0.09 & II & 19.57 $\pm$ 0.03 & - \\
124 & 13 24 35.19 & +50 41 02.31 & G & 0.28711 $\pm$ 0.00006$^{\dag}$ & 162 & 0.72 & 495 $\pm$ 99 & 1.41 $\pm$ 0.28 & 1.14 $\pm$ 0.09 & II & 18.17 $\pm$ 0.01 & - \\
125 & 13 25 54.31 & +55 19 36.23 & G & 1.692 $\pm$ 0.089$^{\mathsection}$ & 87 & 0.75 & 31 $\pm$ 6 & 5.40 $\pm$ 1.32 & 0.85 $\pm$ 0.13 & II & 20.48 $\pm$ 0.03 & - \\
126 & 13 26 14.35 & +49 34 31.52 & G & 0.340 $\pm$ 0.044$^{\mathsection}$ & 169 & 0.85 & 3885 $\pm$ 777 & 14.70 $\pm$ 5.35 & 0.76 $\pm$ 0.09 & II & 19.17 $\pm$ 0.02 & 7 \\
127 & 13 26 15.35 & +48 04 22.67 & Q & 0.59405 $\pm$ 0.00007$^{\dag}$ & 103 & 0.71 & 287 $\pm$ 57 & 4.12 $\pm$ 0.83 & 0.85 $\pm$ 0.09 & II & 20.90 $\pm$ 0.08 & - \\
128 & 13 26 46.56 & +53 58 17.38 & Q & 0.40975 $\pm$ 0.00005$^{\dag}$ & 178 & 1.00 & 444 $\pm$ 89 & 2.65 $\pm$ 0.53 & 0.84 $\pm$ 0.09 & II & 19.56 $\pm$ 0.03 & - \\
129 & 13 27 13.33 & +52 26 49.93 & G & 0.53770 $\pm$ 0.00015$^{\dag}$ & 109 & 0.71 & 10 $\pm$ 3 & 0.11 $\pm$ 0.03 & $<$0.80 & II & 20.31 $\pm$ 0.04 & - \\
130 & 13 29 14.05 & +52 09 38.80 & G & 0.620 $\pm$ 0.070$^{\mathsection}$ & 118 & 0.82 & 25 $\pm$ 5 & 0.40 $\pm$ 0.14 & 0.86 $\pm$ 0.16 & II & 20.92 $\pm$ 0.09 & - \\
131 & 13 29 24.63 & +49 00 15.32 & G & 0.286 $\pm$ 0.081$^{\mathsection}$ & 253 & 1.12 & 67 $\pm$ 13 & 0.17 $\pm$ 0.12 & 0.74 $\pm$ 0.11 & II & 19.08 $\pm$ 0.02 & - \\
132 & 13 30 41.81 & +48 27 54.88 & G & 0.33166 $\pm$ 0.00010$^{\dag}$ & 153 & 0.75 & 47 $\pm$ 10 & 0.16 $\pm$ 0.03 & 0.53 $\pm$ 0.10 & II & 18.38 $\pm$ 0.01 & - \\
133 & 13 31 35.25 & +45 59 55.53 & G & 0.38482 $\pm$ 0.00009$^{\dag}$ & 131 & 0.71 & 115 $\pm$ 23 & 0.57 $\pm$ 0.11 & 0.74 $\pm$ 0.09 & II & 18.06 $\pm$ 0.01 & - \\
134 & 13 32 53.76 & +48 45 32.05 & G & 0.73036 $\pm$ 0.00020$^{\dag}$ & 122 & 0.91 & 65 $\pm$ 13 & 1.34 $\pm$ 0.27 & 0.61 $\pm$ 0.09 & II & 21.27 $\pm$ 0.11 & - \\
135 & 13 32 58.28 & +53 53 55.60 & G & 0.32261 $\pm$ 0.00009$^{\dag}$ & 174 & 0.84 & 199 $\pm$ 40 & 0.70 $\pm$ 0.14 & 0.94 $\pm$ 0.09 & II & 18.58 $\pm$ 0.01 & - \\
136 & 13 33 22.79 & +53 32 50.94 & G & 0.354 $\pm$ 0.034$^{\mathsection}$ & 173 & 0.88 & 423 $\pm$ 86 & 1.75 $\pm$ 0.54 & - & III & 18.90 $\pm$ 0.02 & - \\
137 & 13 34 11.70 & +55 01 24.87 & Q & 1.24470 $\pm$ 0.00043$^{\dag}$ & 91 & 0.78 & 3380 $\pm$ 676 & 294.00 $\pm$ 59.50 & 0.90 $\pm$ 0.09 & II & 17.97 $\pm$ 0.01 & 4 \\
138 & 13 34 18.63 & +48 13 17.08 & Q & 2.20849 $\pm$ 0.00009$^{\dag}$ & 88 & 0.75 & 487 $\pm$ 97 & 140.00 $\pm$ 28.90 & - & II & 18.15 $\pm$ 0.01 & - \\
139 & 13 35 12.01 & +49 44 27.20 & G & 0.436 $\pm$ 0.041$^{\mathsection}$ & 173 & 1.01 & 13 $\pm$ 3 & 0.08 $\pm$ 0.03 & $<$0.71 & II & 19.32 $\pm$ 0.02 & - \\
140 & 13 36 12.67 & +54 47 42.24 & Q & 0.71180 $\pm$ 0.00012$^{\dag}$ & 142 & 1.05 & 95 $\pm$ 19 & 1.97 $\pm$ 0.40 & - & II & 17.72 $\pm$ 0.01 & - \\
141 & 13 36 18.75 & +53 39 52.12 & G & 0.30138 $\pm$ 0.00006$^{\dag}$ & 305 & 1.41 & 49 $\pm$ 10 & 0.14 $\pm$ 0.03 & 0.72 $\pm$ 0.11 & II & 17.91 $\pm$ 0.01 & - \\
142 & 13 36 37.92 & +55 40 33.27 & G & 0.86400 $\pm$ 0.00003$^{\dag}$ & 105 & 0.83 & 291 $\pm$ 58 & 9.85 $\pm$ 1.98 & 0.80 $\pm$ 0.09 & II & 21.16 $\pm$ 0.05 & - \\
143 & 13 38 04.28 & +46 46 41.19 & Q & 1.36900 $\pm$ 0.00030$^{\dag}$ & 140 & 1.21 & 95 $\pm$ 19 & 9.17 $\pm$ 1.87 & - & II & 21.13 $\pm$ 0.06 & - \\
144 & 13 39 22.70 & +50 57 47.40 & G & 0.316 $\pm$ 0.098$^{\mathsection}$ & 174 & 0.83 & 56 $\pm$ 11 & 0.17 $\pm$ 0.13 & 0.66 $\pm$ 0.10 & II & 19.89 $\pm$ 0.02 & - \\
145 & 13 41 03.08 & +49 15 59.99 & G & 0.74672 $\pm$ 0.08486$^{\star}$ & 113 & 0.85 & 334 $\pm$ 67 & 7.64 $\pm$ 2.62 & 0.71 $\pm$ 0.09 & II & 21.63 $\pm$ 0.10 & - \\
146 & 13 42 06.98 & +47 25 53.04 & G & 0.17213 $\pm$ 0.00004$^{\dag}$ & 369 & 1.11 & 366 $\pm$ 73 & 0.30 $\pm$ 0.06 & 0.68 $\pm$ 0.09 & II & 15.92 $\pm$ 0.00 & - \\
147 & 13 43 13.31 & +56 00 08.35 & G & 0.48474 $\pm$ 0.00007$^{\dag}$ & 245 & 1.51 & 174 $\pm$ 35 & 1.62 $\pm$ 0.33 & 0.97 $\pm$ 0.10 & II & 20.00 $\pm$ 0.03 & - \\
148 & 13 44 15.65 & +48 45 48.96 & G & 0.725 $\pm$ 0.160$^{\mathsection}$ & 202 & 1.51 & 495 $\pm$ 99 & 11.50 $\pm$ 6.59 & 0.87 $\pm$ 0.09 & II & 21.66 $\pm$ 0.14 & - \\
149 & 13 44 41.82 & +50 22 57.29 & G & 0.76317 $\pm$ 0.00025$^{\dag}$ & 123 & 0.93 & 61 $\pm$ 12 & 1.58 $\pm$ 0.32 & 0.86 $\pm$ 0.10 & II & 21.82 $\pm$ 0.16 & - \\
150 & 13 44 52.71 & +48 57 49.96 & G & 0.34021 $\pm$ 0.00011$^{\dag}$ & 167 & 0.84 & 36 $\pm$ 7 & 0.14 $\pm$ 0.03 & 0.92 $\pm$ 0.13 & II & 19.18 $\pm$ 0.02 & - \\
151 & 13 45 57.55 & +54 03 16.62 & G & 0.16250 $\pm$ 0.00003$^{\dag}$ & 334 & 0.96 & 2582 $\pm$ 516 & 1.95 $\pm$ 0.39 & 0.89 $\pm$ 0.09 & II & 17.15 $\pm$ 0.01 & 2 \\
152 & 13 46 00.30 & +53 18 41.21 & G & 0.536 $\pm$ 0.096$^{\mathsection}$ & 156 & 1.01 & 10 $\pm$ 2 & 0.12 $\pm$ 0.06 & $<$1.09 & II & 20.61 $\pm$ 0.07 & - \\
153 & 13 46 13.49 & +50 35 07.24 & G & 0.69605 $\pm$ 0.00034$^{\dag}$ & 141 & 1.04 & 12 $\pm$ 3 & 0.22 $\pm$ 0.05 & $<$0.59 & II & 21.49 $\pm$ 0.12 & - \\
154 & 13 46 45.96 & +47 27 19.03 & G & 0.420 $\pm$ 0.049$^{\mathsection}$ & 184 & 1.05 & 148 $\pm$ 30 & 0.91 $\pm$ 0.31 & - & II & 19.83 $\pm$ 0.03 & - \\
155 & 13 47 30.56 & +47 05 09.28 & G & 0.371 $\pm$ 0.047$^{\mathsection}$ & 211 & 1.11 & 77 $\pm$ 15 & 0.38 $\pm$ 0.14 & 1.00 $\pm$ 0.12 & II & 19.53 $\pm$ 0.02 & - \\
156 & 13 48 16.94 & +49 50 24.80 & G & 0.64318 $\pm$ 0.00017$^{\dag}$ & 120 & 0.85 & 8 $\pm$ 2 & 0.13 $\pm$ 0.03 & $<$0.68 & II & 20.24 $\pm$ 0.06 & - \\
157 & 13 48 37.68 & +47 08 00.01 & Q & 0.50724 $\pm$ 0.00009$^{\dag}$ & 119 & 0.75 & 127 $\pm$ 25 & 1.15 $\pm$ 0.23 & 0.65 $\pm$ 0.09 & II & 19.84 $\pm$ 0.05 & - \\
158 & 13 48 53.20 & +46 45 50.63 & Q & 1.66716 $\pm$ 0.00065$^{\dag}$ & 102 & 0.89 & 58 $\pm$ 12 & 11.40 $\pm$ 2.34 & 1.01 $\pm$ 0.11 & II & 20.37 $\pm$ 0.03 & - \\
159 & 13 49 27.92 & +46 20 15.11 & G & 0.42110 $\pm$ 0.00010$^{\dag}$ & 131 & 0.75 & 120 $\pm$ 24 & 0.74 $\pm$ 0.15 & 0.76 $\pm$ 0.09 & II & 19.05 $\pm$ 0.02 & - \\
160 & 13 52 53.13 & +46 25 20.66 & G & 0.471 $\pm$ 0.071$^{\mathsection}$ & 220 & 1.34 & 397 $\pm$ 79 & 3.13 $\pm$ 1.29 & 0.72 $\pm$ 0.09 & II & 20.63 $\pm$ 0.04 & - \\
161 & 13 54 14.72 & +49 13 15.2 & G & 0.43885 $\pm$ 0.13902$^{\star}$ & 123 & 0.72 & 633 $\pm$ 127 & 4.55 $\pm$ 3.53 & 0.91 $\pm$ 0.09 & II & 22.08 $\pm$ 0.18 & - \\
162 & 13 56 28.50 & +52 42 19.23 & G & 0.42709 $\pm$ 0.00011$^{\dag}$ & 297 & 1.71 & 64 $\pm$ 13 & 0.37 $\pm$ 0.07 & 0.50 $\pm$ 0.09 & II & 18.62 $\pm$ 0.02 & - \\
163 & 13 56 35.89 & +56 09 44.77 & G & 0.66300 $\pm$ 0.00021$^{\dag}$ & 111 & 0.80 & 18 $\pm$ 4 & 0.32 $\pm$ 0.06 & 0.76 $\pm$ 0.13 & II & 20.83 $\pm$ 0.07 & - \\
164 & 13 59 51.16 & +47 03 21.03 & G & 0.46224 $\pm$ 0.00010$^{\dag}$ & 140 & 0.85 & 23 $\pm$ 5 & 0.17 $\pm$ 0.03 & 0.55 $\pm$ 0.14 & II & 19.48 $\pm$ 0.03 & - \\
165 & 14 02 55.41 & +51 27 28.60 & G & 0.51796 $\pm$ 0.16062$^{\star}$ & 135 & 0.87 & 304 $\pm$ 61 & 3.07 $\pm$ 2.36 & 0.79 $\pm$ 0.09 & II & 20.93 $\pm$ 0.06 & - \\
166 & 14 03 15.11 & +51 44 44.77 & G & 0.48517 $\pm$ 0.16607$^{\star}$ & 228 & 1.41 & 1596 $\pm$ 319 & 14.30 $\pm$ 12.00 & 0.87 $\pm$ 0.09 & II & 21.07 $\pm$ 0.06 & - \\
167 & 14 04 08.61 & +46 42 39.42 & G & 0.537 $\pm$ 0.060$^{\mathsection}$ & 453 & 2.96 & 224 $\pm$ 45 & 2.43 $\pm$ 0.81 & - & I & 20.34 $\pm$ 0.06 & - \\
168 & 14 05 40.97 & +54 10 55.42 & G & 0.76118 $\pm$ 0.13989$^{\star}$ & 116 & 0.88 & 588 $\pm$ 118 & 14.10 $\pm$ 6.96 & 0.72 $\pm$ 0.09 & II & 20.70 $\pm$ 0.04 & - \\
169 & 14 06 05.42 & +55 47 49.87 & G & 0.78025 $\pm$ 0.06341$^{\star}$ & 114 & 0.87 & 598 $\pm$ 120 & 16.50 $\pm$ 4.67 & 0.86 $\pm$ 0.09 & II & 23.09 $\pm$ 0.40 & - \\
170 & 14 06 12.30 & +53 38 34.64 & G & 0.46455 $\pm$ 0.00010$^{\dag}$ & 117 & 0.71 & 5 $\pm$ 2 & 0.04 $\pm$ 0.02 & $<$0.55 & II & 19.90 $\pm$ 0.03 & - \\
171 & 14 07 18.48 & +51 32 04.63 & G & 0.34048 $\pm$ 0.00003$^{\dag}$ & 177 & 0.89 & 4246 $\pm$ 849 & 16.50 $\pm$ 3.31 & 0.83 $\pm$ 0.09 & II & 18.52 $\pm$ 0.02 & - \\
172 & 14 07 19.96 & +55 06 01.29 & G & 0.33224 $\pm$ 0.00010$^{\dag}$ & 180 & 0.88 & 776 $\pm$ 155 & 2.79 $\pm$ 0.56 & - & II & 17.09 $\pm$ 0.02 & - \\
173 & 14 08 32.49 & +47 38 37.42 & Q & 1.43633 $\pm$ 0.00052$^{\dag}$ & 85 & 0.74 & 282 $\pm$ 56 & 32.50 $\pm$ 6.60 & 0.83 $\pm$ 0.09 & II & 18.90 $\pm$ 0.01 & - \\
174 & 14 09 36.64 & +53 40 27.64 & G & 0.52484 $\pm$ 0.00015$^{\dag}$ & 240 & 1.54 & 819 $\pm$ 164 & 9.06 $\pm$ 1.82 & 0.93 $\pm$ 0.09 & II & 19.98 $\pm$ 0.05 & - \\
175 & 14 11 17.03 & +47 10 29.47 & G & 0.39350 $\pm$ 0.00013$^{\dag}$ & 300 & 1.64 & 62 $\pm$ 12 & 0.30 $\pm$ 0.06 & 0.53 $\pm$ 0.10 & I & 18.78 $\pm$ 0.02 & - \\
176 & 14 11 51.97 & +55 09 48.74 & Q & 1.89670 $\pm$ 0.00019$^{\dag}$ & 99 & 0.86 & 559 $\pm$ 112 & 126.00 $\pm$ 25.80 & 0.85 $\pm$ 0.09 & II & 18.83 $\pm$ 0.01 & - \\
177 & 14 14 08.45 & +48 41 56.11 & G & 1.361 $\pm$ 0.130$^{\mathsection}$ & 107 & 0.92 & 1055 $\pm$ 211 & 93.40 $\pm$ 29.00 & 0.67 $\pm$ 0.09 & II & 17.90 $\pm$ 0.01 & - \\
178 & 14 15 04.70 & +46 34 28.97 & G & 0.220 $\pm$ 0.029$^{\mathsection}$ & 332 & 1.22 & 129 $\pm$ 26 & 0.18 $\pm$ 0.07 & 0.65 $\pm$ 0.09 & II & 18.13 $\pm$ 0.01 & - \\
179 & 14 15 54.37 & +49 09 21.15 & Q & 1.37310 $\pm$ 0.00037$^{\dag}$ & 111 & 0.96 & 141 $\pm$ 28 & 14.40 $\pm$ 2.93 & 0.82 $\pm$ 0.09 & II & 19.03 $\pm$ 0.01 & 2 \\
180 & 14 16 06.64 & +47 09 27.72 & G & 0.534 $\pm$ 0.131$^{\mathsection}$ & 117 & 0.76 & 3 $\pm$ 1 & 0.03 $\pm$ 0.02 & $<$0.51 & II & 20.30 $\pm$ 0.07 & - \\
181 & 14 16 25.89 & +54 25 45.85 & G & 0.24474 $\pm$ 0.00005$^{\dag}$ & 336 & 1.33 & 797 $\pm$ 159 & 1.59 $\pm$ 0.32 & 1.20 $\pm$ 0.09 & II & 18.02 $\pm$ 0.01 & - \\
182 & 14 18 03.59 & +48 36 04.61 & G & 0.51396 $\pm$ 0.00019$^{\dag}$ & 134 & 0.86 & 2 $\pm$ 1 & 0.03 $\pm$ 0.01 & $<$1.46 & II & 20.35 $\pm$ 0.04 & - \\
183 & 14 18 05.89 & +45 49 06.66 & G & 0.85825 $\pm$ 0.00024$^{\dag}$ & 142 & 1.13 & 129 $\pm$ 26 & 4.50 $\pm$ 0.91 & 0.88 $\pm$ 0.10 & II & 22.02 $\pm$ 0.17 & - \\
184 & 14 19 35.98 & +48 37 43.22 & Q & 0.49599 $\pm$ 0.00006$^{\dag}$ & 127 & 0.79 & 552 $\pm$ 110 & 4.77 $\pm$ 0.96 & 0.65 $\pm$ 0.09 & II & 19.11 $\pm$ 0.03 & - \\
185 & 14 20 56.84 & +53 13 07.45 & G & 0.74211 $\pm$ 0.00010$^{\dag}$ & 101 & 0.76 & 153 $\pm$ 31 & 4.19 $\pm$ 0.84 & 1.06 $\pm$ 0.09 & II & 19.82 $\pm$ 0.02 & - \\
186 & 14 21 22.63 & +48 55 22.82 & G & 0.47680 $\pm$ 0.00008$^{\dag}$ & 124 & 0.76 & 661 $\pm$ 132 & 5.82 $\pm$ 1.17 & 0.93 $\pm$ 0.09 & II & 19.79 $\pm$ 0.03 & - \\
187 & 14 23 14.59 & +47 51 26.99 & G & 0.666 $\pm$ 0.110$^{\mathsection}$ & 197 & 1.42 & 122 $\pm$ 24 & 2.18 $\pm$ 0.98 & - & II & 20.74 $\pm$ 0.06 & - \\
188 & 14 25 36.30 & +53 12 48.03 & Q & 0.91084 $\pm$ 0.00020$^{\dag}$ & 139 & 1.12 & 97 $\pm$ 20 & 3.61 $\pm$ 0.73 & - & II & 19.35 $\pm$ 0.01 & - \\
189 & 14 26 03.30 & +51 29 37.38 & G & 0.41955 $\pm$ 0.15104$^{\star}$ & 138 & 0.78 & 471 $\pm$ 94 & 3.01 $\pm$ 2.63 & 0.88 $\pm$ 0.09 & II & 22.88 $\pm$ 0.30 & - \\
190 & 14 27 42.22 & +47 18 47.17 & G & 0.81534 $\pm$ 0.00045$^{\dag}$ & 115 & 0.90 & 16 $\pm$ 3 & 0.36 $\pm$ 0.07 & 0.41 $\pm$ 0.11 & II & 21.72 $\pm$ 0.16 & - \\
191 & 14 28 57.40 & +54 23 52.99 & G & 0.40573 $\pm$ 0.00011$^{\dag}$ & 127 & 0.71 & 66 $\pm$ 13 & 0.40 $\pm$ 0.08 & 0.97 $\pm$ 0.11 & II & 18.54 $\pm$ 0.02 & - \\
192 & 14 28 57.66 & +54 36 27.81 & G & 0.38148 $\pm$ 0.00012$^{\dag}$ & 200 & 1.08 & 251 $\pm$ 50 & 1.46 $\pm$ 0.29 & 1.27 $\pm$ 0.11 & II & 19.16 $\pm$ 0.02 & - \\
193 & 14 29 14.03 & +46 56 11.32 & G & 0.77764 $\pm$ 0.00027$^{\dag}$ & 130 & 0.99 & 20 $\pm$ 4 & 0.43 $\pm$ 0.09 & 0.45 $\pm$ 0.10 & III & 21.93 $\pm$ 0.12 & - \\
194 & 14 29 33.45 & +54 43 35.29 & G & 0.12328 $\pm$ 0.00002$^{\dag}$ & 452 & 1.03 & 797 $\pm$ 159 & 0.32 $\pm$ 0.06 & 0.70 $\pm$ 0.09 & II & 15.89 $\pm$ 0.00 & - \\
195 & 14 29 46.05 & +51 09 58.42 & G & 0.687 $\pm$ 0.161$^{\mathsection}$ & 280 & 2.04 & 530 $\pm$ 106 & 12.50 $\pm$ 7.58 & 1.14 $\pm$ 0.09 & II & 21.90 $\pm$ 0.14 & - \\
196 & 14 30 45.29 & +45 46 33.57 & G & 0.393 $\pm$ 0.042$^{\mathsection}$ & 140 & 0.77 & 133 $\pm$ 27 & 0.72 $\pm$ 0.23 & 0.83 $\pm$ 0.09 & II & 19.76 $\pm$ 0.02 & - \\
197 & 14 30 45.65 & +50 27 10.84 & G & 0.544 $\pm$ 0.058$^{\mathsection}$ & 110 & 0.72 & 109 $\pm$ 22 & 1.22 $\pm$ 0.40 & 0.75 $\pm$ 0.09 & II & 20.06 $\pm$ 0.05 & - \\
198 & 14 31 36.99 & +52 27 24.90 & G & 0.29198 $\pm$ 0.00007$^{\dag}$ & 186 & 0.84 & 371 $\pm$ 74 & 0.97 $\pm$ 0.19 & 0.62 $\pm$ 0.09 & II & 17.57 $\pm$ 0.01 & - \\
199 & 14 31 41.21 & +47 44 35.15 & Q & 1.17148 $\pm$ 0.00052$^{\dag}$ & 91 & 0.77 & 205 $\pm$ 41 & 13.70 $\pm$ 2.78 & - & II & 20.77 $\pm$ 0.04 & - \\
200 & 14 33 44.95 & +48 08 19.45 & G & 0.69759 $\pm$ 0.00023$^{\dag}$ & 108 & 0.79 & 350 $\pm$ 70 & 7.21 $\pm$ 1.45 & 0.82 $\pm$ 0.09 & II & 21.63 $\pm$ 0.13 & - \\
201 & 14 34 41.85 & +48 03 40.54 & Q & 1.61238 $\pm$ 0.00041$^{\dag}$ & 124 & 1.08 & 147 $\pm$ 29 & 20.60 $\pm$ 4.21 & - & II & 20.96 $\pm$ 0.04 & - \\
202 & 14 35 10.3 & +49 48 19.49 & Q & 0.16606 $\pm$ 0.00005$^{\dag}$ & 348 & 1.02 & 71 $\pm$ 14 & 0.05 $\pm$ 0.01 & 0.44 $\pm$ 0.09 & II & 17.14 $\pm$ 0.00 & 10 \\
203 & 14 37 16.47 & +46 30 28.39 & G & 0.683 $\pm$ 0.081$^{\mathsection}$ & 222 & 1.62 & 44 $\pm$ 9 & 0.83 $\pm$ 0.29 & 0.73 $\pm$ 0.11 & III & 21.16 $\pm$ 0.09 & - \\
204 & 14 37 30.77 & +53 35 42.42 & G & 0.50279 $\pm$ 0.00015$^{\dag}$ & 123 & 0.78 & 236 $\pm$ 47 & 2.39 $\pm$ 0.48 & 0.96 $\pm$ 0.09 & II & 20.18 $\pm$ 0.04 & - \\
205 & 14 37 37.66 & +51 44 46.45 & G & 0.49518 $\pm$ 0.21509$^{\star}$ & 207 & 1.30 & 244 $\pm$ 49 & 2.35 $\pm$ 2.48 & 0.93 $\pm$ 0.09 & II & 21.42 $\pm$ 0.07 & - \\
206 & 14 38 04.03 & +46 50 51.20 & G & 0.74546 $\pm$ 0.00023$^{\dag}$ & 115 & 0.87 & 8 $\pm$ 2 & 0.17 $\pm$ 0.04 & 0.57 $\pm$ 0.16 & II & 21.17 $\pm$ 0.10 & - \\
207 & 14 40 07.85 & +55 27 06.08 & G & 0.58442 $\pm$ 0.06106$^{\star}$ & 238 & 1.62 & 93 $\pm$ 19 & 1.19 $\pm$ 0.38 & 0.69 $\pm$ 0.10 & II & 20.93 $\pm$ 0.06 & - \\
208 & 14 40 13.98 & +50 26 01.87 & Q & 1.57000 $\pm$ 0.00108$^{\dag}$ & 143 & 1.24 & 136 $\pm$ 27 & 22.00 $\pm$ 4.49 & 0.96 $\pm$ 0.10 & II & 20.65 $\pm$ 0.05 & - \\
209 & 14 40 17.49 & +53 26 20.29 & G & 0.52672 $\pm$ 0.00013$^{\dag}$ & 195 & 1.26 & 348 $\pm$ 70 & 3.60 $\pm$ 0.72 & - & II & 20.13 $\pm$ 0.04 & - \\
210 & 14 42 15.00 & +47 06 46.09 & G & 0.44503 $\pm$ 0.00011$^{\dag}$ & 121 & 0.71 & 197 $\pm$ 39 & 1.33 $\pm$ 0.27 & 0.65 $\pm$ 0.09 & II & 19.85 $\pm$ 0.03 & - \\
211 & 14 42 58.59 & +52 09 02.07 & G & 0.272 $\pm$ 0.057$^{\mathsection}$ & 289 & 1.24 & 824 $\pm$ 165 & 1.94 $\pm$ 1.02 & 0.88 $\pm$ 0.09 & II & 17.88 $\pm$ 0.01 & - \\
212 & 14 43 08.22 & +51 18 10.01 & G & 0.46603 $\pm$ 0.00011$^{\dag}$ & 152 & 0.92 & 7 $\pm$ 1 & 0.06 $\pm$ 0.01 & $<$1.10 & II & 19.63 $\pm$ 0.03 & - \\
213 & 14 43 32.79 & +45 46 54.56 & G & 0.60401 $\pm$ 0.00023$^{\dag}$ & 117 & 0.81 & 44 $\pm$ 9 & 0.59 $\pm$ 0.12 & 0.61 $\pm$ 0.10 & II & 20.44 $\pm$ 0.07 & - \\
214 & 14 44 07.52 & +49 07 30.78 & Q & 0.91579 $\pm$ 0.00009$^{\dag}$ & 193 & 1.55 & 175 $\pm$ 35 & 7.59 $\pm$ 1.53 & 0.97 $\pm$ 0.09 & II & 22.01 $\pm$ 0.15 & - \\
215 & 14 44 10.50 & +55 47 45.64 & G & 0.33475 $\pm$ 0.00005$^{\dag}$ & 144 & 0.71 & 542 $\pm$ 108 & 1.97 $\pm$ 0.39 & 0.72 $\pm$ 0.09 & II & 18.15 $\pm$ 0.01 & - \\
216 & 14 44 58.09 & +53 09 29.6 & G & 0.38228 $\pm$ 0.13684$^{\star}$ & 167 & 0.90 & 184 $\pm$ 37 & 0.82 $\pm$ 0.71 & 0.44 $\pm$ 0.09 & III & 19.47 $\pm$ 0.03 & - \\
217 & 14 45 20.87 & +54 03 29.62 & G & 0.43345 $\pm$ 0.00013$^{\dag}$ & 162 & 0.94 & 127 $\pm$ 26 & 0.81 $\pm$ 0.16 & 0.66 $\pm$ 0.09 & II & 18.72 $\pm$ 0.02 & - \\
218 & 14 46 07.20 & +48 41 37.79 & G & 0.37676 $\pm$ 0.00006$^{\dag}$ & 146 & 0.78 & 174 $\pm$ 35 & 0.87 $\pm$ 0.17 & 0.89 $\pm$ 0.09 & II & 19.10 $\pm$ 0.02 & - \\
219 & 14 46 48.98 & +47 36 31.43 & G & 0.584 $\pm$ 0.052$^{\mathsection}$ & 274 & 1.86 & 59 $\pm$ 12 & 0.74 $\pm$ 0.22 & 0.66 $\pm$ 0.11 & II & 20.48 $\pm$ 0.05 & - \\
220 & 14 47 25.82 & +56 37 50.13 & G & 0.71650 $\pm$ 0.00025$^{\dag}$ & 105 & 0.78 & 12 $\pm$ 3 & 0.23 $\pm$ 0.05 & $<$0.50 & III & 21.32 $\pm$ 0.09 & - \\
221 & 14 50 02.36 & +54 05 28.27 & G & 0.549 $\pm$ 0.076$^{\mathsection}$ & 126 & 0.83 & 507 $\pm$ 101 & 5.78 $\pm$ 2.26 & - & II & 20.51 $\pm$ 0.05 & - \\
222 & 14 50 57.28 & +53 00 07.76 & Q & 0.91800 $\pm$ 0.00008$^{\dag}$ & 184 & 1.48 & 1407 $\pm$ 281 & 56.50 $\pm$ 11.40 & 0.84 $\pm$ 0.09 & II & 20.59 $\pm$ 0.04 & - \\
223 & 14 51 06.41 & +53 33 53.84 & Q & 0.43220 $\pm$ 0.00007$^{\dag}$ & 246 & 1.43 & 267 $\pm$ 53 & 1.78 $\pm$ 0.36 & 0.80 $\pm$ 0.09 & II & 18.64 $\pm$ 0.01 & 2 \\
224 & 14 51 16.59 & +51 13 33.28 & G & 0.509 $\pm$ 0.050$^{\mathsection}$ & 194 & 1.23 & 99 $\pm$ 20 & 0.95 $\pm$ 0.29 & - & II & 19.92 $\pm$ 0.05 & - \\
225 & 14 51 21.28 & +56 08 35.45 & G & 0.525 $\pm$ 0.073$^{\mathsection}$ & 208 & 1.34 & 57 $\pm$ 12 & 0.57 $\pm$ 0.22 & 0.68 $\pm$ 0.10 & II & 20.73 $\pm$ 0.06 & - \\
226 & 14 52 50.73 & +51 37 01.90 & G & 0.652 $\pm$ 0.126$^{\mathsection}$ & 108 & 0.77 & 51 $\pm$ 10 & 0.97 $\pm$ 0.50 & 1.00 $\pm$ 0.14 & II & 21.64 $\pm$ 0.11 & - \\
227 & 14 57 02.81 & +48 06 46.64 & G & 0.26738 $\pm$ 0.00007$^{\dag}$ & 168 & 0.71 & 5 $\pm$ 1 & 0.01 $\pm$ 0.00 & $<$1.11 & II & 18.00 $\pm$ 0.01 & - \\
228 & 14 57 49.59 & +50 45 59.62 & G & 0.628 $\pm$ 0.146$^{\mathsection}$ & 199 & 1.40 & 55 $\pm$ 11 & 0.90 $\pm$ 0.54 & 0.88 $\pm$ 0.11 & II & 21.58 $\pm$ 0.11 & - \\
229 & 15 00 00.56 & +50 43 55.62 & G & 0.415 $\pm$ 0.087$^{\mathsection}$ & 142 & 0.80 & 12 $\pm$ 2 & 0.07 $\pm$ 0.04 & $<$0.83 & II & 19.70 $\pm$ 0.03 & - \\
230 & 15 01 32.11 & +50 34 55.14 & G & 0.32062 $\pm$ 0.00007$^{\dag}$ & 164 & 0.79 & 32 $\pm$ 6 & 0.10 $\pm$ 0.02 & 0.73 $\pm$ 0.11 & II & 17.66 $\pm$ 0.01 & - \\
231 & 15 01 48.47 & +52 10 47.16 & G & 0.492 $\pm$ 0.058$^{\mathsection}$ & 145 & 0.91 & 10 $\pm$ 2 & 0.09 $\pm$ 0.03 & 0.67 $\pm$ 0.16 & II & 19.68 $\pm$ 0.04 & - \\
232 & 15 03 33.65 & +48 11 17.67 & G & 0.672 $\pm$ 0.095$^{\mathsection}$ & 111 & 0.80 & 80 $\pm$ 16 & 1.81 $\pm$ 0.72 & 1.16 $\pm$ 0.13 & II & 21.52 $\pm$ 0.11 & - \\
233 & 15 04 45.61 & +50 30 08.57 & G & 0.65227 $\pm$ 0.05463$^{\star}$ & 201 & 1.44 & 1295 $\pm$ 259 & 23.90 $\pm$ 6.83 & 0.91 $\pm$ 0.09 & II & 21.89 $\pm$ 0.17 & - \\
234 & 15 06 12.81 & +51 37 07.11 & G & 0.61121 $\pm$ 0.00010$^{\dag}$ & 262 & 1.82 & 574 $\pm$ 115 & 8.38 $\pm$ 1.69 & - & II & 20.25 $\pm$ 0.05 & - \\
235 & 15 06 24.10 & +53 55 02.61 & G & 0.30437 $\pm$ 0.00005$^{\dag}$ & 184 & 0.85 & 118 $\pm$ 24 & 0.33 $\pm$ 0.07 & 0.57 $\pm$ 0.09 & II & 17.95 $\pm$ 0.01 & - \\
236 & 15 18 30.19 & +51 58 17.03 & G & 0.708 $\pm$ 0.168$^{\mathsection}$ & 105 & 0.78 & 51 $\pm$ 10 & 1.17 $\pm$ 0.72 & 0.95 $\pm$ 0.11 & II & 21.84 $\pm$ 0.13 & - \\
237 & 15 18 35.37 & +51 04 10.70 & G & 0.619 $\pm$ 0.138$^{\mathsection}$ & 119 & 0.83 & 371 $\pm$ 74 & 5.63 $\pm$ 3.25 & 0.77 $\pm$ 0.09 & II & 21.14 $\pm$ 0.08 & - \\
238 & 15 18 45.36 & +52 37 08.56 & G & 0.52105 $\pm$ 0.00011$^{\dag}$ & 110 & 0.71 & 312 $\pm$ 62 & 3.66 $\pm$ 0.73 & 1.11 $\pm$ 0.09 & II & 20.24 $\pm$ 0.05 & - \\
239 & 15 23 11.09 & +52 03 03.52 & Q & 0.51667 $\pm$ 0.00010$^{\dag}$ & 132 & 0.84 & 382 $\pm$ 77 & 4.35 $\pm$ 0.87 & 1.09 $\pm$ 0.09 & II & 20.74 $\pm$ 0.06 & - \\

\hline
\end{longtable}
\tablebib{(1)~\citet{amir15}; (2) \citet{amirgrgs}; (3) \citet{cotter96grg}; (4) \citet{koziel11};
(5) \citet{m98}; (6) \citet{Machalski01grgs}; (7) \citet{Nilsson98} ; (8) \citet{rente} ; (9) \citet{thwala19} ; (10) \citet{grscat}}
\end{tiny}
\clearpage
\twocolumn

\clearpage
\onecolumn
\setlength{\tabcolsep}{3.7pt}
\begin{table}
\captionsetup{width=\textwidth}
\caption{LoTSS GRGs in galaxy clusters: Parameters $r_{200}$ and $R_{L*}$ (cluster richness parameter) have been taken from the WHL galaxy cluster catalog and  $M_{200}$ has been computed from Eq.2 
of \citet{whl}. GRGs marked with 
$^{\dag}$ in the redshift column have photometric redshift estimate and the rest have spectroscopic redshifts. $r_{200}$ is the radius within which the mean density of a cluster is 200 times of the 
critical density of the universe, and 
$M_{200}$, is the mass of the cluster within $r_{200}$. $N_{200}$ is the number of galaxies within $r_{200}$. Clusters 6, 7 and 20 are identified from GMBCG cluster catalog \citep{gmbcg}.}
\label{tab:clusters}
\centering
\begin{tabular}{c c c c c c c c c}
\hline\hline \\
No & Cluster Name  & GRG RA & GRG Dec & $z$  & $r_{200}$ & $R_{L*}$ & $M_{200}$ & $N_{200}$ \\ 
\hline \\
  &   &    &     &     & (Mpc)  &   & ($10^{14}$ $M_{\odot}$) &  \\ 
  \hline
1 & WHLJ112126.4+534457 & 11 21 26.44 & +53 44 56.71 & 0.10378 $\pm$ 0.00002 & 1.7 & 69.4 & 4.6 & 46 \\
2 & WHLJ113250.7+505705 & 11 32 50.67 & +50 57 04.68 & 0.35857 $\pm$ 0.00010 & 0.9 & 16.9 & 0.9 & 14 \\
3 & WHLJ113503.2+482612 & 11 35 03.20 & +48 26 12.12 & 0.22597 $\pm$ 0.00006 & 1.1 & 21.7 & 1.2 & 17 \\
4 & WHLJ121900.8+505255 & 12 19 00.76 & +50 52 54.41 & 0.38509 $\pm$ 0.00008 & 0.9 & 17.3 & 0.9 & 14 \\
5 & GMBCG J188.75636+53.29864 & 12 35 01.52 & +53 17 55.09 & 0.34480 $\pm$ 0.00030 & - & - & - & - \\
6 & GMBCG J191.35486+47.11948 & 12 45 25.16 & +47 07 10.11 & 0.51179 $\pm$ 0.00015 & - & - & - & - \\
7 & WHLJ124913.2+500044 & 12 49 13.20 & +50 00 43.68 & 0.34956 $\pm$ 0.00010 & 1.1 & 25.3 & 1.4 & 30 \\
8 &WHL J125142.1+503425 & 12 51 42.04 & +50 34 24.65 & 0.54904 $\pm$ 0.00007 & 1.1 & 24.3 & 1.4 & 11 \\
9 &WHL J131404.6+543938 & 13 14 04.60 & +54 39 37.88 & 0.34645 $\pm$ 0.00008 & 1.2 & 26.1 & 1.5 & 23 \\
10 & WHLJ132229.1+504845 & 13 22 29.07 & +50 48 44.53 & 0.18439 $\pm$ 0.00004 & 0.9 & 15.9 & 0.8 & 11 \\
11 & WHLJ133258.3+535356 & 13 32 58.28 & +53 53 55.60 & 0.32261 $\pm$ 0.00009 & 0.9 & 16.6 & 0.9 & 13 \\
12 & WHLJ133512.0+494427 & 13 35 12.01 & +49 44 27.20 & 0.436 $\pm$ 0.041$^{\dag}$ & 0.9 & 14.7 & 0.7 & 9 \\
13 & WHLJ133618.8+533952 & 13 36 18.75 & +53 39 52.12 & 0.30138 $\pm$ 0.00006 & 0.9 & 13.4 & 0.7 & 11 \\
14 & WHLJ134207.0+472553 & 13 42 06.98 & +47 25 53.04 & 0.17213 $\pm$ 0.00004 & 0.9 & 13.2 & 0.7 & 13 \\
15 & WHLJ142857.4+542353 & 14 28 57.40 & +54 23 52.99 & 0.40573 $\pm$ 0.00011 & 1.2 & 30.7 & 1.8 & 22 \\
16 & WHLJ142933.4+544336 & 14 29 33.45 & +54 43 35.29 & 0.12328 $\pm$ 0.00002 & 1.3 & 34.7 & 2.0 & 29 \\
17 & WHLJ144410.5+554746 & 14 44 10.50 & +55 47 45.64 & 0.33475 $\pm$ 0.00005 & 1.2 & 30.1 & 1.7 & 26 \\
18 & WHLJ144520.9+540330 & 14 45 20.87 & +54 03 29.62 & 0.43345 $\pm$ 0.00013 & 1.0 & 25.1 & 1.4 & 18 \\
19 &  GMBCG J222.81919+51.22588 & 14 51 16.59 & +51 13 33.28 & 0.509 $\pm$ 0.050$^{\dag}$ & - & - & - & - \\
20 & WHLJ150624.1+535503 & 15 06 24.10 & +53 55 02.61 & 0.30437 $\pm$ 0.00005 & 1.0 & 19.1 & 1.0 & 11 \\
\hline
\end{tabular}
\end{table}
\clearpage
\twocolumn

\begin{acknowledgements}
We acknowledge the comments and suggestions of the anonymous referee which has helped us improve the quality of our paper. 
PD and JB gratefully acknowledge generous support from the 
Indo-French Center for the Promotion of Advanced Research (Centre Franco-Indien pour la Promotion de la Recherche Avan\'{c}ee) under programme no. 5204-2 and thank IUCAA and IUCAA Radio Physics Lab (RPL) for financial and logistic support. This research has made use of the Dutch national e-infrastructure with support of the SURF Cooperative (e-infra 180169) and the LOFAR e-infra group. The Jülich LOFAR Long Term Archive and the German LOFAR network, are both coordinated and operated by the Jülich Supercomputing Centre (JSC), and computing resources on the Supercomputer JUWELS 
at JSC were provided by the Gauss Centre for Supercomputing e.V. (\url{www.gauss-centre.eu, grant CHTB00}) through the John von Neumann Institute for Computing (NIC). 
HJR gratefully acknowledges generous support from the European Research Council under the European Unions Seventh Framework Programme (FP/2007-2013)/ERC Advanced Grant NEWCLUSTERS-321271. 
MJH acknowledges support from the UK Science and Technology Facilities Council [ST/R000905/1]. This research has made use of the University of Hertfordshire high-performance computing facility (\url{https://uhhpc.herts.ac.uk/}) and the LOFAR-UK compute facility, located at the University of Hertfordshire and supported by STFC [ST/P000096/1]. HJR and KJD acknowledges support from the ERC Advanced Investigator programme NewClusters 321271.
LOFAR, the Low Frequency Array designed and constructed by ASTRON, has facilities in several countries, which are owned by various parties (each with their own funding), and are collectively operated by the International LOFAR Telescope (ILT) foundation under a joint scientific policy.
We thank the LOFAR Galaxy Zoo team.
We gratefully acknowledge the use of Edward (Ned) Wright's online Cosmology Calculator. This research has made use of the NASA Extragalactic Database (NED) which is operated by the Jet Propulsion Laboratory, California Institute of Technology, under contract with the National Aeronautics and 
Space Administration. This research has also made use of the SIMBAD database, operated at CDS, Strasbourg, France. This publication makes use of data products from the Wide-field Infrared Survey 
Explorer, which is a joint project of the University of California, Los Angeles, and the Jet Propulsion Laboratory/California Institute of Technology, funded by the National Aeronautics and Space 
Administration.
\end{acknowledgements}


\bibliographystyle{aa}


\onecolumn
\begin{appendix}
\section{Appendix}
\label{sec:appendix}

\begin{figure*}[ht]
\centering
\includegraphics[scale=0.4]{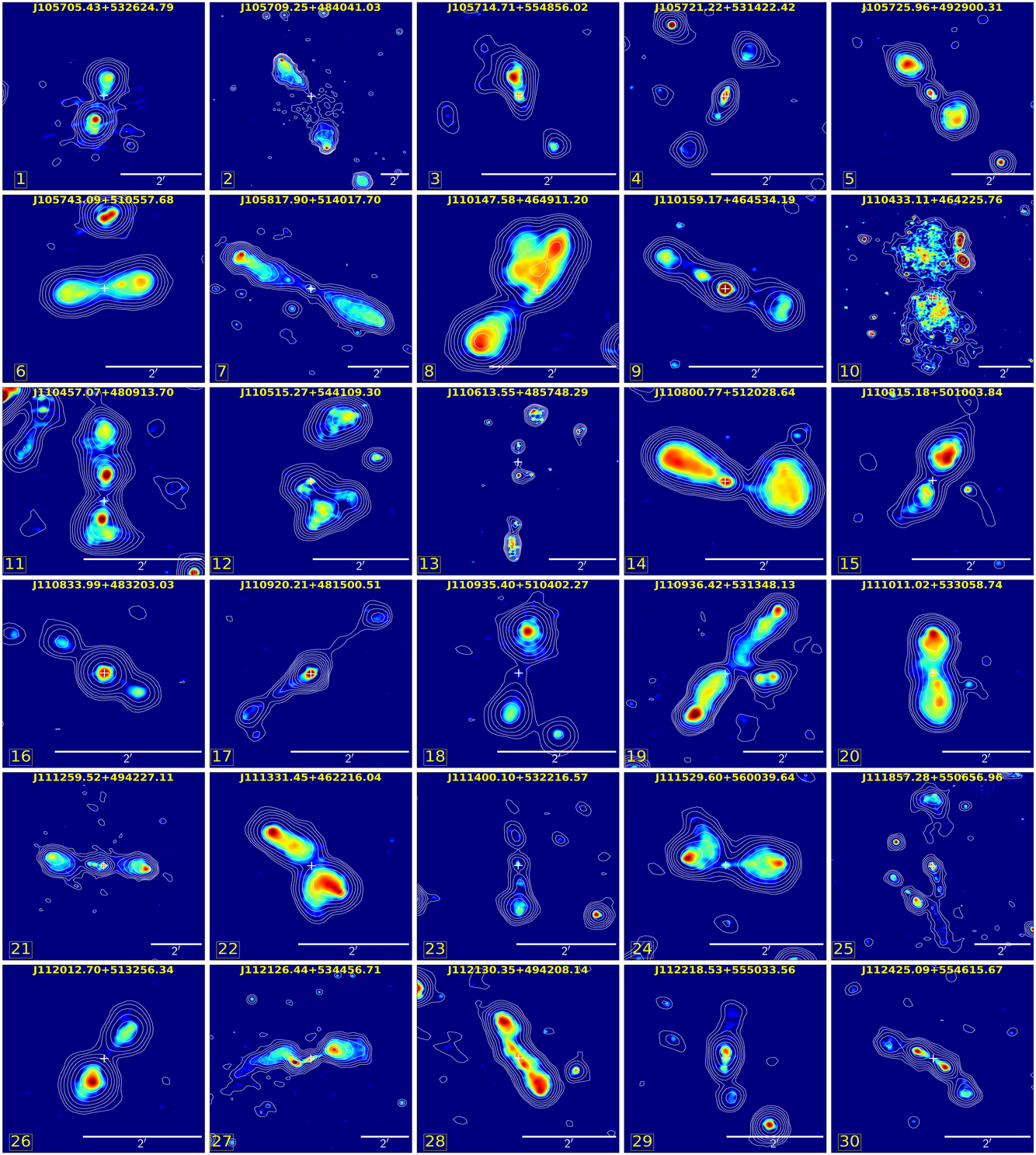} 
\caption{LoTSS high resolution ($\sim$ 6 \arcsec) images 1-30 with LoTSS low resolution ($\sim$ 20 \arcsec) contours overlaid. The cross '+' marker indicates the location of the host galaxy.The 
radio contours are drawn with 8 levels which are chosen by equally (log scale) dividing the data value range above $\sim$ 3$\sigma$, where $\sigma$ is the local RMS of the map.}
\label{fig:lotsshighres1}
\end{figure*}

\newpage
\begin{figure*}[ht]
\centering
\includegraphics[scale=0.4]{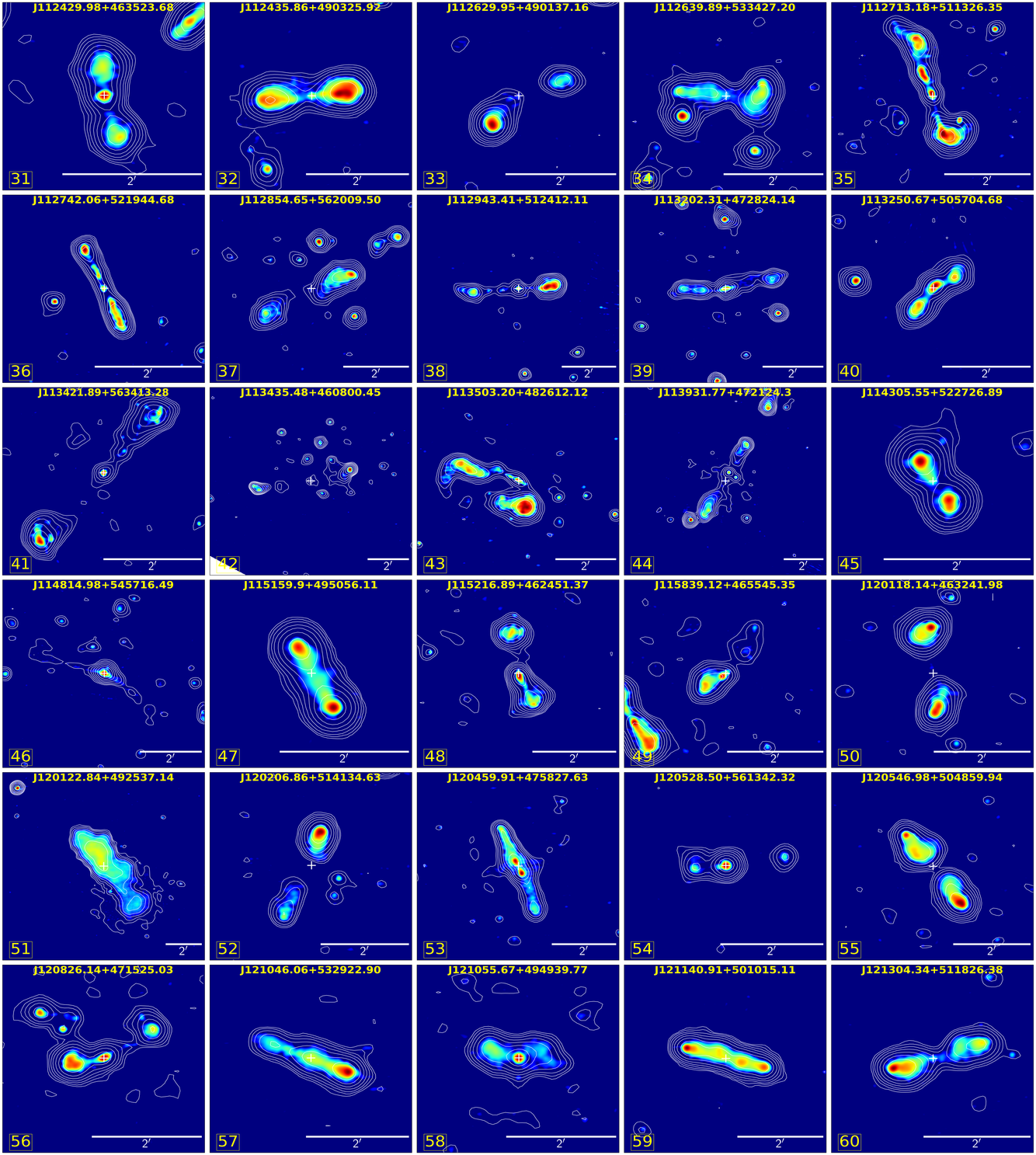}
\caption{LoTSS high resolution ($\sim$ 6 \arcsec) images 31-60 with LoTSS low resolution ($\sim$ 20 \arcsec) contours overlaid.The cross '+' marker indicates the location of the host galaxy. 
The 
radio contours are drawn with 8 levels which are chosen by equally (log scale) dividing the data value range above $\sim$ 3$\sigma$, where $\sigma$ is the local RMS of the map.}
\label{fig:lotsshighres2}
\end{figure*}

\newpage
\begin{figure*}[ht]
\centering
\includegraphics[scale=0.4]{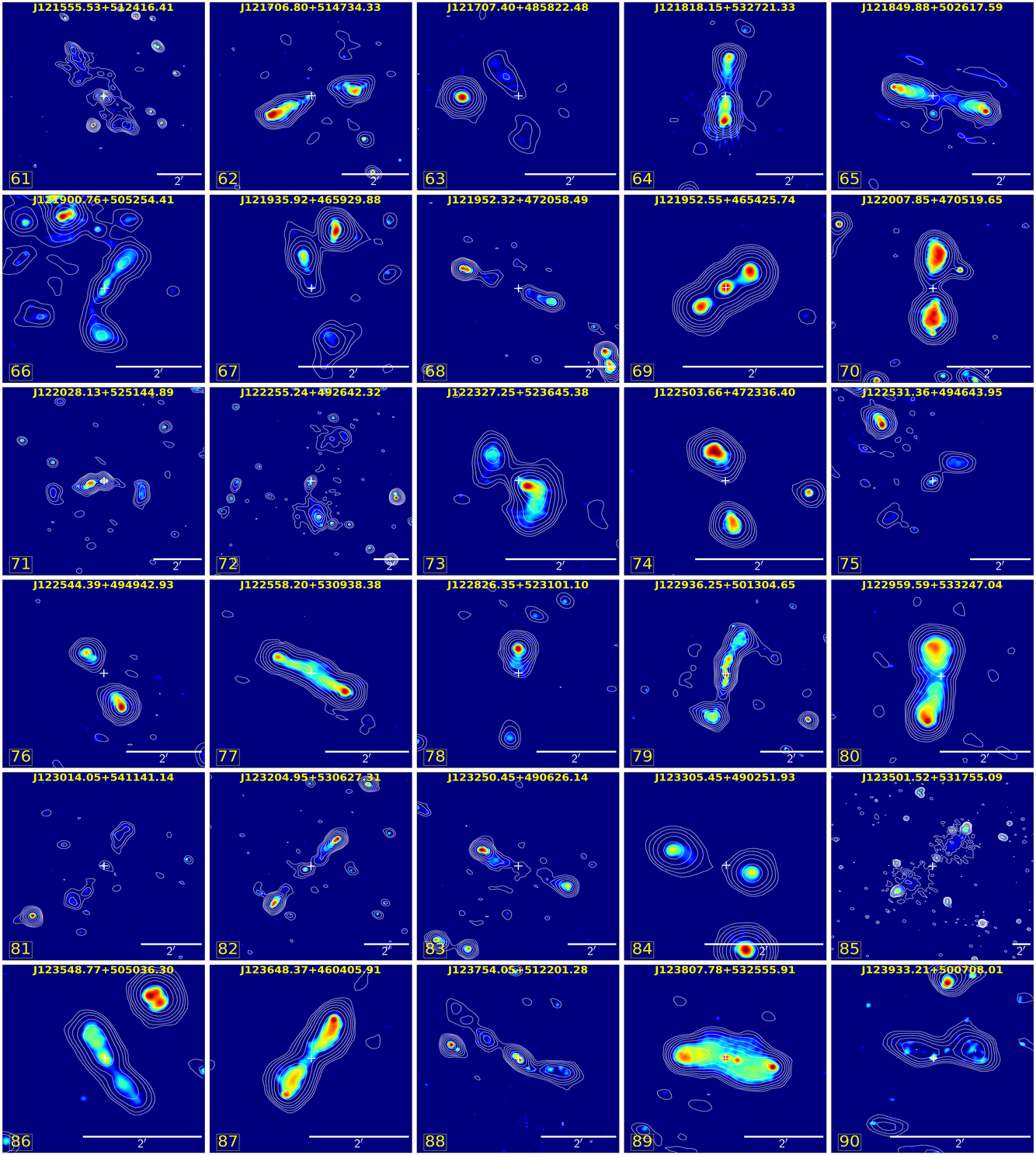} 
\caption{LoTSS high resolution ($\sim$ 6 \arcsec) images 61-90 with LoTSS low resolution ($\sim$ 20 \arcsec) contours overlaid. The cross '+' marker indicates the location of the host galaxy. 
The 
radio contours are drawn with 8 levels which are chosen by equally (log scale) dividing the data value range above $\sim$ 3$\sigma$, where $\sigma$ is the local RMS of the map.}
\label{fig:lotsshighres3}
\end{figure*}

\newpage
\begin{figure*}[ht]
\centering
\includegraphics[scale=0.4]{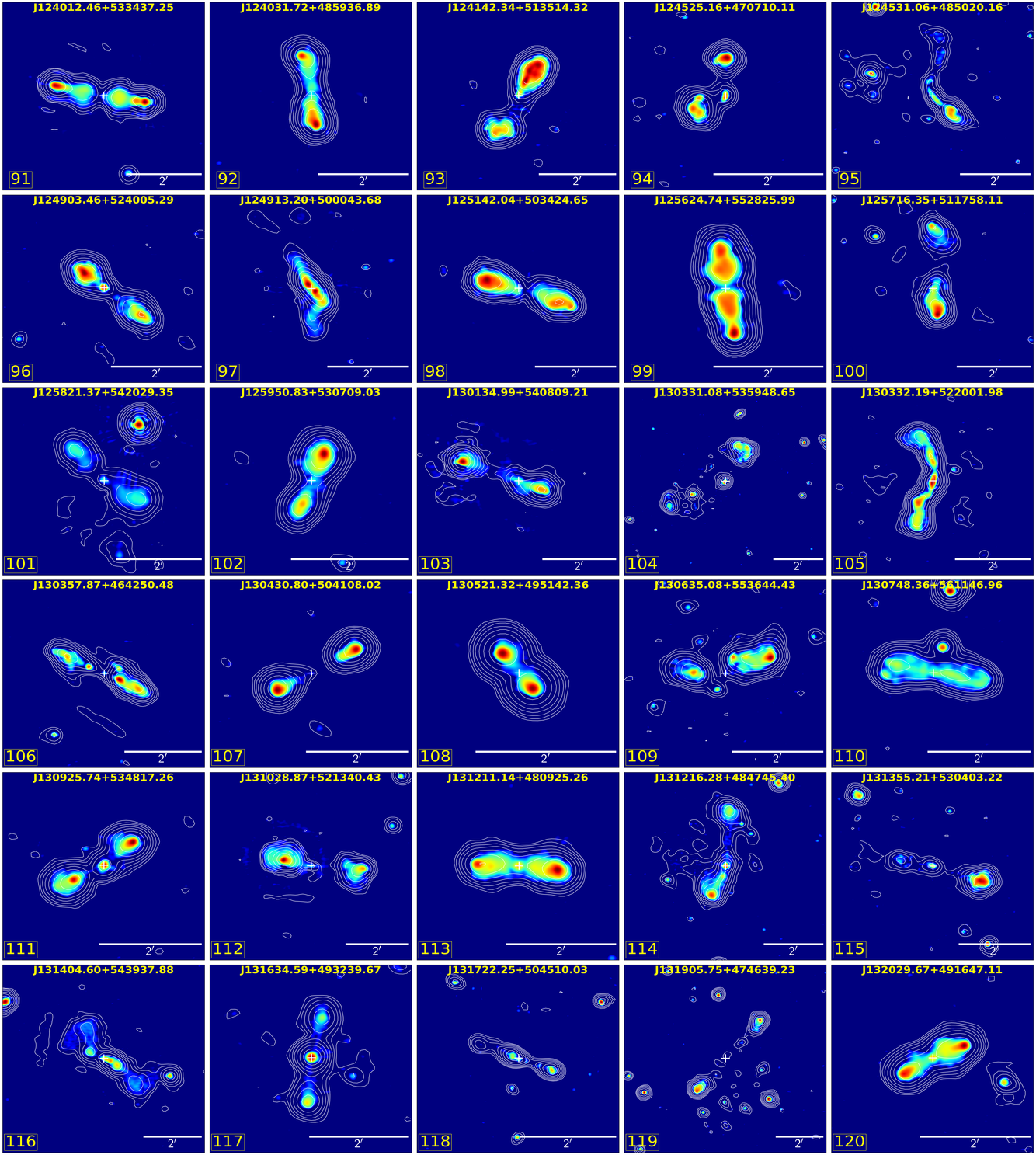}
\caption{LoTSS high resolution ($\sim$ 6 \arcsec) images 91-120 with LoTSS low resolution ($\sim$ 20 \arcsec) contours overlaid. The cross '+' marker indicates the location of the host galaxy. 
The 
radio contours are drawn with 8 levels which are chosen by equally (log scale) dividing the data value range above $\sim$ 3$\sigma$, where $\sigma$ is the local RMS of the map.}
\label{fig:lotsshighres4}
\end{figure*}

\newpage
\begin{figure*}[ht]
\centering
\includegraphics[scale=0.4]{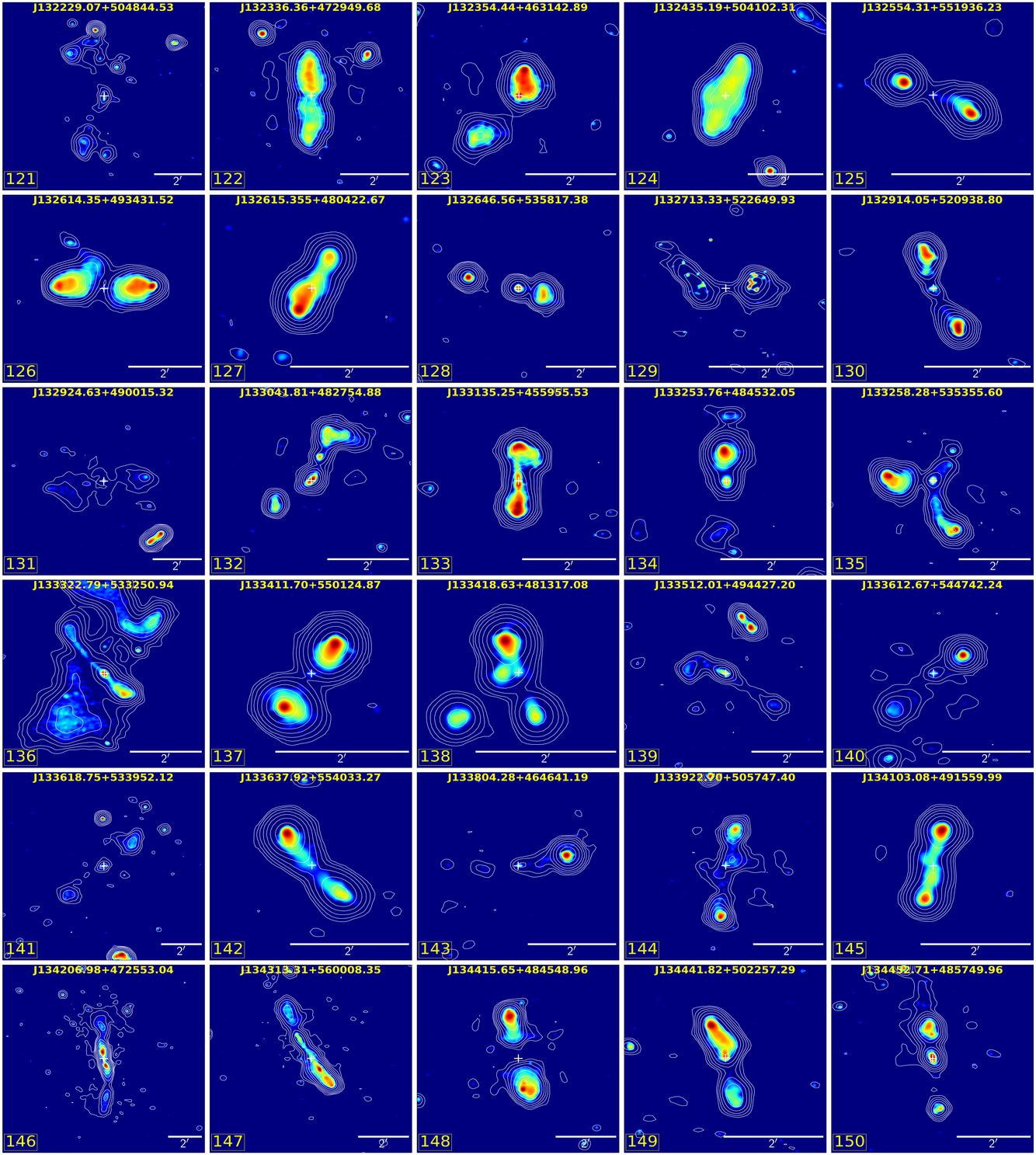} 
\caption{LoTSS high resolution ($\sim$ 6 \arcsec) images 121-150 with LoTSS low resolution ($\sim$ 20 \arcsec) contours overlaid. The cross '+' marker indicates the location of the host 
galaxy. The 
radio contours are drawn with 8 levels which are chosen by equally (log scale) dividing the data value range above $\sim$ 3$\sigma$, where $\sigma$ is the local RMS of the map.}
\label{fig:lotsshighres5}
\end{figure*}

\newpage
\begin{figure*}[ht]
\centering
\includegraphics[scale=0.4]{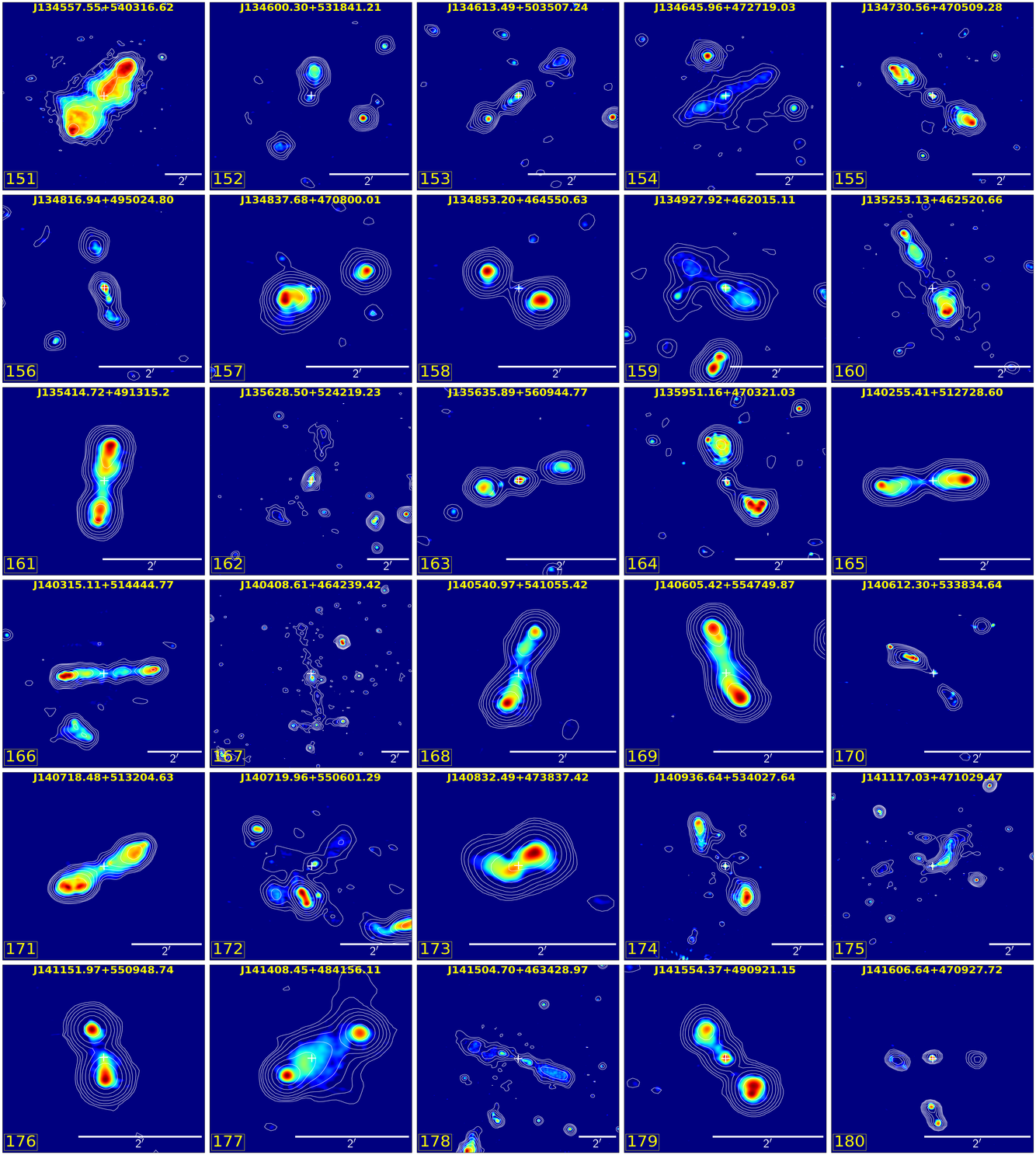}
\caption{LoTSS high resolution ($\sim$ 6 \arcsec) images 151-180 with LoTSS low resolution ($\sim$ 20 \arcsec) contours overlaid. The cross '+' marker indicates the location of the host 
galaxy. The 
radio contours are drawn with 8 levels which are chosen by equally (log scale) dividing the data value range above $\sim$ 3$\sigma$, where $\sigma$ is the local RMS of the map.}
\label{fig:lotsshighres6}
\end{figure*}

\newpage
\begin{figure*}[ht]
\centering
\includegraphics[scale=0.4]{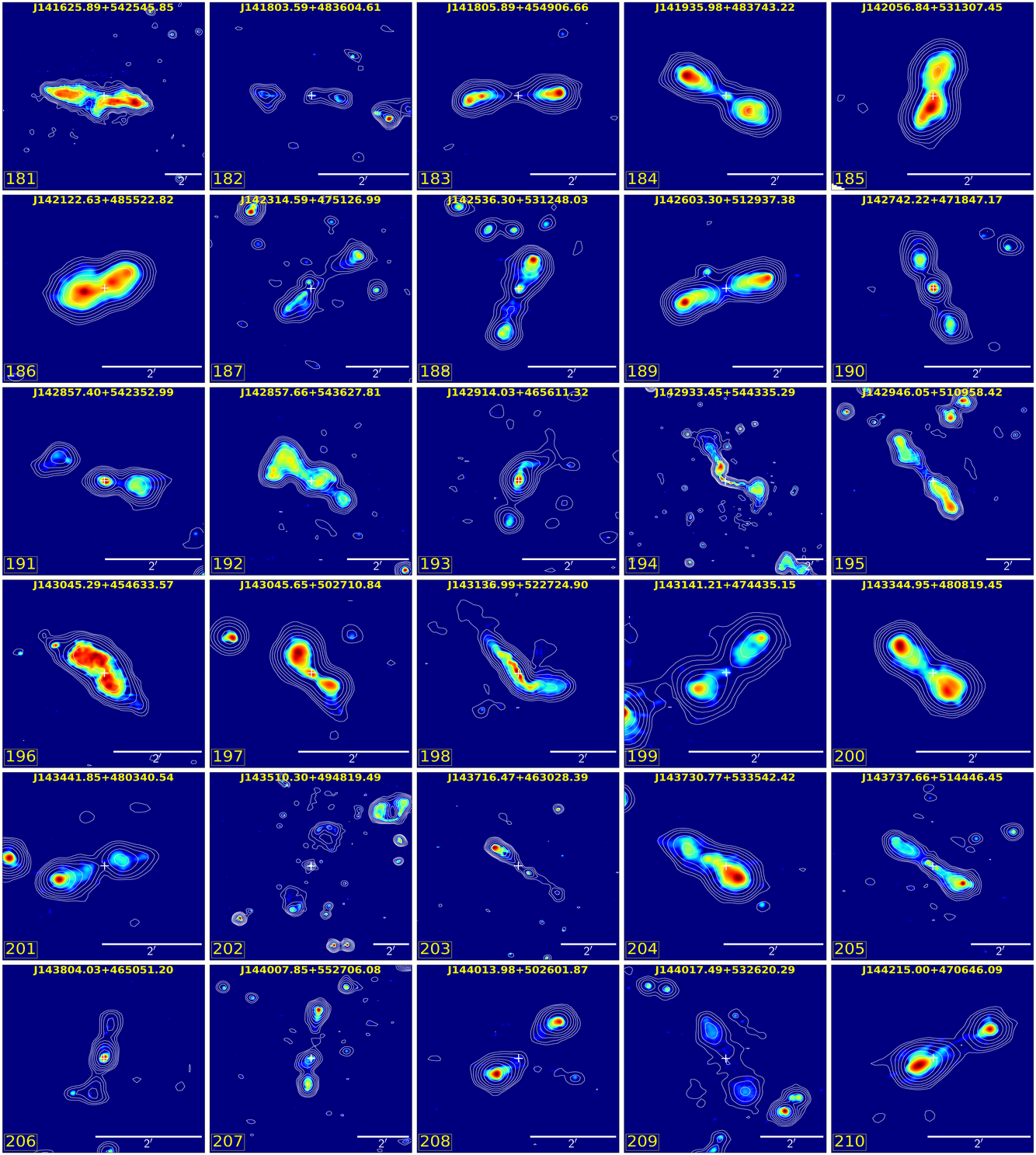} 
\caption{LoTSS high resolution ($\sim$ 6 \arcsec) images 181-210 with LoTSS low resolution ($\sim$ 20 \arcsec) contours overlaid. The cross '+' marker indicates the location of the host 
galaxy. The 
radio contours are drawn with 8 levels which are chosen by equally (log scale) dividing the data value range above $\sim$ 3$\sigma$, where $\sigma$ is the local RMS of the map.}
\label{fig:lotsshighres7}
\end{figure*}

\newpage
\begin{figure*}[ht]
\centering
\includegraphics[scale=0.4]{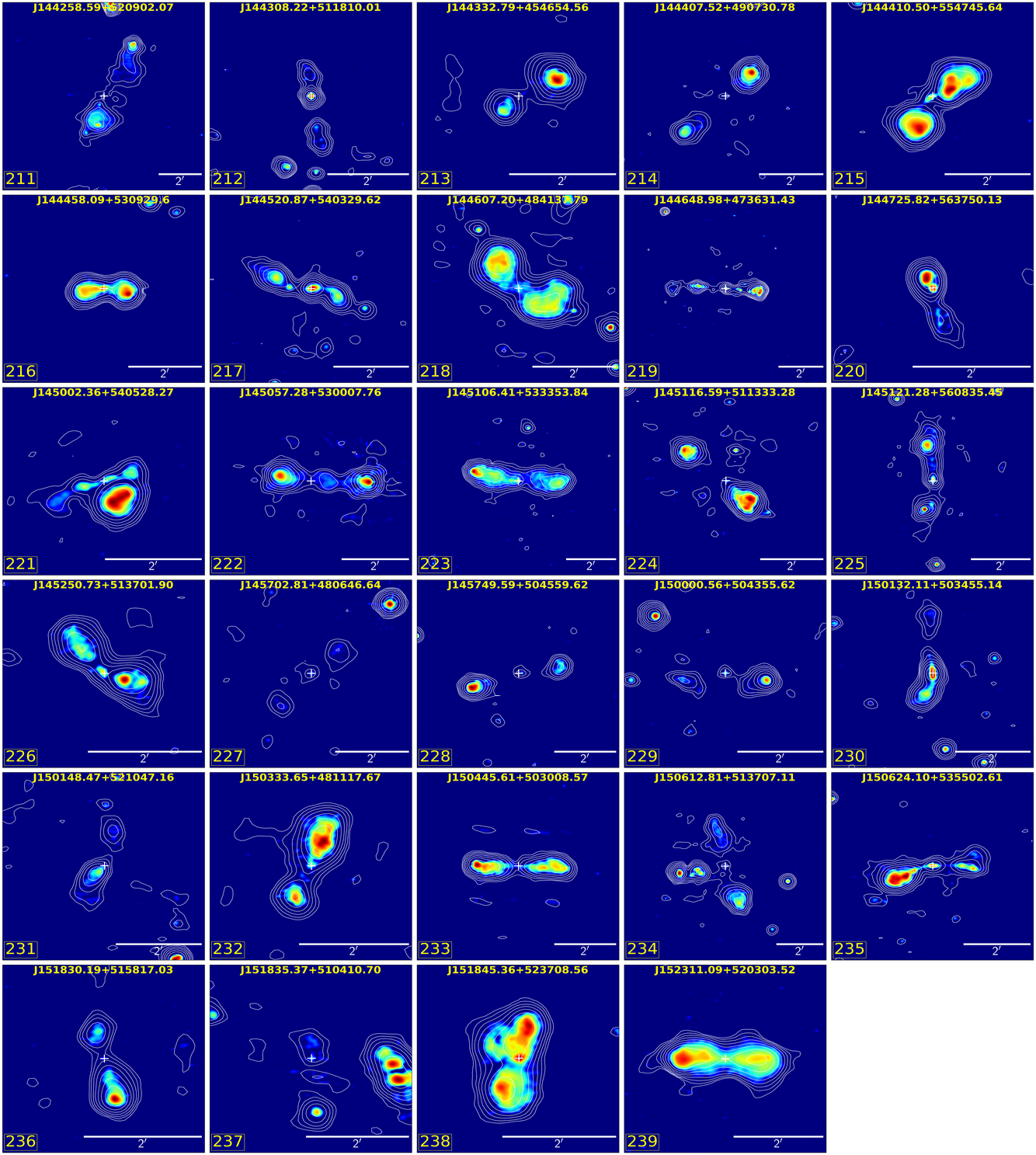}
\caption{LoTSS high resolution ($\sim$ 6 \arcsec) images 211-239 with LoTSS low resolution ($\sim$ 20 \arcsec) contours overlaid. The cross '+' marker indicates the location of the host 
galaxy. The 
radio contours are drawn with 8 levels which are chosen by equally (log scale) dividing the data value range above $\sim$ 3$\sigma$, where $\sigma$ is the local RMS of the map. }
\label{fig:lotsshighres8}
\end{figure*}

\end{appendix}



\end{document}